\documentclass[twocolumn, tighten, numberedappendix, twocolappendix]{aastex63}

\usepackage{amsmath,amstext}
\usepackage[utf8]{inputenc}
\usepackage{graphicx}
\usepackage{ulem}
\providecommand{\sorthelp}[1]{}

\newcommand{\SkyPercentage}{75\%}

\newcommand{\EraOneStart}{2016 September}
\newcommand{\EraOneEnd}{2018 February}

\newcommand{\dg}{$^{\circ}$} 
\newcommand{\degpsec}{$^{\circ}~\mathrm s^{-1}$}
\newcommand{\pseudo}{survey Moon scans}
\newcommand{\bs}{boresight}
\newcommand{\demod}{demodulated}

\usepackage[encapsulated]{CJK}
\usepackage{ucs}
\newcommand{\cntext}[1]{\begin{CJK}{UTF8}{gbsn}#1\end{CJK}}

\usepackage{devanagari}

\shorttitle{CLASS 40 GHz Optical Characterization \& Calibration}
\shortauthors{Xu et al.}

\begin{document}

\title{Two-year Cosmology Large Angular Scale Surveyor (CLASS) Observations:\\ 40~GHz Telescope Pointing, Beam Profile, Window Function, and Polarization Performance}

\author[0000-0001-5112-2567]{Zhilei Xu (\cntext{徐智磊}\!\!)}
\affiliation{Department of Physics and Astronomy, University of Pennsylvania, 209 South 33rd Street, Philadelphia, PA 19104, USA}
\affiliation{Department of Physics and Astronomy, Johns Hopkins University, 3701 San Martin Drive, Baltimore, MD 21218, USA}

\author{Michael~K. Brewer}
\affiliation{Department of Physics and Astronomy, Johns Hopkins University, 3701 San Martin Drive, Baltimore, MD 21218, USA}

\author[0000-0002-2061-0063]{Pedro Flux\'a Rojas}
\affiliation{Instituto de Astrof\'isica, Facultad de F\'isica, Pontificia Universidad Cat\'olica de Chile, Avenida Vicu\~na Mackenna 4860, 7820436, Chile}
\affiliation{Instituto de Astrof\'isica and Centro de Astro-Ingenier\'ia, Facultad de F\'isica, Pontificia Universidad Cat\'olica de Chile, Av. Vicu\~na Mackenna 4860, 7820436 Macul, Santiago, Chile}

\author[0000-0002-4820-1122]{Yunyang Li (\cntext{李云炀}\!\!)}
\affiliation{Department of Physics and Astronomy, Johns Hopkins University, 3701 San Martin Drive, Baltimore, MD 21218, USA}

\author[0000-0003-2838-1880]{Keisuke Osumi}
\affiliation{Department of Physics and Astronomy, Johns Hopkins University, 3701 San Martin Drive, Baltimore, MD 21218, USA}

\author{Basti\'an Pradenas}
\affiliation{Departamento de F\'isica, FCFM, Universidad de Chile, Blanco Encalada 2008, Santiago, Chile}
\affiliation{Department of Physics and Astronomy, Johns Hopkins University, 3701 San Martin Drive, Baltimore, MD 21218, USA}

\author[0000-0001-7941-9602]{Aamir Ali}
\affiliation{Department of Physics,
University Of California,
Berkeley, CA 94720, USA}
\affiliation{Department of Physics and Astronomy, Johns Hopkins University,
3701 San Martin Drive, Baltimore, MD 21218, USA}

\author[0000-0002-8412-630X]{John~W. Appel}
\affiliation{Department of Physics and Astronomy, Johns Hopkins University, 
3701 San Martin Drive, Baltimore, MD 21218, USA}

\author[0000-0001-8839-7206]{Charles~L. Bennett}
\affiliation{Department of Physics and Astronomy, Johns Hopkins University, 
3701 San Martin Drive, Baltimore, MD 21218, USA}

\author[0000-0001-8468-9391]{Ricardo Bustos}
\affiliation{Facultad de Ingenier\'ia, Universidad Cat\'olica de la Sant\'isima Concepci\'on, Alonso de Ribera
2850, Concepci\'on, Chile}

\author[0000-0003-1127-0965]{Manwei Chan}
\affiliation{Department of Physics and Astronomy, Johns Hopkins University, 3701 San Martin Drive, Baltimore, MD 21218, USA}

\author[0000-0003-0016-0533]{David T.~Chuss}
\affiliation{Department of Physics, Villanova University, 800 Lancaster Avenue, Villanova, PA 19085, USA}

\author{Joseph Cleary}
\affiliation{Department of Physics and Astronomy, Johns Hopkins University,
3701 San Martin Drive, Baltimore, MD 21218, USA}

\author[0000-0002-0552-3754]{Jullianna Denes~Couto}
\affiliation{Department of Physics and Astronomy, Johns Hopkins University, 3701 San Martin Drive, Baltimore, MD 21218, USA}

\author[0000-0002-1708-5464]{Sumit Dahal ({\dn \7{s}Emt dAhAl})}
\affiliation{Department of Physics and Astronomy, Johns Hopkins University, 3701 San Martin Drive, Baltimore, MD 21218, USA}

\author{Rahul Datta}
\affiliation{Department of Physics and Astronomy, Johns Hopkins University, 3701 San Martin Drive, Baltimore, MD 21218, USA}

\author{Kevin~L. Denis}
\affiliation{Goddard Space Flight Center, 8800 Greenbelt Road, Greenbelt, MD 20771, USA}

\author{Rolando D\"unner}
\affiliation{Instituto de Astrof\'isica and Centro de Astro-Ingenier\'ia, Facultad de F\'isica, Pontificia Universidad Cat\'olica de Chile, Av. Vicu\~na Mackenna 4860, 7820436 Macul, Santiago, Chile}

\author[0000-0001-6976-180X]{Joseph R.~Eimer}
\affiliation{Department of Physics and Astronomy, Johns Hopkins University,
3701 San Martin Drive, Baltimore, MD 21218, USA}

\author[0000-0002-4782-3851]{Thomas~Essinger-Hileman}
\affiliation{Goddard Space Flight Center, 8800 Greenbelt Road, Greenbelt, MD 20771, USA}
\affiliation{Department of Physics and Astronomy, Johns Hopkins University,
3701 San Martin Drive, Baltimore, MD 21218, USA}

\author{Dominik Gothe}
\affiliation{Department of Physics and Astronomy, Johns Hopkins University,
3701 San Martin Drive, Baltimore, MD 21218, USA}

\author[0000-0003-1248-9563]{Kathleen Harrington}
\affiliation{Department of Physics, University of Michigan, Ann Arbor, MI, 48109, USA}
\affiliation{Department of Physics and Astronomy, Johns Hopkins University,
3701 San Martin Drive, Baltimore, MD 21218, USA}

\author[0000-0001-7466-0317]{Jeffrey Iuliano}
\affiliation{Department of Physics and Astronomy, Johns Hopkins University,
3701 San Martin Drive, Baltimore, MD 21218, USA}

\author{John Karakla}
\affiliation{Department of Physics and Astronomy, Johns Hopkins University,
3701 San Martin Drive, Baltimore, MD 21218, USA}

\author[0000-0003-4496-6520]{Tobias A.~Marriage}
\affiliation{Department of Physics and Astronomy, Johns Hopkins University, 
3701 San Martin Drive, Baltimore, MD 21218, USA}

\author{Nathan J.~Miller}
\affiliation{Department of Physics and Astronomy, Johns Hopkins University, 
3701 San Martin Drive, Baltimore, MD 21218, USA}
\affiliation{Goddard Space Flight Center, 8800 Greenbelt Road, Greenbelt, MD 20771, USA}

\author[0000-0002-5247-2523]{Carolina N\'u\~{n}ez}
\affiliation{Department of Physics and Astronomy, Johns Hopkins University,
3701 San Martin Drive, Baltimore, MD 21218, USA}

\author[0000-0002-0024-2662]{Ivan L.~Padilla}
\affiliation{Department of Physics and Astronomy, Johns Hopkins University,
3701 San Martin Drive, Baltimore, MD 21218, USA}

\author[0000-0002-8224-859X]{Lucas Parker}
\affiliation{Space and Remote Sensing, MS B244, Los Alamos National Laboratory, Los Alamos, NM 87544, USA}
\affiliation{Department of Physics and Astronomy, Johns Hopkins University,
3701 San Martin Drive, Baltimore, MD 21218, USA}

\author[0000-0002-4436-4215]{Matthew A. Petroff}
\affiliation{Department of Physics and Astronomy, Johns Hopkins University,
3701 San Martin Drive, Baltimore, MD 21218, USA}

\author[0000-0001-5704-271X]{Rodrigo Reeves}
\affiliation{CePIA, Departamento de Astronomia, Universidad de Concepcion, Concepcion, Chile}

\author[0000-0003-4189-0700]{Karwan Rostem}
\affiliation{Goddard Space Flight Center, 8800 Greenbelt Road, Greenbelt, MD 20771, USA}

\author{Deniz Augusto~Nunes~Valle}
\affiliation{Department of Physics and Astronomy, Johns Hopkins University,
3701 San Martin Drive, Baltimore, MD 21218, USA}

\author[0000-0002-5437-6121]{Duncan J.~Watts}
\affiliation{Department of Physics and Astronomy, Johns Hopkins University, 3701 San Martin Drive, Baltimore, MD 21218, USA}

\author[0000-0003-3017-3474]{Janet L.~Weiland}
\affiliation{Department of Physics and Astronomy, Johns Hopkins University, 3701 San Martin Drive, Baltimore, MD 21218, USA}

\author[0000-0002-7567-4451]{Edward J.~Wollack}
\affiliation{Goddard Space Flight Center, 8800 Greenbelt Road, Greenbelt, MD 20771, USA}

\correspondingauthor{Zhilei Xu}
\email{zhileixu@sas.upenn.edu}

\collaboration{36}{CLASS Collaboration}


\begin{abstract}
The Cosmology Large Angular Scale Surveyor (CLASS) is a telescope array that observes the cosmic microwave background (CMB) over \SkyPercentage{} of the sky from the Atacama Desert, Chile, at frequency bands centered near 40, 90, 150, and 220~GHz. CLASS measures the large angular scale ($1^\circ\lesssim\theta\leqslant 90^\circ$) CMB polarization to constrain the tensor-to-scalar ratio at the $r\sim0.01$ level and  the optical depth to last scattering to the sample variance limit. This paper presents the optical characterization of the 40~GHz telescope during its first observation era, from \EraOneStart{} to \EraOneEnd{}. High signal-to-noise observations of the Moon establish the pointing and beam calibration. The telescope \bs{} pointing variation is $<0.023^\circ$ ($<1.6$\% of the beam's full width at half maximum (FWHM)). We estimate beam parameters per detector and in aggregate, as in the CMB survey maps. The aggregate beam has an FWHM of  $1.579^\circ\pm.001^\circ$ and a solid angle of $838 \pm 6\ \mu{\rm sr}$, consistent with physical optics simulations. The corresponding beam window function has a sub-percent error per multipole at $\ell < 200$. An extended $90^\circ$ beam map reveals no significant far sidelobes. The observed Moon polarization shows that the instrument polarization angles are consistent with the optical model and that the temperature-to-polarization leakage fraction is $<10^{-4}$ (95\% C.L.). We find that the Moon-based results are  consistent with measurements of M42, RCW~38, and Tau~A from CLASS's CMB survey data. In particular, Tau~A measurements establish degree-level precision for instrument polarization angles. 
\end{abstract}

\keywords{\href{http://astrothesaurus.org/uat/799}{Astronomical instrumentation (799)}; \href{http://astrothesaurus.org/uat/322}{Cosmic microwave background radiation (322)}; \href{http://astrothesaurus.org/uat/435}{Early Universe (435)}; \href{http://astrothesaurus.org/uat/1146}{Observational Cosmology (1146)}; \href{http://astrothesaurus.org/uat/1127}{Polarimeters (1127)}; \href{http://astrothesaurus.org/uat/1692}{The Moon (1692)}}

\section{Introduction}
\label{sec:intro}
\begin{figure*}[ht!]
    \centering
    \includegraphics[width=1.0\linewidth]{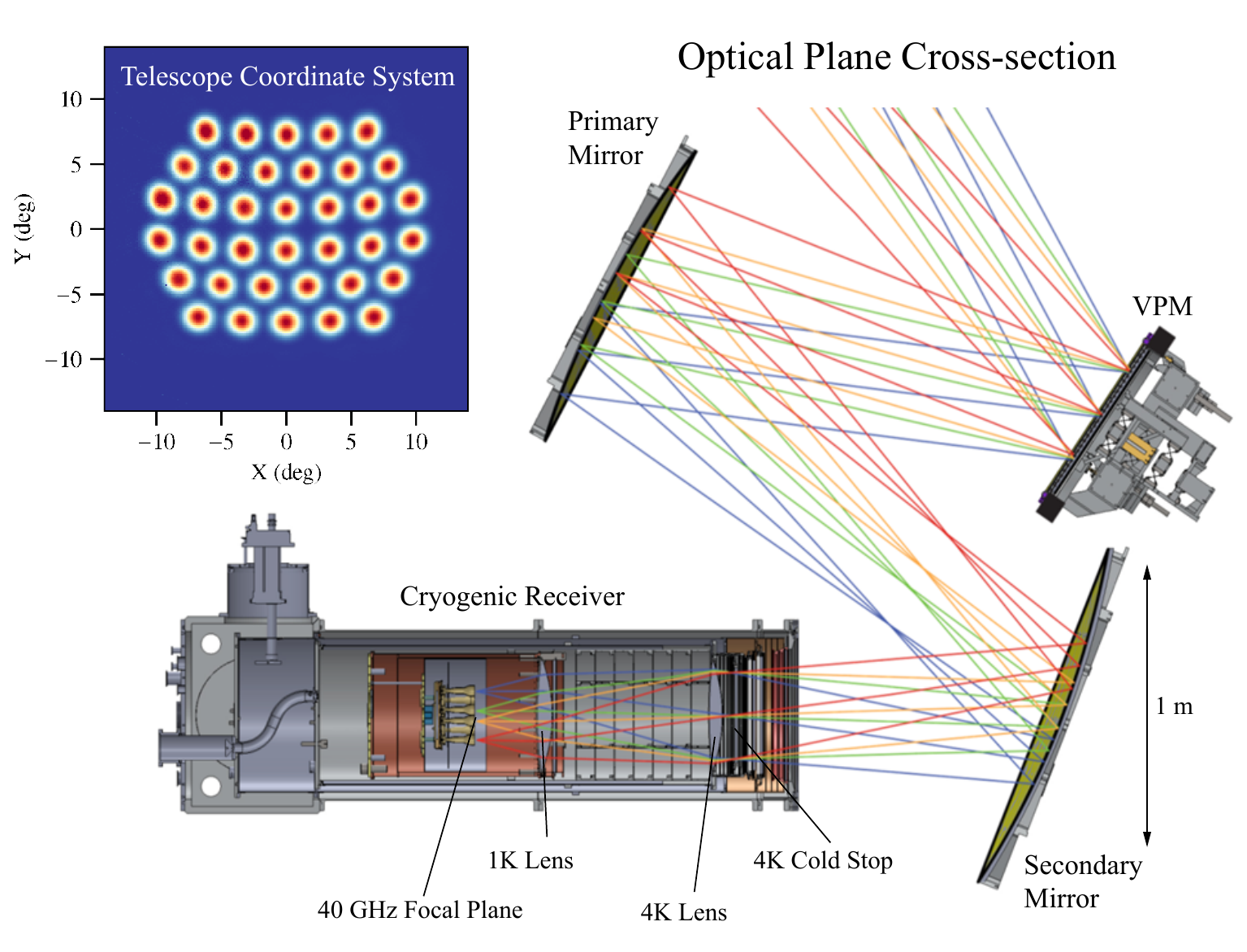}
    \caption{A cross-section of the 40~GHz telescope along the optical plane is shown with major components. Rays in four colors (blue, green, orange, and red) show how light travels through the telescope to four different feedhorns. The VPM is the first optical element. Mirrors produce an image of the cold stop near the VPM. Plastic lenses then focus the light onto 36 feedhorn-coupled dual-polarization detectors with speed $F=2$. A scale of one meter is shown at the bottom right of the figure. In the top left, the $19^\circ\times14^\circ$ measured focal-plane beam map is shown in the telescope coordinate system, with x(y)-axis pointing to the right (top). The beams have a characteristic FWHM of 1.5$^\circ$ and are separated by $3.5^\circ$ ($2.4\cdot F\lambda)$, consistent with the original design \citep{eime12}. }
    \label{fig:tele_beam}
\end{figure*}

Since its discovery by \citet{1965CMB_discovery}, the relic $2.728\pm0.004$\,K  cosmic microwave background (CMB) blackbody radiation~\citep{1996CMB_temperature} has been foundational to the hot Big Bang paradigm of an expanding universe. The 100~$\mu$K temperature anisotropy has provided the strongest constraints on this paradigm, elucidating the constituents and expansion history of the universe and establishing a standard model of cosmology~\citep[e.g.,][]{1996COBE_anisotropy, benn13, Hinshaw13,plan19v}. 
CMB polarization measurements can be decomposed into E modes, due to scalar and tensor perturbations, and B modes, due to tensor perturbations and conversion of E modes through gravitational lensing (``lensing B modes'') \citep{kami97,zald97}. 
Measurements of the E-mode polarization have further supported the standard model \citep[e.g.,][]{dasi, readhead04,Hinshaw13,loui17,henn18,keck18,kusa18,plan19v}. 
There is a focus on  measuring CMB lensing \citep[e.g.,][]{pola14,das14,bice16,omor17,plan18viii}, including lensing B modes \citep[e.g.,][]{keis15,loui17,pola17, keck18}, and on measuring primordial tensor B modes \citep[e.g.,][]{keck18, gual18, kusa18, pola19, spt19}. 
Lensing provides improved constraints on the sum of neutrino masses \citep{Neutrino_Allison}, and the tensor B modes would provide evidence for primordial gravitational waves of quantum origin, serving as evidence for and as a characterization of inflation \citep{guth81,mukh81,sato81,albr82,lind82,kami16}. Recently, ground-based and balloon-borne projects have begun targeting CMB polarization on the largest angular scales ($\theta>10^\circ$, $\ell<30$) that have so far only been probed from space \citep[e.g.,][]{gand16, ogur16, buzz17, geno17, appe19}. These measurements constrain both tensor B modes and the optical depth to reionization through the E modes.

Within this landscape of CMB measurements, Cosmology Large Angular Scale Surveyor (CLASS) is a telescope array that maps microwave polarization over 75\% of the sky from Cerro Toco in the Atacama Desert of Chile at frequency bands centered near 40, 90, 150, and 220~GHz \citep{essi14, harr16}. The 40~GHz CLASS telescope has been observing since 2016. The 90~GHz telescope was deployed and started observing in 2018 \citep{daha18}. The dual-band telescope---covering frequency bands of 150 and 220~GHz---was deployed in 2019 and has started collecting data \citep{daha19}. Multifrequency observations enable CLASS to distinguish the CMB from Galactic foregrounds \citep{watt15}. CLASS uses rapid front-end polarization modulation to recover the polarization signal at up to $90^\circ$ scales ($\ell \geqslant 2$)~\citep{mill16, harr18}. This measurement will constrain the tensor B modes at the tensor-to-scalar ratio $r\approx0.01$ level~\citep{watt15}. CLASS will measure the reionization optical depth $\tau$ to near the cosmic variance limit~\citep{2018_tau_Duncan}. Combining the CLASS optical depth measurement with higher-resolution CMB data and Baryon Acoustic Oscillation (BAO) measurements will improve constraints on the sum of neutrino masses~\citep{Neutrino_Allison, 2018_tau_Duncan}. CLASS will also provide the deepest wide-sky-area Galactic microwave polarization maps to date for studies of the interstellar medium.

A critical component of all CMB measurements is a detailed calibration of the telescope's optical response \cite[e.g.,][]{page03,hass13,pan18, bice19}. Of particular utility are the absolute pointing, angular response (i.e., beam function), and polarization angle associated with each detector in a telescope's focal plane. The observed signal is the true signal convolved with the telescope beam pattern. In order to recover the true signals from the sky, an accurate calibration of the beam properties is critical. Misestimation of these properties leads to systematic errors---including window function miscalibration, temperature-to-polarization leakage,  and E/B mode mixing---that degrade the accuracy of the measurement.  

This paper describes the optical characterization of the CLASS 40~GHz telescope \citep{eime12} and is one in a series of results based on data taken with the 40~GHz telescope from \EraOneStart{} to \EraOneEnd{} (``Era 1''). Herein, we discuss the telescope's pointing and beam calibration, beam window function, and polarization response. Other Era~1 papers address telescope calibration, efficiency, and sensitivity \citep{appe19}; circular polarization \citep{Padilla2019, Petroff2019}; polarization modulation and instrument stability (Harrington et al. 2019, in preparation); and polarization maps, angular power spectra, and large angular scale recovery (Eimer et al. 2020, in preparation).

This paper is structured as follows. Section~\ref{sec:instrument}  introduces the CLASS instrument and survey. 
Section~\ref{sec:lunar_thermal_model} describes the thermal model of the Moon as our optical calibration source at 40~GHz. The Moon data and the time-ordered data analysis method are also described in this section. The pointing analysis is discussed in Section~\ref{sec:pointing_analysis}, including the analysis method, results, and comparison to simulations. The first half of Section~\ref{sec:intensity_beam} describes the intensity-beam analysis, including the main beam and far-sidelobe maps to 90\dg{}; the latter half discusses the beam profile and the window function for cosmological analysis. Polarization measurements of the Moon are discussed in Section~\ref{sec:moon_pol}, including the simulated and measured Moon polarization patterns, estimates of  detector polarization angles, and intensity-to-polarization leakage estimate. Finally, we compare the Moon-based results with the measurements of unresolved sources in the CMB survey maps in Section~\ref{sec:comparison_survey_map}.

\section{Instrument and Observations}
\label{sec:instrument}

To achieve its science goals, CLASS must address systematic effects on long timescales and at large angles to unprecedented levels. Therefore, two central goals of the CLASS telescopes are (1) to limit intensity-to-polarization leakage by rapidly modulating the CMB polarization with the first optical element and (2) avoid far sidelobes and other systematic effects by propagating well formed beams with low distortion and high spill efficiency through the telescope. To achieve these goals, we use the telescope design described in detail by \cite{eime12}. Here, we summarize the optical design along with other aspects of the instrument and observations relevant to our measurements. 

The 40~GHz telescope design is shown in Figure~\ref{fig:tele_beam}. The first optical element that the polarized sky signal encounters is the Variable-delay Polarization Modulator (VPM) \citep{eime11,chus12,harr18}. The VPM consists of a 60~cm mirror that moves with its surface parallel to a wire grid. In this way, the VPM serves as an actively tuned reflective waveplate that modulates the polarized signal at 10~Hz, much faster than the atmospheric and instrumental drifts. Since the VPM is the first element in the optical chain, any instrument-introduced polarization signals are not modulated. Thus, the VPM limits temperature-to-polarization leakage, particularly from the brighter unpolarized atmospheric signal becoming polarized through reflections in the telescope. Front-end modulation with the VPM is foundational to recovering signals at the largest angular scales. 

After being modulated, the polarized signal is reflected by the primary and secondary mirrors into the cryogenic receiver to the cold stop through an ultra-high molecular weight polyethylene vacuum window and infrared filters.\footnote{Details of the vacuum window and filtering are given by \citet{essi14} and \citet{ 2018Cryostat_Jeff}.} The two mirrors produce an image  of the VPM at the 30~cm cold stop. Therefore, the entrance pupil of the telescope nearly coincides with the VPM, which means all of the beams formed at the focal plane have a similar illumination of the VPM up to an angle and plane of incidence. This entrance pupil placement also prevents the VPM from changing the telescope pointing during modulation. The size of the entrance pupil is $\sim$30~cm. Therefore, the 60~cm VPM is significantly underilluminated, protecting against systematic errors arising from unwanted diffraction and other systematic effects at the edge of the modulator.

After the cold stop, two high-density polyethylene lenses feed detectors in the focal plane with an $f$-number of 2 ($F=2$ with $f \approx 60$\,cm). In the focal plane, the beam-forming elements are single-moded, smooth-walled feedhorns \citep{zeng10}. The feeds illuminate the edge of the cold stop (corresponding to $F=2$) at $-10$~dB, resulting in high spill efficiency and low levels of unwanted diffraction as the beam propagates through the telescope. The feeds are spaced by 38~mm ($2.4\cdot F\lambda$). At this spacing, the field of view (FOV) with 36 beams is $19^\circ \times 14^\circ$, as shown in Figure \ref{fig:tele_beam}. The beams have a characteristic full width at half maximum (FWHM) of $1.5^\circ$ and are separated by $3.5^\circ$.

At the base of each feed is a microfabricated sensor that separates the two linear polarization states, defines the passband, and detects the power in each polarization with transition edge sensors (TES) \citep{chuss12a,rost12,2014_40GHz_Detector_John, appe19}. The 36 feedhorns are coupled pairwise to 72 TESs, one for each polarization state. The entire detector-feedhorn assembly is cooled to  $\sim$40~mK by a dilution refrigerator \citep{2018Cryostat_Jeff}. During the Era~1 observation campaign, eight sensors were nonoperational (but were recovered after Era 1). The remaining 64 sensors were optically sensitive, and all of the feedhorns were coupled to at least one sensor. The detectors saturate at an additional antenna temperature of $T=55$~K beyond normal atmospheric loading. The telescope (including detector) efficiency is $\eta=0.48$. The detector noise equivalent temperature (NET) is $248$~$\mu{\rm K_{RJ}}\sqrt{{\rm s}}$, and the telescope bandpass is from 32.3 to 43.7\,GHz, centering around 38\,GHz \citep{appe19}. 

The four CLASS telescopes are supported by two three-axis mounts. The two mounts are independent and identical, providing azimuth, elevation, and \bs{} rotations. 
The mounts rotate 720\dg{} in azimuth, and from 20\dg{} to 90\dg{} in elevation. The azimuth and elevation rotations together enable the telescope to point freely on the sky. 
However, polarization is a spin-2 field. The detectors only measure its projection onto one orientation at a time. Measuring many projections onto different orientations helps recovering the spin-2 polarization field accurately. 
In order to measure the polarization signal projected onto different orientations, \bs{} rotation is included as the third axis of the mount. 
This \bs{} rotation keeps the telescope \bs{} pointing unchanged while rotating the detector polarization direction on the sky within a 90\dg{} range. 
The telescope \bs{} angle is changed every day, cycling through seven angles ($-45$\dg{}, $-30$\dg{}, $-15$\dg{}, 0\dg{}, 15\dg{}, 30\dg{}, 45\dg{}) each week. This scan strategy is designed to provide even coverage of the seven \bs{} angles.

CLASS nominally observes 24 hr per day, 365 days per year. During  Era~1, approximately 60\% of the calendar time was spent on CMB observations with all systems in operation.  
During CMB observations, the telescopes stay at 45\dg{} elevation and scan azimuthally across 720\dg{} at the speed of 1\degpsec{}. When the Sun is up, we avoid it by 20\dg{} from the telescope boresight pointing, reducing the azimuthal range to less than the nominal 720\dg{}.
As the Earth spins, the telescopes cover $\sim$\SkyPercentage{} of the sky every day with large-scale scan cross-linking. 
Aside from the CMB observations, 3\% of the calendar time was dedicated to scanning calibration sources (primarily the Moon). 
In order to emulate the CMB observations, the  calibration observations are generally conducted at the same 45\dg{} elevation.\footnote{Scans at different elevations are used for a full pointing solution (Section \ref{subsec:pointing_analysis_method}).}
During these scans, the telescope maintains the elevation at 45\dg{} and scans across the source azimuthally. 
Since the focal plane is $\sim$10\dg{} in radius, the azimuthal scans cover $\pm$13\dg{} on the sky, centered on the Moon, so that beams at the edge of the FOV are measured at least to 3\dg{} in all directions. 
Furthermore, \bs{} rotations help to probe every beam out to 10\dg{} in all directions.

The Moon is the primary calibration source  for the 40~GHz  telescope; therefore, no attempts were made to avoid the Moon during normal CMB observations. 
Aside from dedicated Moon scans as described above, the optical performance was checked when the telescope sees the Moon during normal CMB observations. 
We call these types of Moon scans \textit{\pseudo{}}. The dedicated calibration scans together with the \pseudo{} are all included in the following analysis unless stated otherwise.

\section{Calibration with the Moon}
\label{sec:lunar_thermal_model}

The Moon spans ${\sim}0.5$\dg{} in the sky, one-third of the CLASS 40~GHz beam FWHM. Simulations show that the 1.5\dg{} beam is enlarged by $<2$\% after being convolved with the Moon. Therefore, the angular size of the Moon is small enough to be chosen as the primary calibration source for the CLASS 40~GHz telescope. In Section \ref{sec:comparison_survey_map}, we consider other unresolved sources of polarized and unpolarized emission with sufficient signal-to-noise in the preliminary Era~1 40~GHz survey maps, namely, Taurus A (hereafter Tau A), the Orion nebula (M42), and RCW 38.

\subsection{Moon-intensity Model}
\label{subsec:moon_intensity_model}
The Moon is the second brightest microwave source on the sky after the Sun. We simulated the Moon's microwave brightness temperature and polarization signal based on measurements made at 37\,GHz by the \textit{Chang'E} lunar satellite mission~\citep{2012ChangE_data}. \textit{Chang'E} measured the microwave brightness temperature at different lunar latitudes (0\dg{}, $\pm 20$\dg{}, $\pm 40$\dg{}, $\pm 60$\dg{}) across 360\dg{} lunar hour angles. Since the lunar hour angles are defined by solar illumination, the apparent Moon brightness temperature properties are a time-independent function of the lunar hour angles. The changes we observe from the Earth result from the variation in the section of the lunar hour angles facing the Earth. 

\begin{figure}
    \centering
    \includegraphics[width=1\linewidth]{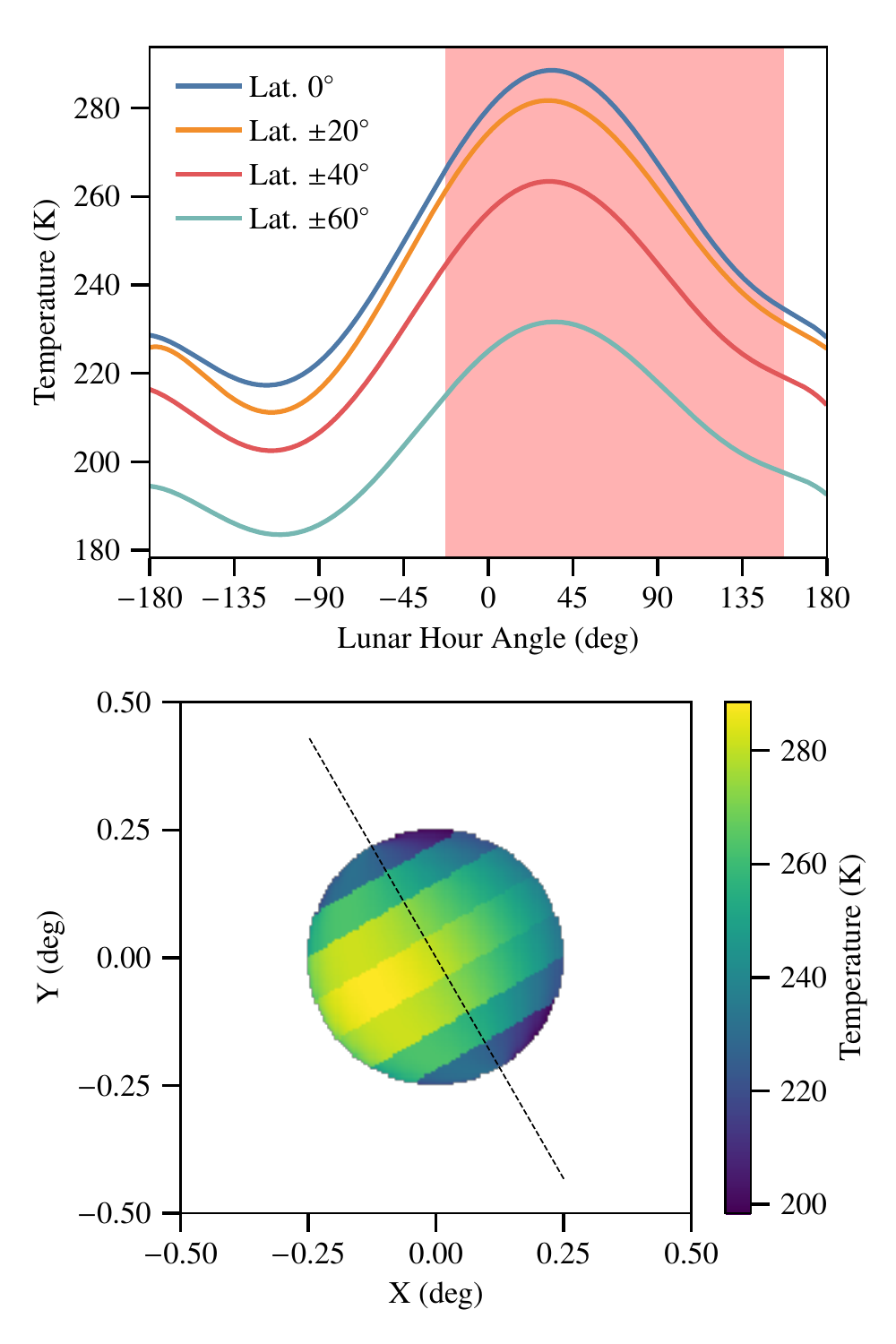}
    \caption{The Moon brightness temperature model. The upper panel shows the measured lunar-hour-angle thermal variation at different latitudes by \citet{2012ChangE_data}. Note that the temperature does not peak at 0\dg{} lunar hour angle, which is the center of the Sun illumination. This is because of the thermal lag of the lunar regolith. The shaded red region shows the section of lunar hour angles facing the Earth at the simulated time. Viewed from the Earth, the Moon brightness temperature model is simulated as shown in the bottom panel. The angular diameter of the Moon is set to 0.5\dg{}.  The thermal model is separated by different latitude bands. The orientation of the Moon in the telescope is also calculated and illustrated by the orbit axis in a dashed line.}
    \label{fig:moon_sim_temp}
\end{figure}

The lunar brightness temperature model is constructed by using the measured lunar hour-angle brightness temperature variation at different 20\dg{}-wide latitude bands, including ($-$10\dg{}, $+$10\dg{}), ($\pm$10\dg{}, $\pm$30\dg{}), ($\pm$30\dg{}, $\pm$50\dg{}), and ($\pm$50\dg{}, $\pm$90\dg{)}. Lunar phase is determined by the fractional illumination of the Moon presented to the Earth, which eventually results in different observed radiation amplitudes. The brightness temperature variations across different lunar hour angles are measured by the \textit{Chang'E} satellite. The variations at different latitudes and the Moon brightness temperature model are presented in Figure~\ref{fig:moon_sim_temp}. The Earth--Moon distance changes the apparent size of the Moon, equivalently changing its solid angle. The brightness temperature model and the solid angle enables us to simulate the expected intensity amplitude of the Moon at any given time. The lunar phase cycle has a period of 29.5 days, while the angular size change has a period of 27 days. With the two factors modulating the amplitude of the Moon, we observed a 414 days beat pattern on top of the monthly ($\sim$28 days) oscillation \citep{appe19}.

Since the Moon emission is not an isotropic disk, its orientation relative to the telescope must be accounted for in the simulation. The orientation of the Moon is characterized by the lunar orbit axis. We first calculated the orientation of this axis on the sky and then accounted for the telescope \bs{} rotation to yield the Moon orientation with respect to the telescope's view. 

The simulated Moon intensity map is then convolved with the 40\,GHz beam pattern. The peak Moon antenna temperature is estimated (and observed) to be $\sim$20\,K, well within the antenna temperature saturation limit of 55\,K. The detector response stays within the linear range throughout the lunar observations. Given the detector noise level at $\sim$250\,$\mathrm{\mu K \sqrt{s}}$ \citep{appe19} and that a single pass of the Moon takes $\sim$1\,s, the measurement noise is estimated as 
\begin{equation}
    N_{\textrm{Moon}} = \frac{250\,\mathrm{\mu K \sqrt{s}}}{\sqrt{\mathrm{1\,s}}} = 250\, \mathrm{\mu K}.
\end{equation}
The Moon antenna temperature for the 40\,GHz telescope is approximately $T_\textrm{Moon} \approx 20~\mathrm{K}$; hence, the signal-to-noise ratio (S/N) is:
\begin{equation}
    \textrm{SNR} = \frac{T_\textrm{Moon}}{N_\textrm{Moon}} = \frac{20\,\mathrm K}{250\,\mathrm{\mu K}} \approx 8\times 10^4.
\end{equation}

\subsection{Moon Data}
\label{subsec:moon_data}
As described in Section~\ref{sec:instrument}, the telescope scans azimuthally over the Moon as it rises or sets, passing through the constant scan elevation. For a 45\dg{} elevation Moon scan, when the Moon is in the elevation range from  32\dg{} to 58\dg{}, the telescope scans azimuthally $\pm$18.4\dg{} ($\pm$13\dg{} on the sky), centering on the instantaneous azimuthal position of the Moon. The scan speed is chosen to be 1\degpsec{} (in azimuth), matching the CMB scans. 
This scan speed provides a short enough turnaround time to give sufficiently dense sampling as the Moon rises or sets. Two dedicated Moon scans, rising and setting, can be executed on days when the Moon transits higher than 60\dg{} elevation.

During the initial commissioning period and whenever a change was made to the telescope that required recalibration of the pointing, dedicated Moon scans were performed frequently, covering the full range of boresights and including extra elevation range. These dedicated scans enabled us to quickly understand several basic properties of the instrument, including pointing and beams. Once the pointing and beams were well determined, the frequency of the dedicated Moon scans was reduced. Instead, the instrument properties were checked with \pseudo{} during CMB observations. Both the dedicated Moon scans and the \pseudo{} are analyzed through the same algorithm. Unless otherwise indicated, the term ``Moon scan'' refers to both dedicated and survey Moon scans.

In Era~1, 822 Moon scans (including 304 dedicated Moon scans and 518 \pseudo{}) were performed at different \bs{} angles. 
The \bs{} angle distribution is shown in Table~\ref{tab:moon_scan_boresight}. 
The boresight angle for each Moon scan is the same as the CMB observation \bs{} angle of the day. We aimed to have an even distribution over the seven \bs{} angles, as in the CMB observations. This goal was achieved during  Era~1. The higher weight on zero \bs{} angle is due to the initial commissioning observations, which were primarily performed at $0^{\circ}$ \bs{} rotation angle.

\begin{deluxetable}{ccc}
    \tablecaption{Boresight angle distribution over Moon scans.\label{tab:moon_scan_boresight}}
    \tablehead{\colhead{Boresight Angle} & \colhead{Moon Scan Count} & \colhead{Percentage}}

    \startdata
    $-45$\dg{}          & 131 & 15.9\% \\
    $-30$\dg{}          & \phn96 & 11.7\% \\
    $-15$\dg{}          & \phn94 & 11.4\% \\
    $\phn\phn0$\dg{}    & 198 & 24.1\% \\
    $+15$\dg{}          & \phn96 & 11.7\% \\
    $+30$\dg{}          & \phn93 & 11.3\% \\
    $+45$\dg{}          & 114 & 13.9\% \\
    \enddata
\end{deluxetable}

\subsection{Time-ordered Data Treatment}
\label{subsec:tod_treatment}

During Moon scans, we collect time-ordered data (TOD) for each detector at the rate of ${\sim}200$\,Hz. The raw data, which are proportional to current through the TES, are read out with a superconducting quantum interference device (SQUID) multiplexing system \citep{rein03} using a flux-locked loop implemented by a Multi-Channel Electronics (MCE) system \citep{MCE_Battistelli2008}. The raw data are  converted into units of optical power using the most recent current--voltage ($I$--$V$) curve calibration \citep{appe19}.

The MCE applies an anti-aliasing Butterworth filter prior to downsampling the raw data output. The filter is deconvolved in analysis, which  removes the associated phase shift. The thermo-electric response of the detector is modeled as a single-pole filter with a single time constant. The detector time constant is closely tracked by the phase delay between the VPM motion and corresponding signal in the TOD \citep{appe19}. We deconvolve the filtering associated with this electro-thermal response of the detector as well.

\begin{figure*}[ht!]
    \centering
    \includegraphics[width=1.0\linewidth]{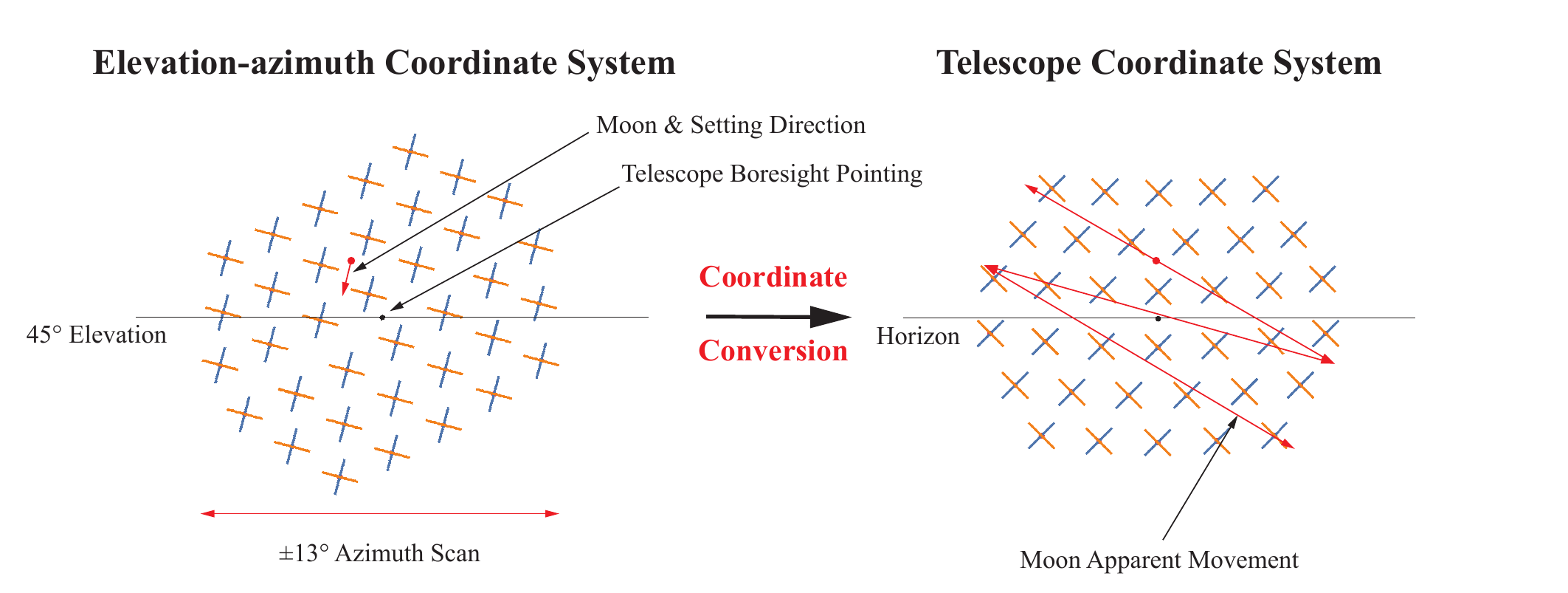}
    \caption{Moon scan illustration and coordinate-system conversion. Left panel: the detector array is presented, centered at the scan elevation with a \bs{} rotation. Each cross in the detector array represents a detector pair sensitive to $\pm45$\dg{} polarization directions. The Moon slowly rises or sets (sets in this example) as the telescope scans $\pm 13$\dg{} along azimuthal directions. Both the Moon positions and the telescope \bs{} pointing are described in the ``Elevation-azimuth Coordinate System.'' Right panel: the Moon positions are converted into the ``Telescope Coordinate System,'' where the telescope \bs{} pointing is the origin and the $x$, $y$ axes are defined similarly to azimuth and elevation in the elevation-azimuth coordinate system. In the telescope coordinate system, every detector is fixed at set $\Delta x$ and $\Delta y$ angular offsets while the Moon appears to zigzag across the array. Note: the spacing of the zigzag paths is exaggerated.}
    \label{fig:sky_tele_conv}
\end{figure*}

After the two rounds of deconvolution, the processed time-ordered data are scrutinized for glitches, which may arise from a detector losing flux-lock, from SQUID $V$--$\Phi$ jumps, from cosmic-rays, and from other non-idealities. These glitches are fixed if possible (say interpolating one data point from one cosmic-ray hit). Otherwise, the data are rejected for subsequent analysis. Details on data processing will be presented in a companion paper (Parker et al. 2020, in preparation).

\section{Pointing Analysis}
\label{sec:pointing_analysis}
The pointing of each detector is determined by two quantities: the telescope \bs{} pointing and the detector pointing offsets from the telescope \bs{} pointing. The telescope \bs{} pointing defines the central location and orientation of the telescope's field of view, parameterized by azimuth, elevation, and \bs{} rotation angles. To specify the pointing offset of a detector in reference to the telescope \bs{} pointing, a new spherical coordinate system is defined with the telescope \bs{} pointing at the origin (new $\mathrm{azimuth = 0}$, $\mathrm{elevation = 0}$) and $x$, $y$ axes  defined similarly to azimuth and elevation in an elevation-azimuth coordinate system (new \bs~$\mathrm{angle = 0}$). This coordinate system is called the \textit{telescope coordinate system}. In other words, the telescope coordinate system is locked to the \bs{} pointing and rotation, and so the pointing offsets for individual detectors are easily defined at fixed locations in the new system (Figure~\ref{fig:sky_tele_conv} and top left of Figure~\ref{fig:tele_beam}). The fixed offsets serve as a fiducial reference to calculate the pointing of individual detectors given the telescope \bs{} pointing.

\subsection{Pointing Analysis Method}
\label{subsec:pointing_analysis_method}
During Moon scans, both the Moon and the telescope \bs{} pointing move in the local elevation-azimuth coordinate system. First, the Moon positions in the elevation-azimuth coordinate system are transformed into the telescope coordinate system using spherical geometry. 
In the telescope coordinate system, only the Moon moves during Moon scans, as shown in Figure~\ref{fig:sky_tele_conv}. 
In the telescope coordinate system, each detector pointing is set where its response to the Moon emission peaks. 
The detector pointings are described by the angular offsets along the two axes in the telescope coordinate system $\Delta x$, $\Delta y$. The Moon signal is modeled with a two-dimensional Gaussian profile, characterized by its amplitude $A$; the FWHM along major and minor axes $\textrm{FWHM}_{\textrm{major}}$, $\textrm{FWHM}_{\textrm{minor}}$; and the rotation angle $\theta$. More details on the parameter can be found in Section~\ref{subsec:instrument_beam}. With these six parameters, a series of time-ordered data are simulated to compare with the measured time-ordered data. Optimized values of the six parameters are obtained by minimizing the sum of the squared difference between the simulated time-ordered data and the measured ones. This time-stream analysis is used for pointing and initial characterization of the main beam and intensity calibrations.

\begin{figure*}[ht!]
    \centering
    \includegraphics[width=1.0\linewidth]{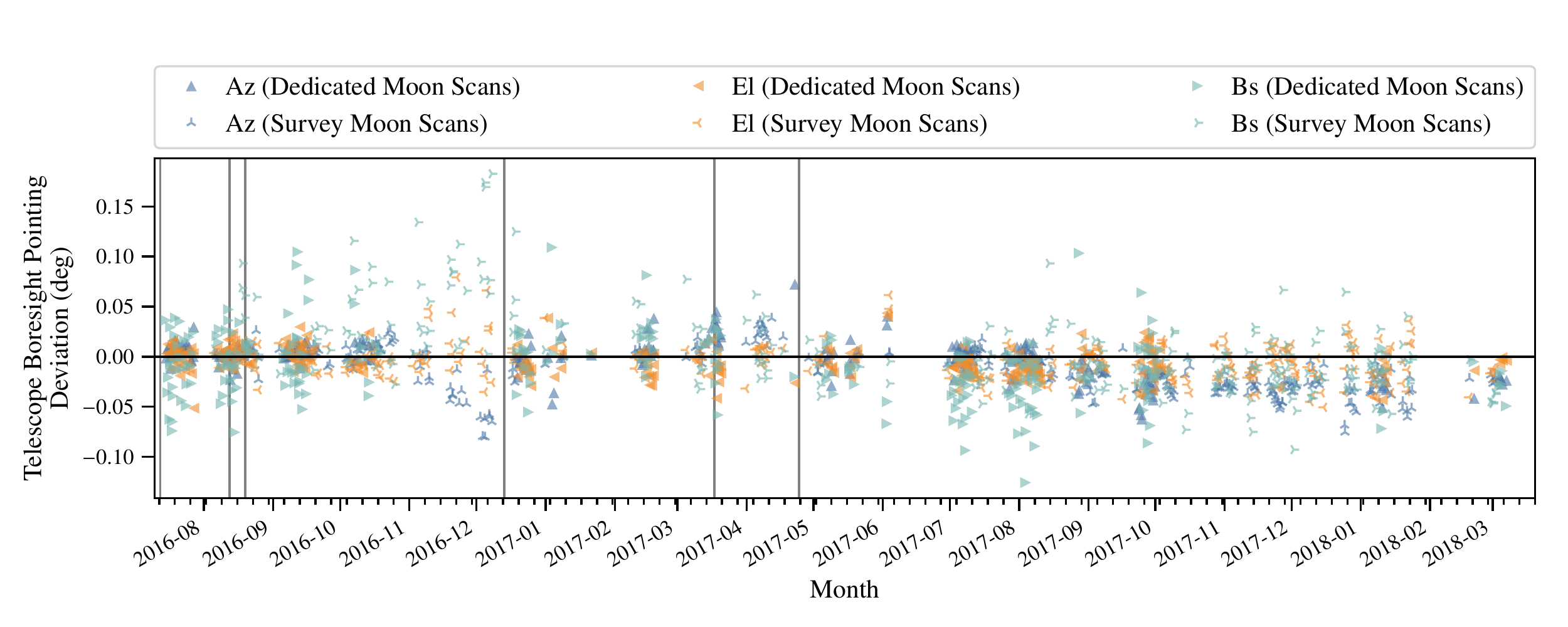}
    \caption{CLASS 40~GHz telescope \bs{} pointing deviation in Era~1. The $x$-axis shows time (with the major ticks showing months and minor ticks showing weeks); the $y$-axis shows the telescope \bs{} pointing deviation, including azimuth, elevation, and \bs{} axes. Filled triangle symbols represent the results from the dedicated Moon scans. Three-pointed stars represent the results from \pseudo{}. For the \pseudo{}, less time was spent on the Moon, resulting in less constraining power. Different colors show the deviations for the three different coordinates: azimuth (blue), elevation (orange), and \bs{} (cyan) angles. Vertical lines delineate the start of different pointing models. The pointing model was unchanged for the second half of Era~1. Also, while many dedicated Moon scans were taken during the beginning of Era~1, we reduced the frequency of dedicated Moon scans after obtaining a more stable understanding of the instrument.}
    \label{fig:bs_pointing}
\end{figure*}

The measured detector pointing offsets from all of the detectors form an array pattern in the telescope coordinate system. 
The array pattern should be leveled and centered at the origin. 
Any deviation indicates an offset between the telescope encoder readings and the true telescope \bs{} pointing. 
The leveling is related to the \bs{} rotation, while the centering is related to the azimuth and elevation positions. 

The telescope \bs{} pointing deviation information is used to establish a telescope pointing model, which is the tool used to transform the telescope mount encoder readings into the telescope \bs{} pointing. Ideally, the encoder readings could be directly interpreted as the telescope \bs{} pointing. In practice, various effects, such as telescope base tilt and structural sag, can produce offsets between the encoder readings and the actual \bs{} pointing. The pointing model captures these effects, allowing a precise reconstruction of the telescope \bs{} pointing. Details of the pointing model will be explained in a companion paper (Parker et al. 2020, in preparation).  To solve for a pointing model, pointing measurements are required at different telescope pointings in azimuth, elevation, and \bs{} angle. The pointing model needs to be renewed from time to time, especially when a hardware modification is conducted on the mount. In Era~1, six pointing models were constructed with six batteries of Moon observations. The time spans for the six pointing models can be found in Figure~\ref{fig:bs_pointing}.

After the telescope pointing models are established, the telescope coordinate system is updated with the improved \bs{} pointing. The detector offsets are then re-calculated in the updated telescope coordinate system. Since the pointing models include the telescope \bs{} pointing deviations, the array pointing pattern should be centered and leveled in the updated telescope coordinate system. The updated detector pointing offsets are fixed in the updated telescope coordinate system, where the detector pointing offset reference is generated. With a good understanding of the telescope \bs{} pointing and detector pointing offsets, the pointing of each detector is reconstructed on the sky.

Each Moon scan provides a telescope \bs{} pointing and a complete set of detector pointing offsets. From the Moon intensity simulation, the phase of the Moon could change the pointing estimate at a $3'$ level. Since it is common for all the detectors, it primarily changes the telescope \bs{} pointing estimate. Using our Moon thermal model, this effect is considered and removed for each Moon scan analysis. After the correction, the measured deviation from the pointing model is the telescope \bs{} pointing deviation, which can be decomposed into azimuth, elevation, and \bs{} angle components. With over 800 Moon scans in Era~1, we are able to closely monitor the telescope \bs{} pointing. Beyond that, the detector pointing offsets are also measured relative to \bs{} pointing. In theory, this analysis method ensures the detector offsets are fixed in the telescope coordinate system. In practice, the measured detector pointing offsets are not fixed across different Moon scans. 
The uncertainty of the offsets is estimated from the scatter of the measurements. 
For relative pointing offsets of individual detectors, only dedicated Moon scans are used because \pseudo{} do not provide sufficient data
per detector.

\begin{figure}
    \centering
    \includegraphics[width=1\linewidth]{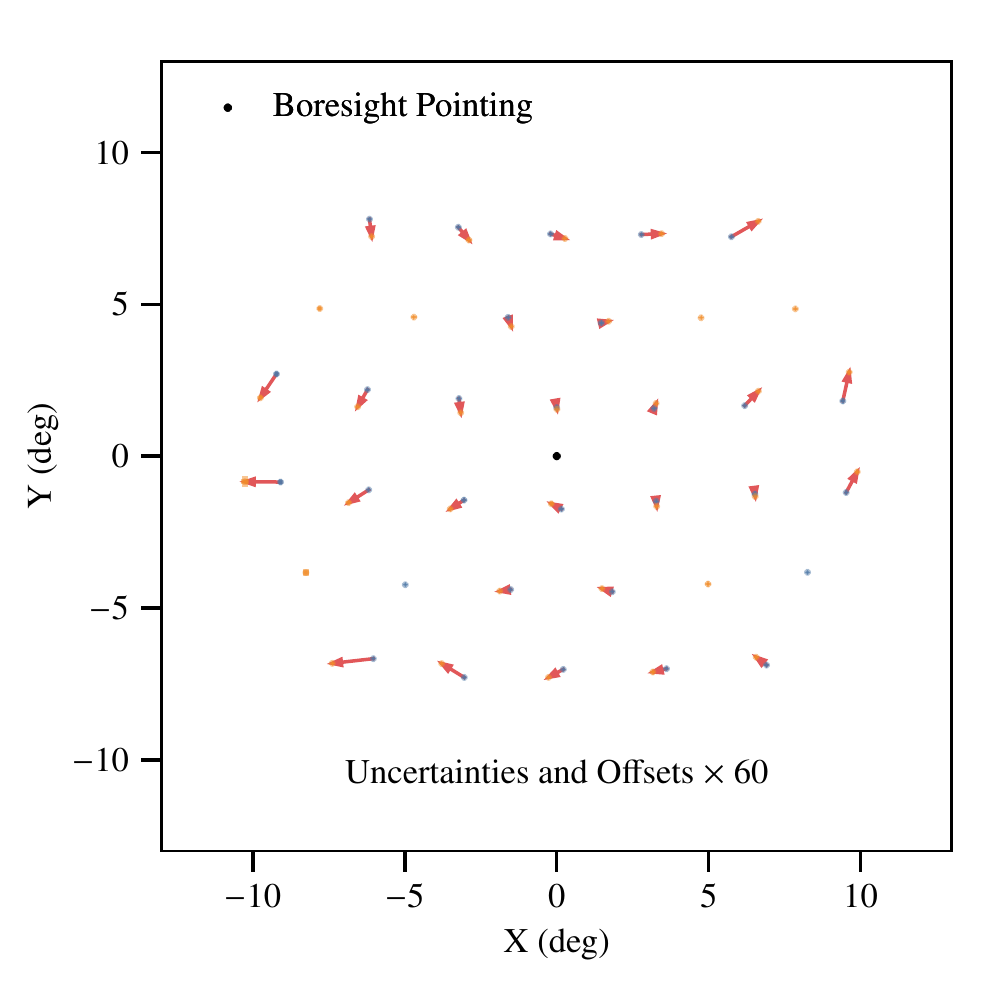}
    \caption{Measured detector pointing offsets. The measured offsets, uncertainties of the offsets, and differential pointing in paired detectors are presented in this plot. Detectors sensitive to $-45$\dg{} ($+45$\dg{}) polarization are shown in blue (orange) symbols. Uncertainties along $x$ and $y$ directions are shown as error bars for each detector. Even though the displayed uncertainties are multiplied by a factor of 60, they are still too small to be seen. The differential pointing vectors point from $-45$ detectors to $+45$ detectors are plotted. The length of the vectors are also multiplied by a factor of 60. Most of the differential pointings are well within $0.5'$. The telescope \bs{} pointing (array center) is indicated by a black dot in the center.}
    \label{fig:array_pointing}
\end{figure}

\begin{figure}
    \centering
    \includegraphics[width=1\linewidth]{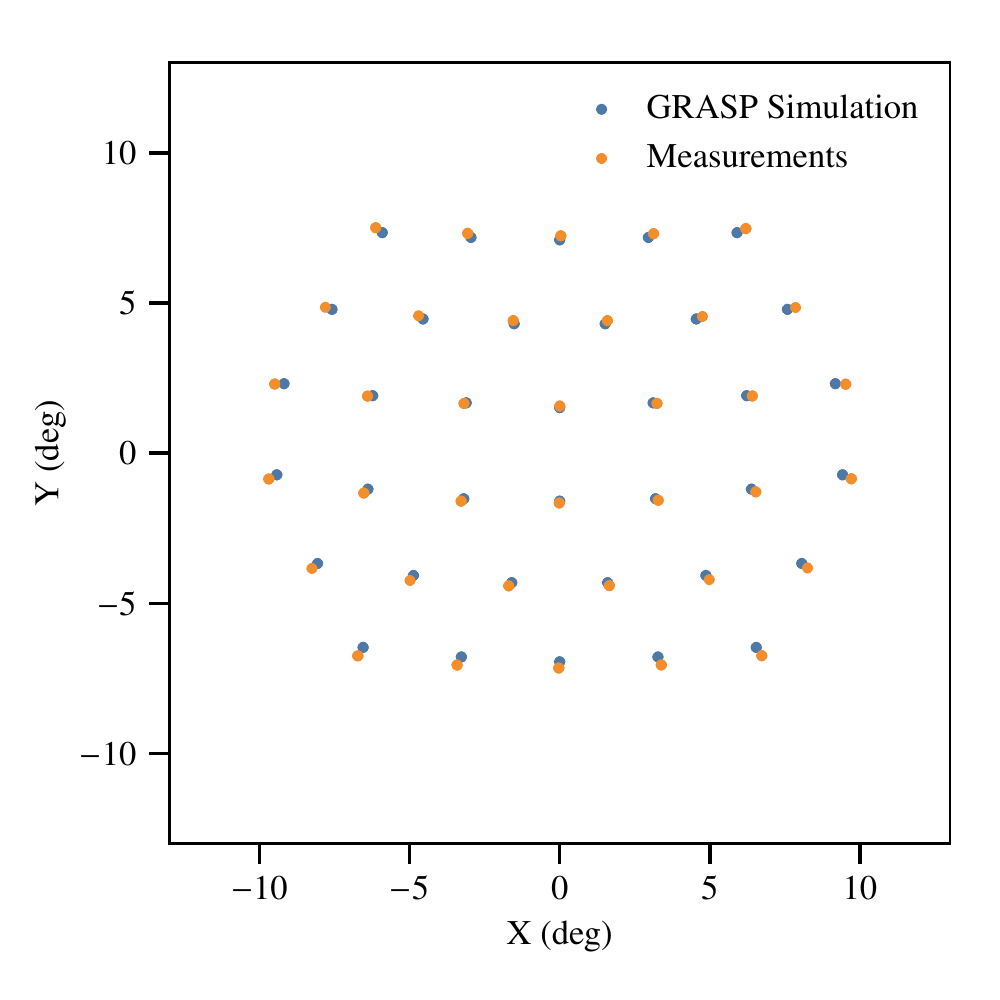}
    \caption{Measured detector pointing offsets compared to the GRASP simulation. The simulated results (blue dots) and measured results (orange dots) are both presented in this figure. A slight magnification in the measured pattern is seen, compared to the simulated result.}
    \label{fig:pointing_comparison}
\end{figure}

\subsection{Pointing Results and Comparison to Simulation}
\label{subsec:pointing_results}
The telescope \bs{} pointing deviation from the corresponding pointing model is fitted with azimuth, elevation, and \bs{} components. Figure~\ref{fig:bs_pointing} presents the deviation components as a function of time in Era~1. Results from dedicated Moon scans and \pseudo{} are distinguished in the plot. Since the sampling density was sparse during \pseudo{}, corresponding to one or two passes for one beam, larger uncertainties are expected compared to the dedicated Moon scans. For this reason, the \pseudo{} are only suitable for pointing consistency checks, and we estimate the telescope \bs{} pointing with only the dedicated Moon scans. The calculated standard deviations are $1.07'$, $0.84'$, and $2.04'$ for azimuth, elevation, and \bs{} angle, respectively. Assuming the furthest distance from one detector to the array center is 10\dg{} for the \bs{} component calculation, adding the three components in quadrature gives the pointing uncertainty at $1.4'$. Considering the 1.5\dg{} beam at 40~GHz, this only represents 1.6\% of the beam size.

Measured detector pointing offsets are shown in Figure~\ref{fig:array_pointing}, with uncertainties given by the standard error of the mean from the dedicated Moon scans. The standard errors are computed along the azimuth and elevation directions in the telescope coordinate system. For the majority of the detectors, the standard errors are within $2''$. Differential pointing within detector pairs in a single feedhorn is also a critical parameter for polarization signal recovery.  Across the focal plane, the differential pointing is normally $<0.5'$, and the positions of each detector are well measured, with uncertainties $<2''$. The detector pointing offsets for each detector are tabulated in Appendix~\ref{app:beam_parameters}.

Together, the collective ``beam jitter'' from the \bs{} pointing and detector offset pointing uncertainties results in an effective broadening of the 1.5\dg{} beam in the survey maps by $\sim 0.3$\%. While essentially negligible, this broadening is accounted for in the cosmological analysis.

The detector pointing offsets were simulated by the General Reflector Antenna Software Package (GRASP). Appendix~\ref{app:grasp_em} provides more details of the simulation, including the simulation method and the instrumental model. The input instrument model is the instrument design; any difference between the simulated results and measured results could be due to imperfect construction and alignment or approximations in simulation. Figure~\ref{fig:pointing_comparison} shows the pointing comparison between the simulated and the measured results. The measured detector pointing offset pattern demonstrates a small magnification compared to the simulation. If we use the angular distance from the detector pointing offset to the array center as a metric, the average magnification for all of the detectors is around 2.5\%. 
This effect will be further discussed in
Section~\ref{subsec:instrument_beam}.

\section{Intensity Beam Mapping}
\label{sec:intensity_beam}
Moon scans enable the calibration of the peak and angular response of each detector on the sky (i.e., absolute calibration and the beam). The absolute calibration was used to measure an overall telescope efficiency of 48\% \citep{appe19}. Characterization of the beams for each detector provides important information on whether the instrument is properly constructed and aligned. The CMB signal from the sky is convolved by the instrument beam before measured by the detectors. Accurate beam reconstruction provides key information to recover the true CMB signal from the measurements.

From the beam map, a beam profile $b(\theta)$ can be obtained. We can then calculate the beam profile's harmonic transform $b_\ell$, whose square is the beam window function. The beam window function, together with other window functions due to filtering and map pixelization, makes up the overall power-spectrum window function $w_\ell$ that has to to be accounted for to recover the true CMB angular power spectrum \citep{page03}. For CLASS, we must understand the beam out to large angles to properly calibrate the window function to low $\ell$.

\subsection{Beam Analysis Method}
\label{subsec:beam_analysis_method}
In the telescope coordinate system, the Moon appears to zigzag across the array during Moon scans (Figure~\ref{fig:sky_tele_conv}). Given a certain detector pointing offset, we define a natural coordinate system for a beam map, the \textit{detector coordinate system}, with the detector at the origin, the $y$-axis pointing along the local meridian in the telescope coordinate system, and the $x$-axis pointing to the right, perpendicular to the $y$-axis.

The amplitude of the two-dimensional Gaussian, fit in TOD space (Section~\ref{subsec:pointing_analysis_method}), provides an initial estimate of the profile peak value, which is also used to normalize the TOD such that the peak of the Moon signal is unity. The normalized TOD from all of the Era~1 Moon scans are then combined into a beam map in the detector coordinate system. The pixel size used for the beam map is 0.05\dg{}. Given the high S/N ratio of the combined beam maps, we can characterize the instrument beam properties with high fidelity; see Appendix~\ref{app:beam_parameters} for details.

For each detector, we use this beam map as a revised model to re-fit the TOD and thus obtain an improved estimation of the peak value. In this step, the beam parameters become deviations from the fiducial beam map, including the peak value, scale factors along the major and minor axes, and a correction to the major axis orientation. The measured beam properties are then corrected with the fitted parameters from the fiducial values. We iterate this process until the fitted amplitude value converges. Then the detector-specific beam map is saved for the subsequent analysis.

\subsection{Instrument Beam and Comparison to Simulation}
\label{subsec:instrument_beam}
Beam maps are generated for each detector and each Moon scan. A selection function rejects suboptimal beam maps according to several criteria, including weather conditions, detector noise level, and detector stability. For most detectors, more than 600 beam maps---out of the 800 Moon scans---are accepted. The accepted beam maps are normalized by the fitted peak amplitude from the TOD analysis before they are stacked to form an aggregate beam map for one detector. 

\begin{figure}
    \centering
    \includegraphics[width=1.0\linewidth]{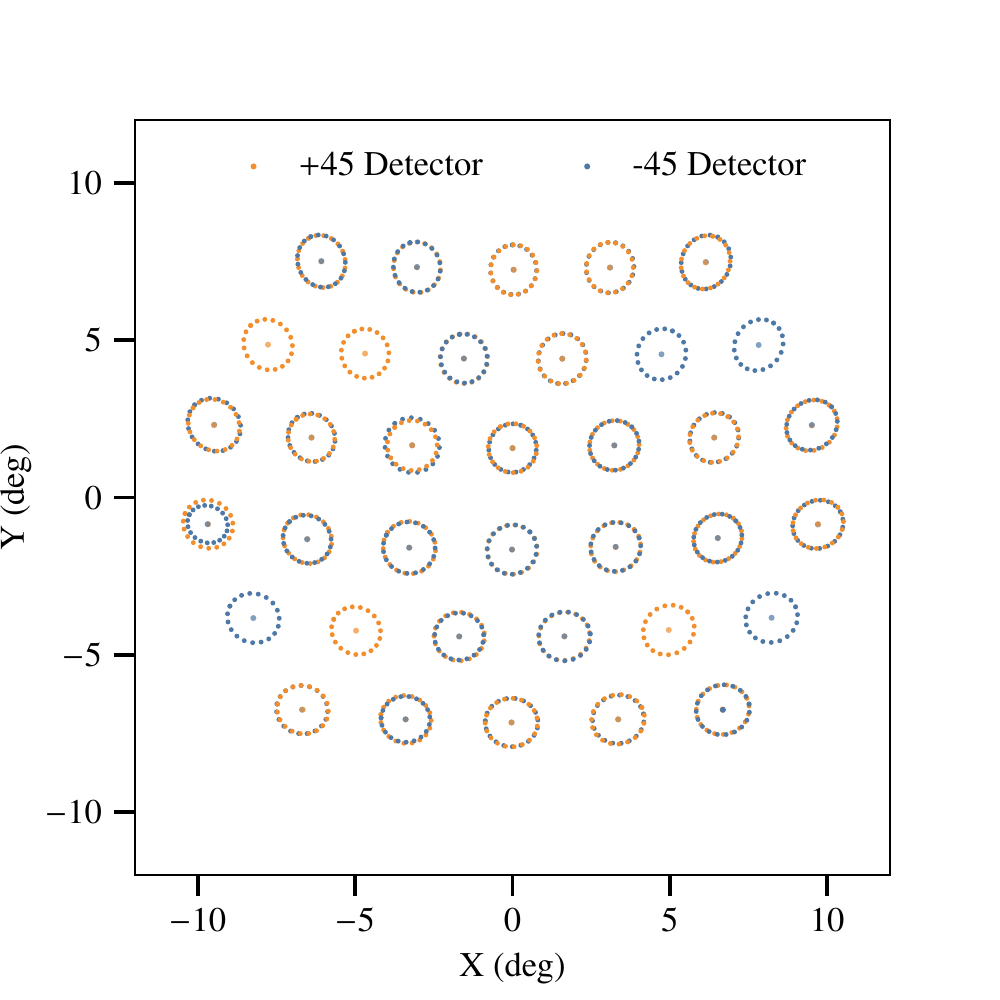}
    \caption{Main beams for all of the detectors. The FWHM of the main beams is shown with the dotted ellipses. The orange dots represent the results from the $+45^\circ$ detectors, and the blue dots represent those from the $-45^\circ$ detectors. Central beams are highly circular, while some eccentricity is apparent at the edge of the focal plane. More details of the statistics can be found in Figure~\ref{fig:beam_para_hist}.}
    \label{fig:beam_eccentricity}
\end{figure}

There are different ways to stack individual maps depending on the treatment of the \bs{} angle. If we stack the maps directly in the detector coordinate system, where the \bs{} rotation effect is  removed, the stacking procedure maintains the pixel positions fixed in reference to the instrument. This stacked map depicts the beam map directly associated with the instrument, the so called \textit{instrument beam map}. The top left part of Figure~\ref{fig:tele_beam} shows instrument beam maps for each detector superposed with their pointing offsets on the focal plane. The instrument beam maps in Figure~\ref{fig:tele_beam} are the average of the beams from the two linearly polarized ($\pm45$\dg{}) detectors associated with each feedhorns. Beam parameters are then measured from each instrument beam, including the FWHM along the major axis $\textrm{FWHM}_{\textrm{major}}$, the FWHM along the minor axis $\textrm{FWHM}_{\textrm{minor}}$, and the angle between the major axis and the $x$-axis in the detector coordinate system $\theta$.
As the Moon is not a perfect point source for the 40~GHz telescope, the measured FWHMs will be slightly enlarged. We simulated this effect in different conditions, by varying parameters including the FWHM along the major/minor axes and the phase of the Moon. The simulation results are then used to correct the convolution effect of the Moon.
The instrument beam parameter measurements are detailed in Appendix~\ref{app:beam_parameters}, together with the tabulated measured values.  Figure~\ref{fig:beam_eccentricity} shows  the beam for each detector with an ellipse constructed from the three fitted FWHM values. Negligible differential beams are observed in the majority of the detector pairs. 

\begin{figure}
    \centering
    \includegraphics[width=1.0\linewidth]{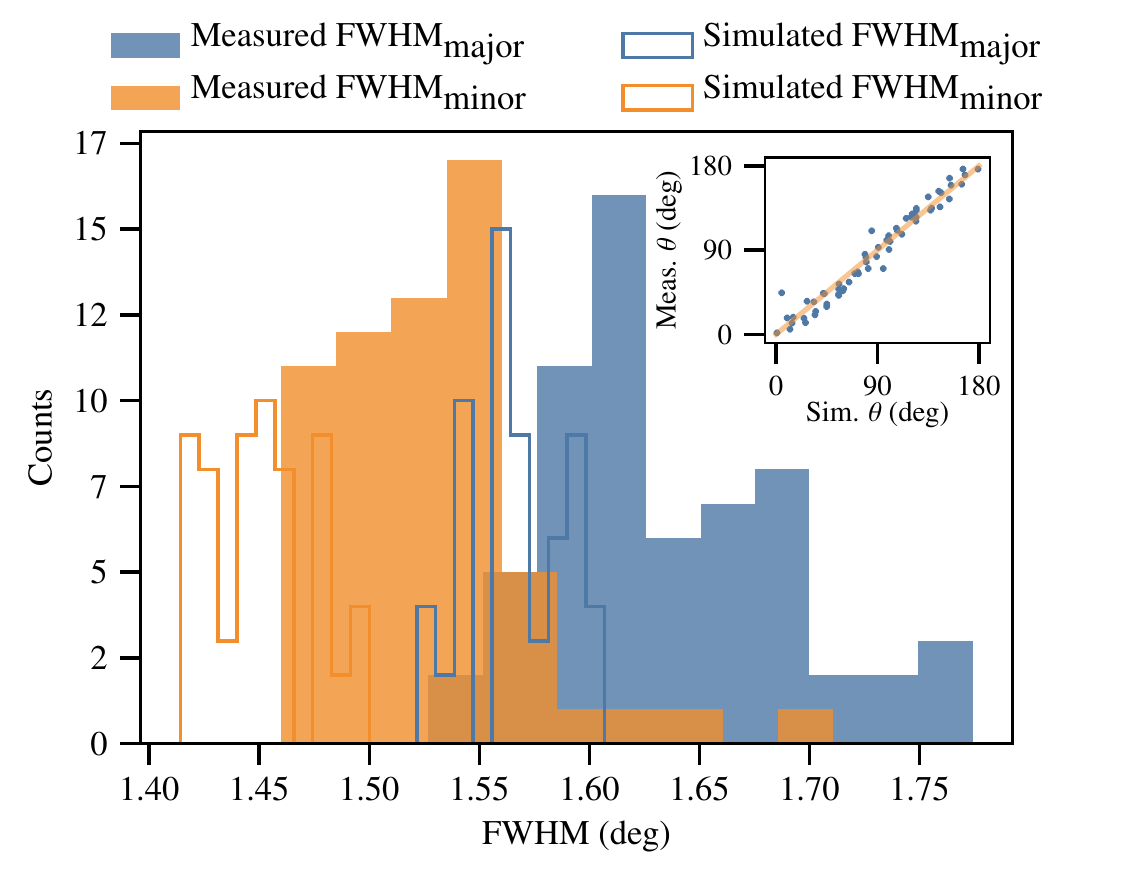}
    \caption{Measured beam parameters compared to their simulated counterparts. This plot shows the three beam parameters: $\textrm{FWHM}_{\textrm{major}}$, $\textrm{FWHM}_{\textrm{major}}$, and the rotation angle $\theta$. The main plot shows the histograms for the FWHM values. The bar histograms represent measured results, and the step histograms represent those from simulation. Different colors represent the results from either major or minor axis, as labeled above the plot. The measured results are slightly larger than the simulated results at $\sim$5\% level. The inset plot shows the comparison of the rotation angle $\theta$ for each detector. Each blue point represents the result from one detector. A one-to-one line is also drawn, demonstrating that the measurement is consistent with the simulation.}
    \label{fig:beam_para_hist}
\end{figure}

The GRASP simulation computes main beams for all the 40~GHz detectors; see Appendix~\ref{app:grasp_em} for more details. To compare with the beam parameters derived from the data, we applied the same algorithm to measure the beam parameters of the simulated instrument beam maps for all of the detectors. 
We then compare the measured and simulated parameters, which provides critical information about whether the instrument was built and aligned as designed. This comparison is summarized in Figure~\ref{fig:beam_para_hist}. For both major and minor beam axes, the measured FWHM values are systematically greater than the simulated values. The average linear enlargement is around 5\%. However, the rotation angle $\theta$ is consistent between the measurement and the simulation. The magnification observed in the pointing analysis (Section~\ref{subsec:pointing_results}) likely shares a cause with the beam enlargement we observe here. Aside from limits of the simulation, there are several possible explanations for these modest differences: imperfect alignment of the optical system could effectively change the focal length of the telescope, and thermal gradients in the lenses could, in combination with thermal contraction, deform the lens out of its ideal shape. However, the differences between measurements and simulations are small and are well characterized. As long as we use the measured parameters for subsequent analysis, these effects will not impact the telescope's ability to achieve our scientific goals.

\begin{figure}
    \centering
    \includegraphics[width=.9\linewidth]{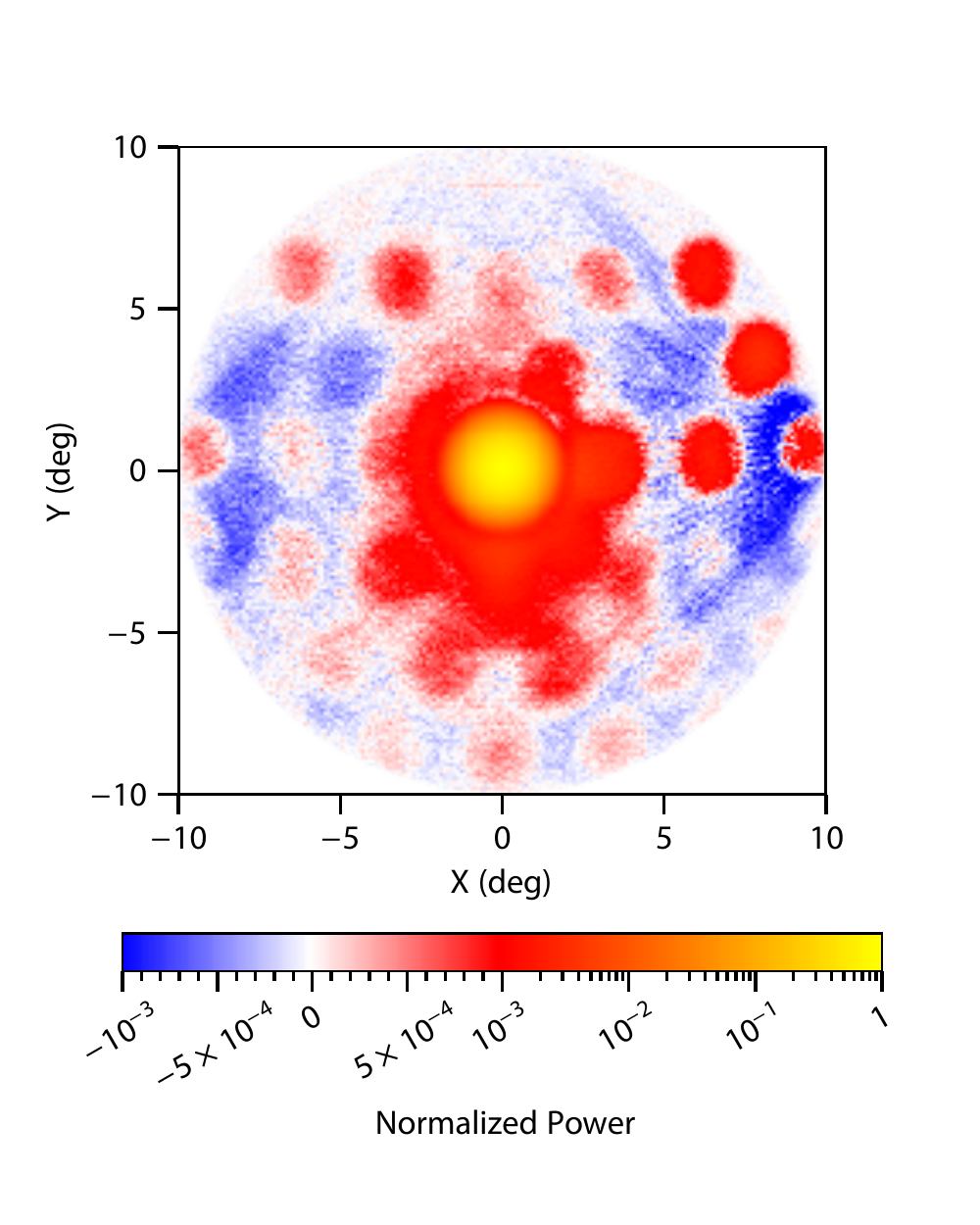}
    \caption{A moon beam map for a single detector.  The color map consists of two parts: one covers the majority of the solid angle from the normalized peak and first sidelobe to $10^{-3}$ in a logarithmic scale; the other emphasizes $10^{-3}$ to $-10^{-3}$ to show the detector cross-talk residual. The individual mini-beams resembling the focal plane come from electrical cross-talk and possibly optical ghosting. This pattern has both positive and negative amplitudes at $<10^{-3}$ levels. This level of cross-talk, revealed here through the $\mathrm{S/N}\approx10^5$ Moon measurement, was expected and is common in CMB experiments. The level has been reduced through a background subtraction, and future analysis will further mitigate this effect. Figure \ref{fig:cos_beam} shows how the telescope scan symmetrizes these features.}
    \label{fig:cross_talk}
\end{figure}

The stacked instrument beam maps extend to at least 10\dg{} for all detectors, including the edge detectors. This is because the \bs{} rotation enables each detector to sample the Moon at different angles. We observe positive and negative signals around three orders of magnitude below the beam peak (see Figure~\ref{fig:cross_talk} for a typical instrument beam). The shape of the signals resemble the focal-plane pattern. We have confirmed the existence of cross-talk between detectors at the percent level in the TOD. This level is characteristic of the readout system. Another possible explanation is optical ghosting, where light reflected by feedhorns is returned off metalized filters or filter/lens mounts. However, unlike cross-talk, we cannot conclusively say that ghosting plays a role. The impact of the cross-talk is reduced in the Moon beam maps due to its extended nature and the fact that we detrend the maps at a 10$^\circ$ radius. It is also reduced due to the impact of ``flux-jump'' corrections in the MCE readout. 
The MCE SQUID readout operates in a flux-locked-loop, where variations on the SQUID input from the TES current is actively canceled by a feedback current sourced by the MCE~\citep{rein03}. However, when the change in the input is large, as is the case when looking directly at the moon, the allowed error of the flux-locked loop may be exceeded. In this case, the MCE relocks the SQUID in the next flux quantum to reduce this error and the corresponding feedback current. This relock is accounted for by the MCE for the associated detector. On the other hand, currents induced in adjacent detectors change with no accounting in the MCE. We take care to fix these jumps in the data processing, but the overall effect is to reduce the impact of the feedback by up to a factor of two. Associated uncertainties are captured in our simulations, discussed further in Section~\ref{subsec:beam_profile_and_window_function}.

\begin{figure}
    \centering
    \includegraphics[width=.9\linewidth]{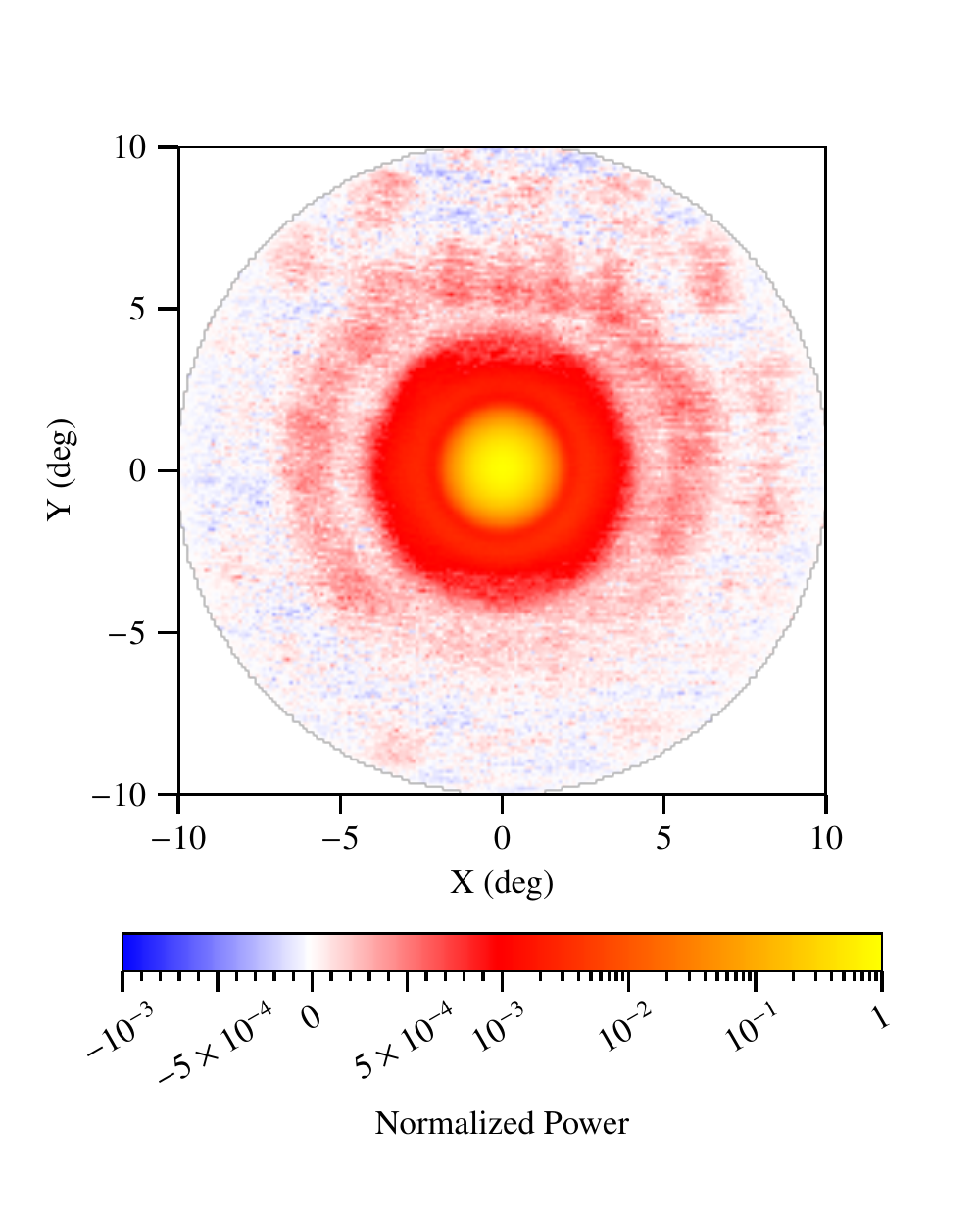}
    \caption{The cosmology beam map. The cosmology beam map is shown in two color scales as in Figure~\ref{fig:cross_talk}. The cross-talk pattern and the central hazing are significantly mitigated and symmetrized due to the telescope scan pattern. From 1 to $10^{-3}$ (normalized at the peak), the map is shown in a logarithmic scale; from $10^{-3}$ to $-10^{-3}$, the map is shown in a linear scale to capture negative values. The map has a resolution of 0.04\dg{}. The beam map is 10\dg{} in radius, almost as wide as the focal plane. The high intensity of the Moon provides a high signal-to-noise at the 50~dB level. The third and fourth ``sidelobes'' are due to electrical cross-talk and possibly optical ghosting spread in circular patterns from the \bs{} rotation.}
    \label{fig:cos_beam}
\end{figure}

\subsection{Cosmology Beam}
Stacking individual beam maps in their detector coordinate systems generates the instrument beam map. The instrument beam map provides information on the instrument optical performance, but it is not the natural beam map for cosmological studies. This is because the effective beam in the survey map, which we will call the \emph{cosmology beam map}, is a superposition of beams from different detectors rotated to different angles with respect to the local celestial meridian.

The daily \bs{} rotation causes the telescope to scan the sky at a different orientation angle every day. Together with the sky rotation, each point of the celestial sky is observed at different azimuthal positions with different \bs{} angles. Therefore, the cosmology beam should be the average of the instrument beams rotated to different \bs{} angles. The weight for each \bs{} angle should be determined by the observation fraction at that \bs{} angle. Since the Moon scans use the same \bs{} angle as the CMB observation of the day, we use the Moon scan \bs{} angle distribution to approximate that of the CMB observation. In practice, the time-ordered data of each Moon scan were rotated by the corresponding \bs{} angle before being binned into beam maps. These beam maps from different Moon scans were stacked together to form an intermediate detector-specific cosmology beam map. Those detector-specific cosmology beam maps were then stacked together to form a full-array cosmology beam map, which is suitable for cosmological analysis.

The 10\dg{} radius cosmology beam map is shown in Figure~\ref{fig:cos_beam}. This beam map contains 64 detector-specific cosmology beam maps. The beam map is normalized at the peak. The fractional uncertainty at the peak is at the $<10^{-5}$ level, providing a $>10^{5}$ S/N measurement of the cosmology beam map. The central beam shows a circular pattern because the stacking procedure averages out the eccentricity. An initial sidelobe is present at -25~dB. A third sidelobe is visible at -35~dB. Beyond this, the uncertainty from residual cross-talk obscures the beam features. We discuss these features further for the beam profile and beam window function below.

\subsection{Far-sidelobe Study}
\label{subsec:far_sidelobe}
Far sidelobes are studied, leveraging the Moon as a bright source. We use the CMB survey data to map the Moon around each detector within a radius of 5\dg{}--90\dg{}.\footnote{We started from 5\dg{} radius because our destriping map maker is not designed to handle point-like sources with S/N of $10^5$.} A destriping technique, used in generating the survey maps, allows us to recover features in the far-sidelobe maps \citep{dela98, buri99, kurk09, mill16}. The destriped far-sidelobe maps are made in the detector coordinate system.  The next step is to aggregate all of the detector far-sidelobe maps into a cosmology far-sidelobe map. We rotated each of the detector far-sidelobe maps to the seven \bs{} angles and stacked the rotated maps into detector-specific cosmology far-sidelobe maps. Then, we stacked the resulting maps from different detectors to form the \textit{cosmology far-sidelobe map} with even distribution across the seven \bs{} angles.

\begin{figure}
    \centering
    \includegraphics[width=1.0\linewidth]{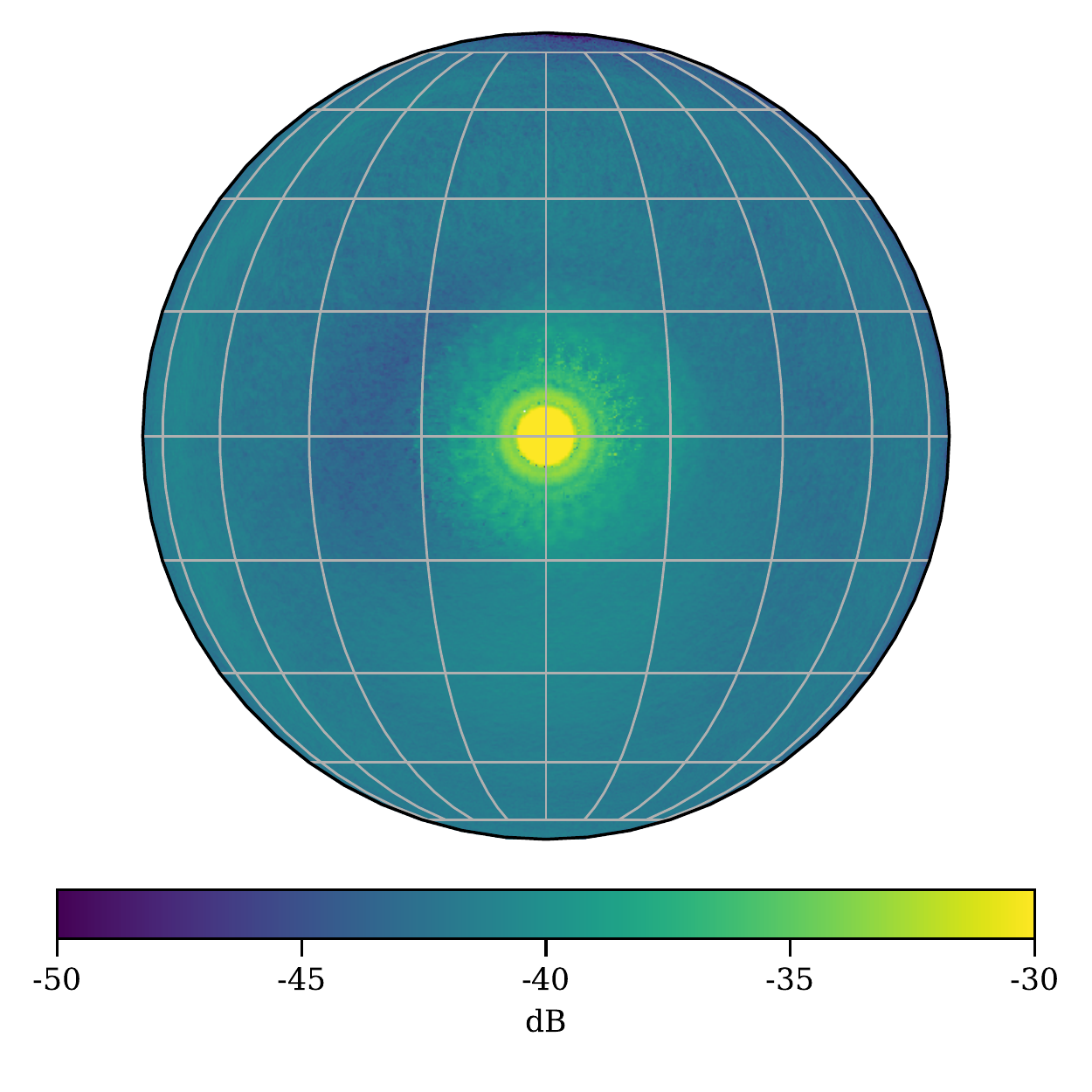}
    \caption{The extended cosmology beam map from Moon observations. The map covers up to 90\dg{} in radius. Large-scale structures are only visible at a $-50$\,{dB} level.}
    \label{fig:far_sidelobe}
\end{figure}

 Recall that the cosmology beam map covers up to 10\dg{} in radius and so has $5$\dg{} of overlap with the cosmology far-sidelobe map. We stitched the two maps together by adjusting the beam map zero-level until the 10\dg{} cosmology beam map matched the far-sidelobe map in the overlapping annulus from 5\dg{} to 10\dg{} in radius. The stitched map is effectively a beam map that extends to 90\dg{}, called the \textit{extended cosmology beam map}, as shown in Figure~\ref{fig:far_sidelobe}. This is the effective extended beam map for the cosmological analysis, containing both the main (and near sidelobe) beam information as well as the far-sidelobe information. The result shows that the far-sidelobe features are below $-45$~dB on large scales.

\subsection{Deconvolution and Beam Profile}
\label{subsec:beam_profile_and_window_function}

Strictly speaking, the 10\dg{} cosmology beam map shows the telescope beam convolved with the Moon. To remove the effect of the finite size of the Moon, we performed a simple deconvolution. The 10\dg{} cosmology beam map and a Moon map (a uniform 0.5\dg{}-diameter disk) were Fourier transformed in two-dimensions. Then, we divided the transformed beam map by the transformed Moon map to get the deconvolved beam information in Fourier space. To avoid numerical instability, we only included the information with scales larger than 0.5\dg{}. In two-dimensional Fourier space, we only included the modes within 3.2 inverse degree around the central base mode. We found that the deconvolved 1.5\dg{} 40~GHz beam maps were insensitive to variations in the 0.5\dg{} cutoff, so discarding information below 0.5\dg{} should not affect the subsequent analysis. We obtained the deconvolved beam in real space via the inverse Fourier transform. Deconvolution was applied to the cosmology beam map for measuring the deconvolved beam profile and the solid angle. In a separate analysis, we also forward-modeled the Moon-convolved cosmology beam profile with a set of Hermite Polynomials convolved by the Moon. We then removed the effect of the Moon in the fitted model to back out the deconvolved beam profile. This independent pipeline yields consistent results. Details on this method are presented in Appendix~\ref{app:beam_modeling}. 

\begin{figure}
    \centering
    \includegraphics[width=1.\linewidth]{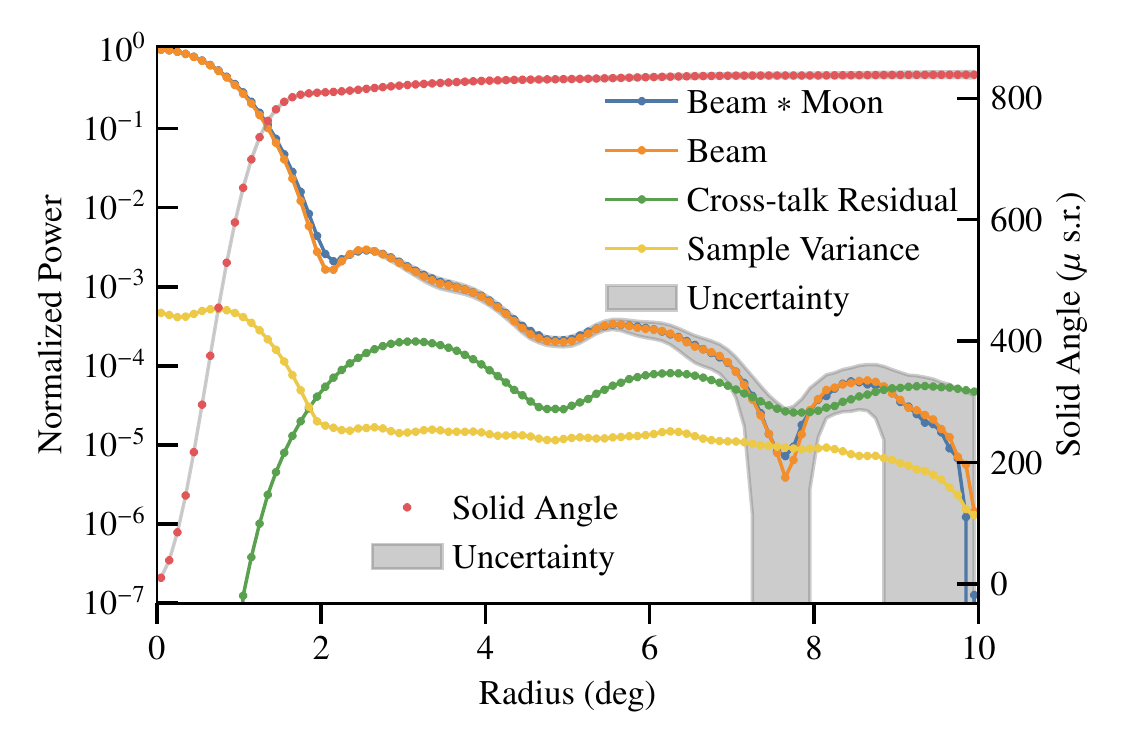}
    \caption{Beam profile and solid-angle measurements. The left axis shows the radial profile of the cosmology beam. The blue line represents the beam convolved with the Moon, while the orange line represents the deconvolved beam. In the main beam, there are small differences, which become negligible further out. The FWHM reduces by 0.02\dg{} after deconvolution, which is $\sim1.3$\% of the beam. The uncertainty of the radial profile is shown with a gray band. Because of the small values of the uncertainties, the gray band is only visible at the smallest profile values. The uncertainties are also broken down to two components: bootstrap sample variance (yellow line) and cross-talk residual variance (green line). The right axis shows the solid angle enclosed within different radii. The red data points represent the measurements at different radius values. The gray band shows the uncertainties of the measurements. Because of the small uncertainty values, the uncertainty band is again difficult to see. According to the figure, the beam encloses most of the power within a 4\dg{} radius. At the 10\dg{} radius, the solid angle is measured as $838 \pm 6\ \mu$sr.}
    \label{fig:profile_solid_angle}
\end{figure}

Once the deconvolved cosmology beam map was created, we reduced it to a one-dimensional radial profile. We computed the average of data binned in radial annuli with 0.1\dg{} width. A bootstrap method was used to estimate the uncertainties of the binned values. We used beam maps from all of the dedicated Moon scans for all of the detectors as the parent sample. Then, 100 cosmology beam maps were stacked from 100 bootstrap resamplings of the parent sample. The choice of bootstrap number, 100, was studied, and we found that the statistics converge well before this sample size. The 100 cosmology beam maps were then deconvolved before the binned profile was measured on them. The measured radial profiles for the beam maps (with and without the Moon convolved) are presented in Figure~\ref{fig:profile_solid_angle}. Also shown are the measurement error estimated through the sample variance of the 100 bootstrap-generated beam profiles.

Besides the bootstrap sample variance, Figure~\ref{fig:cross_talk} shows the existence of cross-talk that was first discussed in Section~\ref{subsec:instrument_beam} in the context of the beam map, and which may lead to the unaccounted systematic errors. To estimate this effect, we simulated detector-specific cross-talk maps with cross-talk coefficients measured between detectors from the Moon TOD. Each detector-specific cross-talk map spans beyond 10\dg{} in radius, including the cross-talk features from all other detectors. Then, we measured the profile from the stacked cross-talk map after stacking the detector-specific cross-talk maps. We found that the cross-talk signal manifests as a relatively flat profile extending to $>10^\circ$ with amplitude (5--10)$\times10^{-4}$ relative to the peak amplitude. Therefore, the $10^\circ$ aperture (background subtraction) imposed by the beam map pipeline largely removes this component from the data, leaving residuals at the (5-10)$\times10^{-5}$ level. This sets the amplitude of the additional beam profile uncertainty from cross-talk, which we take conservatively to be fully correlated across all angles. In practice, the cross-talk profile was added into the 100 bootstrap profiles with randomized amplitudes, normalized at the peak around 6\dg{}. The randomized amplitudes were drawn from a normal distribution with $\sigma = 7.6\times10^{-5}$, the average of the aforementioned residual level. After injecting the randomized cross-talk profile, the profile and solid-angle information was measured on the 100 updated profile bootstraps. Uncertainties for each radial bin were then estimated as the updated sample variance of the 100 updated profile bootstraps.

Figure \ref{fig:profile_solid_angle} also shows the enclosed solid angle within different radii, calculated from the measured radial profile. The uncertainty of the solid-angle values was also estimated along the bootstrap procedure, with the cross-talk effect included. The solid angle at 10\dg{} is measured to be $838 \pm 6\  \mu$sr for the deconvolved cosmology beam.

\subsection{Beam Window Function}
\label{subsec:beam_window_function}
The CMB maps are conventionally transformed into spherical harmonics space for analysis. The multipole number $\ell$ in spherical harmonics encodes space information. With measured the deconvolved beam profile, we then calculated its harmonic transform $b_\ell$ and the associated beam window function $b_\ell^2$. 
The beam window function together with other window functions---including filter window function, pixel window function---form the overall window function $w_\ell$ for cosmological analysis. With the overall window function $w_\ell$, the observed power spectrum is expressed as $\widetilde{C}_{\ell}= w_{\ell}C_{\ell}$. 
In the following text, we reserve the notation of $w_\ell$ for the overall window function and refer to the beam window function as $b_\ell^2$ explicitly. 

For a solid-angle-normalized circularly symmetric beam $b(\theta)$, its spherical harmonic representation reduces to
\begin{equation}
\label{equ:beam_transform}
    b_{\ell} = \int d\Omega\; b(\theta)P_{\ell}(\cos \theta),
\end{equation}
where the $P_\ell$ is the $\ell$th Legendre polynomial. The beam window function is computed as the square modulus of the beam transform as $b_\ell^2$.

To estimate the uncertainties in the 10\dg{} beam window function, we used the same 100 bootstrap samples in the previous section for the 10\dg{} beam. Then, the beam transform and the beam window function are calculated from each of the beam profiles. The uncertainties on the beam window function are then estimated as the sample variance of the simulated beam window functions. We find that the profile uncertainty associated with the cross-talk residual dominates the window function error. To estimate the additional uncertainty associated with the profile from 10\dg{} to 90\dg{}, we computed the $>10^\circ$ beam window function at the profile's upper and lower error limits and estimated the uncertainty as the difference between the two. (We have found that this produces an upper limit on the actual uncertainty.) The uncertainties from this range were then added in quadrature to those of the 10\dg{} beam window function. The result is a negligible increase in the beam window error.

\begin{figure}
    \centering
    \includegraphics[width=1.\linewidth]{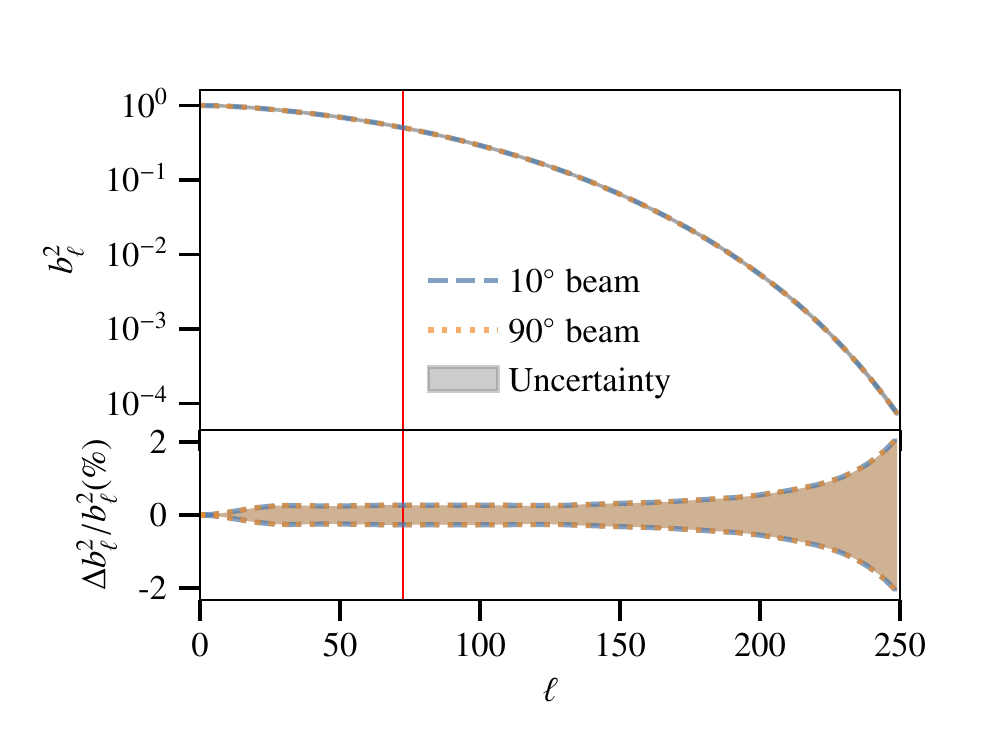}
    \caption{Beam window functions and uncertainties. Results from both the 10\dg{} beam and the 90\dg{} beam are shown in blue (dashed lines) and orange (dotted lines), respectively.
    The beam window functions are shown in the upper panel. The beam window functions are normalized by the corresponding solid angles. The uncertainties are displayed in a gray band, which is thinner than the line width. The relative uncertainties $\Delta b_\ell^2/b_\ell^2$ are plotted in the bottom panel. The relative uncertainties are $<1$\% until $\ell\approx200$ and rise to 2\% at $\ell=250$. The vertical red line shows the half-power position at $\ell = 72$. The consistency between the 90$^\circ$ and 10$^\circ$ measurements shows that far sidelobes do not impact the beam window function.}
    \label{fig:window_function}
\end{figure}

The results for the beam window function and the uncertainties are shown in Figure~\ref{fig:window_function}. Both the results from the 10\dg{} beam and the 90\dg{} beams are shown but are too similar to distinguish. Since Equation~\ref{equ:beam_transform} integrates from 0\dg{} to 180\dg{} and neither the 10\dg{} nor 90\dg{} beam profile covers the entire range, we effectively zero-pad beyond the beam profile range out to 180\dg{}. The results from the 10\dg{} and 90\dg{} beam are normalized by their solid angles. The two results are consistent, demonstrating that the far-sidelobe structure from 10\dg{} to 90\dg{}  does not affect the beam window function. 

Both of the normalized beam window functions start from unity at low $\ell$ and gradually decrease as $\ell$ increases. The relative uncertainty stays below 1\% within $\ell=200$. At higher multipoles, the beam window function drops down to $10^{-4}$ at $\ell=250$, with a relative uncertainty of around 2\%. The beam window function reaches the value of 0.5 at $\ell \approx 72$.

\section{Moon Polarization Analysis}
\label{sec:moon_pol}
The Moon signal has faint polarized features, mainly from the refracted thermal radiation. The lunar regolith is not totally opaque, so thermal emission travels through some depth of the regolith on its way into space, and refracts on the surface in a way that introduces polarization, shown in the top part of Figure~\ref{fig:moon_pol_mechanism}. The Moon polarization signal has been observed and used for calibration by other experiments~\citep{popp2002, bisc2011}.

According to the Fresnel equations, refracted radiation from the Moon has net polarization if the incidence angle is not zero. 
The polarization fraction increases from zero at the center to maximum at the limb of the Moon disk.
If the Moon were a spherical dielectric at a uniform temperature, no polarization should be observed at the center because of the circular symmetry, even for a beam larger than the size of the Moon. Since the CLASS 40~GHz beam is only three times the diameter of the Moon, the telescope beam profile applies a significant gradient across the size of the Moon. When not pointing at the center of the Moon, the beam gradient averages out a net polarization signal from the polarized limb of the Moon. This signal forms a quadrupole pattern in Stokes~$U$ or Stokes~$Q$, aligned with the telescope polarization direction. 
However, the Moon is not at a uniform temperature, so a net polarization (a combination of monopole and dipole) is observed in most cases. Therefore, the observed Moon polarization signal from one Moon scan is the combination of the monopole, dipole, and quadrupole components.

Observing the polarization signal from the Moon is not only interesting for lunar science; it also serves as a useful calibration method for the CLASS 40\,GHz telescope. CLASS is designed to measure the polarization component of the CMB anisotropy, which is at least three orders of magnitude lower than the CMB temperature component. Meanwhile, the Moon polarization signal is expected to be less than three orders of magnitude lower than the brightness temperature signal (see Section~\ref{subsec:moon_pol_model}). Therefore, Moon observations, demonstrating that polarization signals at a level of $10^{-3}$ can be isolated, are a stepping stone to measuring the polarization signal in the CMB. 

Detector polarization angle determines how we transform the observed linear polarization Stokes parameters ($Q$ and $U$) to the coordinate-invariant E and B components. A suboptimal calibration on the detector polarization angle will mix the E and B components. Considering that the E component is much brighter than the B component, a small E-to-B leakage could surpass the real B component. The Moon, as a polarized source, can be used to constrain the detector polarization angle at the 1\dg{} level.

\subsection{Moon Polarization Model}
\label{subsec:moon_pol_model}

\begin{figure}
    \centering
    \includegraphics[width=1.0\linewidth]{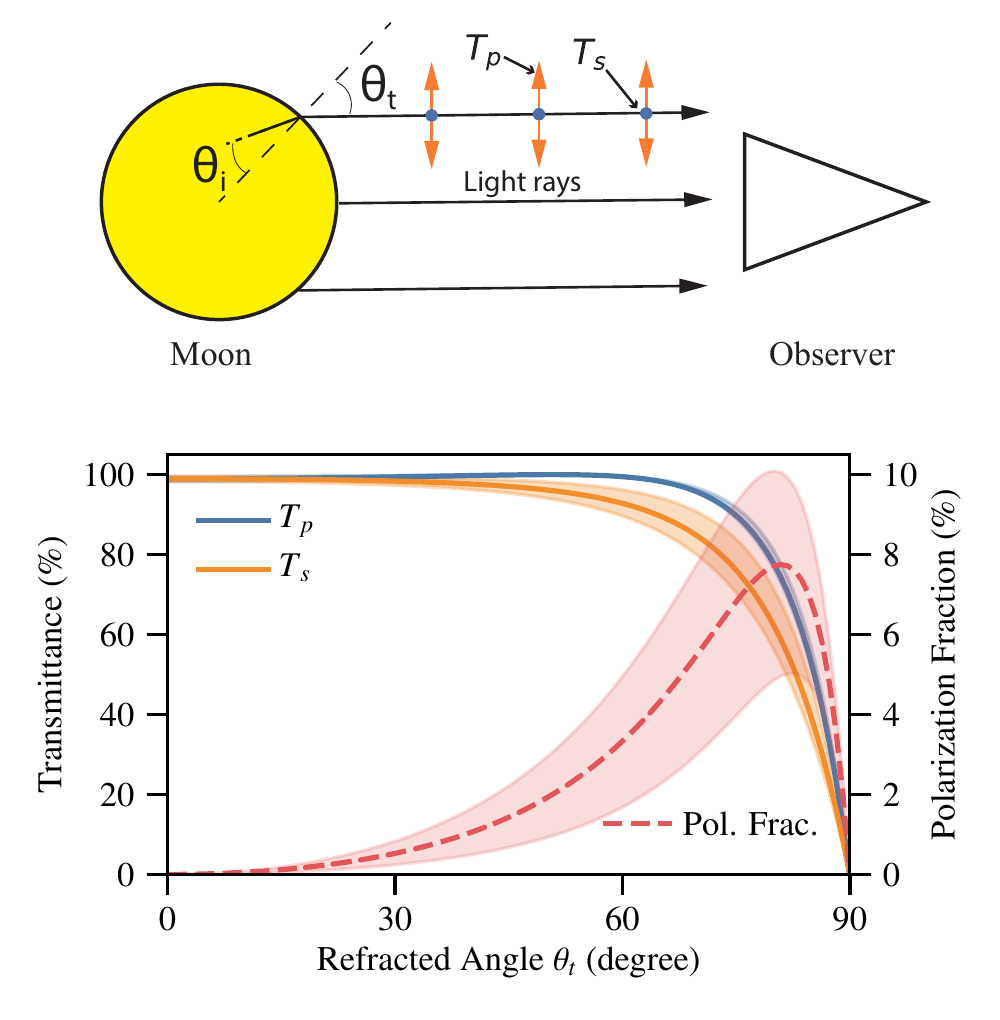}
    \caption{The Moon polarization mechanism. The upper panel shows a schematic diagram of the Moon polarization. Radiation coming from the lunar regolith is refracted at the lunar surface and received by observers. Assuming the Moon has a smooth surface, the incident angle and refracted angle are denoted $\theta_i$ and $\theta_t$, respectively. The transmitted radiation is decomposed into two orthogonal polarization directions, $T_s$ and $T_p$. $T_s$ represents the linear polarization perpendicular to plane of incidence, while $T_p$ represents the polarization parallel to that plane. Transmittance of $T_p$ (blue solid line) and $T_s$ (orange solid line) are shown in the lower panel as a function of the refracted angle $\theta_t$. The excess of $T_p$ results in a net linear polarization signal along the radial direction, viewed by observers.  The red dashed line shows the Moon polarization fraction, relative to the intensity power. The refracted angles from 0\dg{} to 90\dg{} can be mapped to radii on the Moon disk. The curve implies that the lunar signal is unpolarized at the center, while the polarization fraction gradually increases to the limb of the Moon until it drops to zero at the edge. The shaded regions around the three curves show the range of each variable when setting the effective dielectric constant $\epsilon$ from 1.3 to 1.7 \citep{loso67}.}
     \label{fig:moon_pol_mechanism}
\end{figure}

Thermal radiation from the Moon in the microwave bands is not significantly polarized except for at the limb. The incident thermal radiation is slightly polarized when leaving the lunar regolith. The transmitted radiation contains net linear polarization along the plane of incidence. The polarization fraction depends on the refracted angle off the Moon's surface. According to the Fresnel equations, two orthogonal polarization components can be parameterized as:

\begin{align}
    \label{eq:Fresnel}
    T_{p}(\theta_{t}) &= 1 - \left|{\frac {\sqrt{\varepsilon}\; \cos \theta _{t}\;-\;\cos \theta _{i}}{\sqrt{\varepsilon}\; \cos \theta _{t}\;+\;\cos \theta _{i}}}\right|^{2}, \\
    T_{s}(\theta_{t}) &= 1 - \left|{\frac {\sqrt{\varepsilon}\; \cos \theta_{i}\;-\; \cos\theta _{t}}{\sqrt{\varepsilon}\; \cos \theta _{i}\;+\;\cos \theta _{t}}}\right|^{2},
\end{align}
where $\epsilon$ is the dielectric constant of the lunar regolith, and $T_p$ and $T_s$ are the transmitted power of radiation with the polarization perpendicular and parallel to the plane of incidence, respectively. Incident ($\theta_{i}$) and refracted ($\theta_{t}$) angles are the angles that light rays make relative to the normal of the interface surface. To illustrate the variables mentioned above, a schematic is shown in the top part of Figure~\ref{fig:moon_pol_mechanism}.

The dielectric constant of the lunar regolith $\epsilon$ has been measured, especially from the lunar samples brought back by the Apollo program~\citep{olho1973, olho1975, apollo2012}.
The roughness of the lunar surface reduces the coherence from a smooth-surface lunar model, as assumed in Figure~\ref{fig:moon_pol_mechanism}. This tends to reduce the inferred dielectric constant of the lunar regolith from the measured physical value. \citet{loso67} measured the Moon polarization properties at 37.5\,GHz with a 22\,m radio telescope and concluded that the inferred dielectric constant for a smooth-surface lunar model is \begin{equation}
\label{equ:epsilon}
    \epsilon = 1.5 \pm 0.2.
\end{equation}

In the smooth-surface lunar model, the normal direction is determined at each point of the lunar sphere. The incident and refracted angles are related by Snell's law $\sqrt{\varepsilon}\sin{\theta_{i}} = \sin{\theta_{t}}$, where $\varepsilon$ is the dielectric constant of the Moon's regolith. The different amplitudes between $T_p$ and $T_s$ generate a net linear polarization. The polarization fraction, a function of the refraction angle $\theta_{t}$ only, is defined as
\begin{equation}
    p(\theta_{t}) = \frac{1}{2}\left|{T_p(\theta_{t}) - T_s(\theta_{t})}\right|,
\end{equation}
where $1/2$ comes from the fact that unpolarized light from the regolith can be evenly divided into two orthogonal linear polarization states. At the center where $\theta_t = 0^\circ$, $p(\theta_t)$ equals zero, indicating there is no net polarization at the center of the Moon. The trends of variables $T_p$, $T_s$, and $p$ are shown in the lower part of Figure~\ref{fig:moon_pol_mechanism}. The polarization fraction $p$ increases from the lunar center to the limb~\citep{2012MoonPol_Measurements}. Also shown are the shaded regions for each of the curve. The shaded regions are calculated by varying the effective dielectric constant $\epsilon$ from 1.3 to 1.7 (Equation~\ref{equ:epsilon}). The shaded region around the polarization fraction demonstrates that the value can vary by $\pm40\%$ around the mean, due to the uncertainty of the effective dielectric constant. 

\begin{figure}
    \centering
    \includegraphics[width=1.\linewidth]{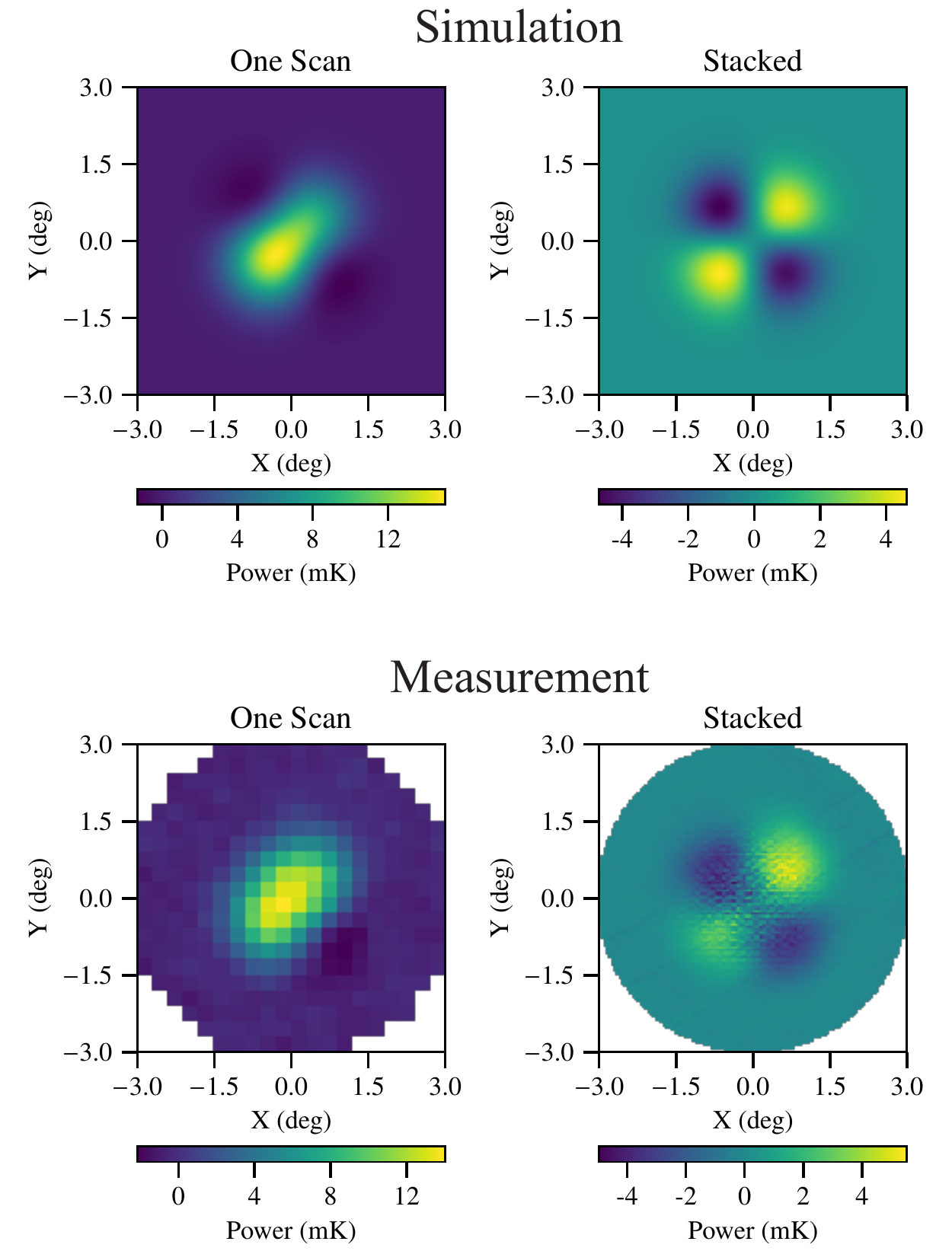}
    \caption{Moon polarization (Stokes $U$) maps. The top two maps are from our simulations. The left map shows the simulated Stokes $U$ map for one specific Moon scan. These simulated maps were generated for selected dedicated Moon scans in Era~1, given the time and the telescope pointing information during each Moon scan. Those maps were then stacked to form the map shown on the right. The bottom two plots show the measured results from one of the central detector pairs. Following the same format, the measurement from the same Moon scan is shown on the left, and the overall stacked map is shown on the right. Significant monopole and dipole components are seen in the single-scan map; both the shape and the amplitude of the pattern are consistent between the simulation and the measurement. The pixel size was chosen to be 0.3\dg{} because of the sparse sampling from one single Moon scan. The stacked map was formed from stacking over 200 scans for the same detector pair. The monopole and dipole components are significantly reduced from averaging over different scans. The pixel size also decreases to 0.05\dg{} because of the increased amount of data.}
     \label{fig:moon_sim_meas}
\end{figure}

The polarization intensity is calculated by multiplying the brightness temperature (from Section~\ref{subsec:moon_intensity_model}) and the polarization fraction at each location of the Moon. 
The polarization information is decomposed into the Stokes parameters for measurement. 
In each detector coordinate system, the CLASS telescopes are sensitive to the $\pm 45$\dg{} polarization directions, equivalent to Stokes $U$. 
Accounting for the CLASS 1.5\dg{} beam by convolution, we can obtain the simulated Moon Stokes~$U$ maps for the CLASS 40~GHz telescope. 
If the Moon had a perfectly uniform thermal distribution, the observed Stokes $U$ maps should have a quadrupole pattern without any monopole or dipole components. However, because of the nonuniform lunar thermal properties, the polarization signal is not completely canceled out at the center of the Moon, creating monopole and dipole polarization components, as shown in the top left map of Figure~\ref{fig:moon_sim_meas}. The monopole and dipole components in Stokes $U$ are variable depending on the angle between the lunar thermal distribution and the detector polarization direction, whose value depends on the Moon phase, the Moon orientation on the sky, and the telescope \bs{} angle. However, the quadrupole component is native to the detector polarization angle, independent of the aforementioned factors.

The Moon was observed at different phases and orientations on the sky, with the telescope at different \bs{} angles. Therefore, when we stack the polarization maps from different dedicated Moon scans, the monopole and dipole components are significantly averaged down while the quadrupole component stays. We simulated Moon polarization maps for selected dedicated Moon scans in Era~1 (selection details are elaborated in Section~\ref{subsec:moon_pol_map}), and stacked them together as shown in the top right plot of Figure~\ref{fig:moon_sim_meas}. As expected, the monopole and dipole components are significantly reduced in the stacked map. 

To estimate the effects from the dielectric constant uncertainty, simulations were performed with different effective dielectric constant values, ranging from 1.3 to 1.7 (Equation~\ref{equ:epsilon}) at a step of 0.05. At different effective dielectric constant values, the shapes of the map features are maintained while the amplitudes vary. The amplitudes (monopole, dipole, and quadrupole) vary by approximately $\pm40\%$ around the central value corresponding to $\epsilon = 1.5$. We use the central value $\epsilon=1.5$ in the following analysis, realizing that the simulated polarization amplitudes have $\sim40\%$ systematic uncertainties. 

At $\epsilon = 1.5$, the amplitude of the quadrupole observed by the 40~GHz telescope is simulated to be around $5$~mK, three orders of magnitude lower than the temperature signal at $\sim$20\,K. In addition, the orientation of the quadrupole pattern is directly related to the polarization angle of the detectors.

\subsection{Polarization Data Processing}
\label{subsec:moon_data_processing}
With the CLASS optical design, the sky polarization signal is first modulated by the VPM. The modulator has a reflective mirror moving behind a static wire grid to inject a phase delay $\phi$ between the two orthogonal polarization states~\citep{chus12, harr18}. For a single-frequency, the phase delay is expressed as
\begin{equation}
\label{equ:vpm_phase_delay}
    \phi = \frac{4 \pi \nu}{c}\, z \, \cos{\theta},
\end{equation}
where $\nu$ is the frequency, $c$ is the speed of light, $z$ is the distance between the wire grid and the reflective mirror, and $\theta$ is the incidence angle to the VPM. The reflective mirror moves at a frequency of 10~Hz, modulating the phase delay $\phi$ at the same frequency. The VPM radiation transfer function can be expressed as a function of the phase delay $\phi$:
\begin{equation}
    \begin{pmatrix}
    I'\\
    Q'\\
    U'\\
    V'\\
    \end{pmatrix}
    =
    \begin{pmatrix}
    1 & 0 & 0 & 0 \\
    0 & 1 & 0 & 0 \\
    0 & 0 & \cos{\phi} & \sin{\phi} \\
    0 & 0 & -\sin{\phi} & \cos{\phi} \\
    \end{pmatrix}
    \begin{pmatrix}
    I\\
    Q\\
    U\\
    V\\
    \end{pmatrix},
\end{equation}
where $I,\ Q,\ U,\ V$ are the Stokes parameters for the incoming radiation while $I',\ Q',\ U',\ V'$ are the Stokes parameters for the radiation leaving the VPM \citep{mill16, harr18}. The transfer function depicts how the VPM transfers the sky polarization signal to the modulated TOD, and, thus, how to recover the original polarization signal. The above calculation is only for a simplified single-frequency model; a more realistic study is further described in a companion paper (Harrington et al. 2019, in preparation). 

The \demod{} TOD contain the polarization signal from the sky. We analyze the demodulated TOD in detector pairs to remove common modes. Different detectors tend to have slightly different gains. So, before analyzing the demodulated TOD in detector pairs, the gains are first balanced according to the measured Moon intensity from the corresponding Moon scans. Gain-balanced \demod{} TOD from paired detectors are analyzed together to form pair-differenced \demod{} TOD for each feedhorn. The pair-differenced \demod{} TOD are then projected to form Moon polarization maps, similar to the intensity maps described in Section~\ref{sec:intensity_beam}. Note that the beam maps are detector-centered, while the maps used to study the Moon are Moon-centered.
Only dedicated Moon scans are used for the following analysis since the \pseudo{} do not provide sufficient sampling around the Moon.

\subsection{Moon Polarization Maps}
\label{subsec:moon_pol_map}

The CLASS detectors are oriented to be sensitive to $\pm$45\dg{} linear polarization directions (Stokes $U$). These angles are with respect to the optical plane, as shown in Figure~\ref{fig:tele_beam}. Therefore, the Moon polarization maps are presented in Stokes $U$ in the following text. Moon polarization maps from one dedicated scan normally show significant monopole and dipole components, as predicted by the simulation. The bottom left map in Figure~\ref{fig:moon_sim_meas} shows the measured map from one observation, matching the simulated map above. Even though the measured map has a lower resolution compared to the simulated map, both the shape and the amplitude of the pattern are consistent between the simulation and the measurement. This verifies the fidelity of the Moon polarization model.

Next, the measured Moon polarization maps were stacked. According to the simulation, stacking boosts the quadrupole component while suppresses the monopole and dipole components. The Moon polarization maps from the dedicated Moon scans are selected according to several criteria, including VPM status, observation elevation, and noise level. On average, ${\sim}200$ polarization maps are available for each detector pair. The maps from different scans are then stacked for each detector pair. Figure~\ref{fig:moon_sim_meas} shows the stacked Moon polarization map on the bottom right. The monopole and dipole components are significantly reduced in the stacked map, matching the simulation result. Meanwhile, the quadrupole pattern emerges, with the amplitude of  $\sim$5\,mK. Both the shape and the amplitude of the quadrupole are consistent with the simulation.

\subsection{Polarization Angle Determination}
\label{subsec:pol_ang}
Although the nominal detector polarization angles are $\pm$45\dg{}, the realized directions are usually not exactly at those values because of optical distortion and assembly misalignment. This polarization direction determines our interpretation of the polarization data from the sky. 
However, the relative 90\dg{} angle between pair detectors is set by microfabrication to very high precision.
Misunderstanding of the direction results in mixing different polarization components, namely E and B mode polarization. Since the E modes are orders of magnitude stronger than the B modes, the mixing of them would impair our ability to detect the primordial B modes in the CMB. 

The CLASS telescopes are designed to allow use of removable wire-grid polarization calibrators. During the calibration operation, a calibrator is installed in front of the VPM, at the bottom opening of the forebaffle. The wire-grid partially polarizes the incoming signal along the axis aligned with the wire direction, the relative angle of which is known to sub-degree precision. The wire grid is rotatable, providing polarization signals with tunable linear polarization direction. Details on this calibration operation will be described in a later companion paper. However, polarization angles measured through this method are in the near-field region, whereas the polarization angles on the sky are in the far-field region. Although the far-field polarization angles can be simulated using the measured near-field polarization angles, ideally these angles are measured directly, such as by observing celestial objects.

The orientation of the Moon polarization quadrupole pattern can be used to determine the telescope far-field polarization angle. Gauss--Hermite decomposition is an effective tool to extract quadrupole components in a map. Gauss--Hermite patterns form a complete and orthogonal basis for a two-dimensional beam map~\citep{bicep15,2016HWP_Tom}, with an analytical expression as
\begin{equation} 
\begin{split}
    \label{eq:gauss_hermit}
    f_{i,j} (\theta, \phi)
    =& \left( \frac{\textrm{exp} [- \theta^2 / (2 \sigma^2)]}{\sqrt{2^{i+j} i! j! \pi \sigma^2}} \right) \\
    &\times H_i \left( \frac{\theta \cos \phi}{\sigma} \right)
    H_j \left( \frac{\theta \sin \phi}{\sigma} \right) ,
\end{split}
\end{equation}
where $\theta \  \textrm{and} \  \phi$ describe the map in a polar coordinate system, $H_i \textrm{ and } H_j$ are Hermite polynomials, and $\sigma = \textrm{FWHM}/\sqrt{8 \ln 2}$ is the Gaussian width of the beam.

\begin{figure}
    \centering
    \includegraphics[width=1.0\linewidth]{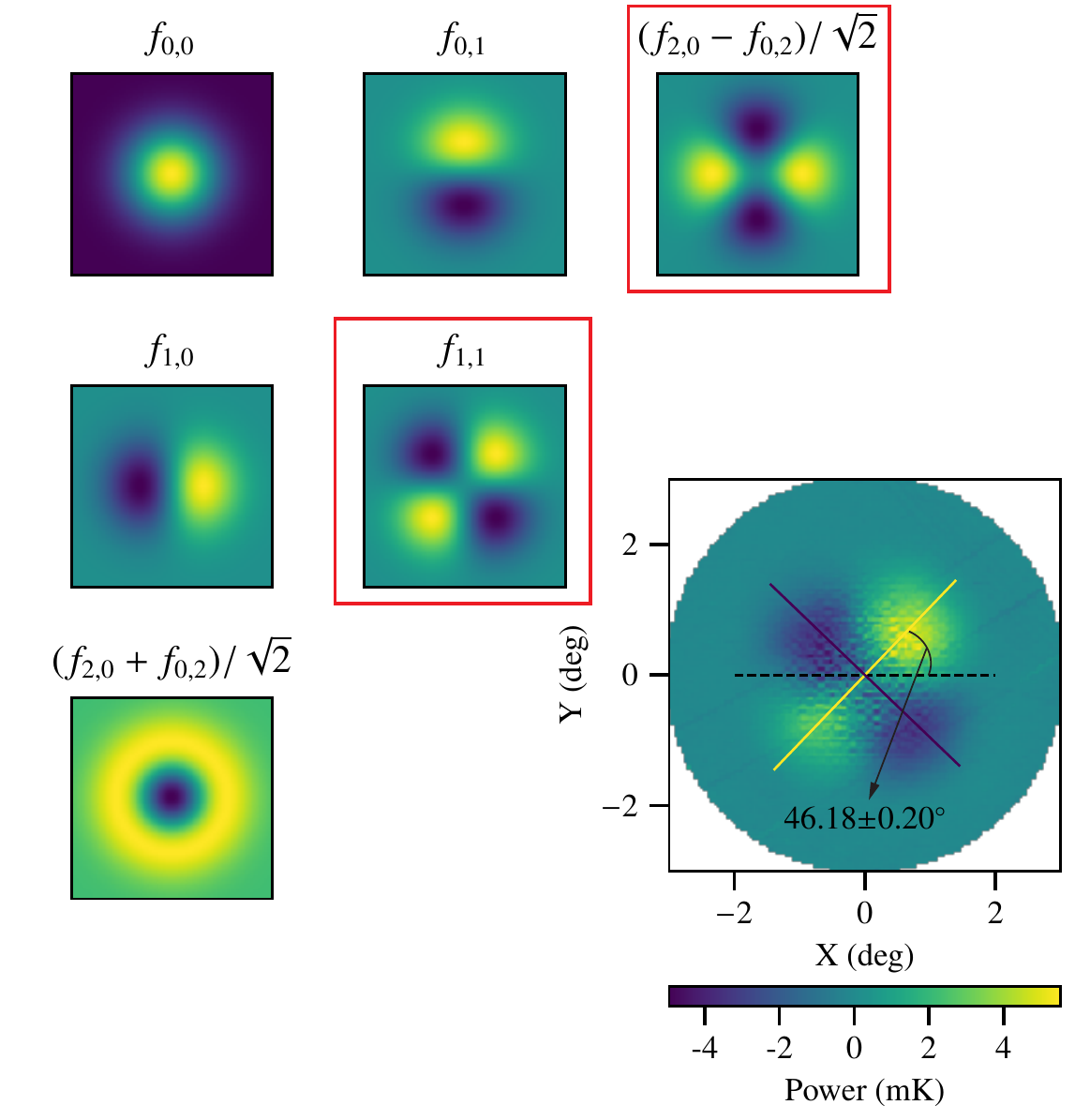}
    \caption{On the top left part, basic Gauss--Hermite patterns are shown with $f_{i, j}$ defined in Equation~\ref{eq:gauss_hermit}. The two quadrupole components are emphasized with red boxes. The sum of the indices $i + j$ is defined as the order of the patterns. Polarization angle determination for the stacked Moon polarization map in Figure~\ref{fig:moon_sim_meas} is shown in the bottom right. Applying the Gauss--Hermite separation method, the polarization angle for this detector pair is determined to be $46.18\pm0.20$\dg{}, with the uncertainty estimated by bootstrapping. Auxiliary lines are shown to represent the orientation of the quadrupole pattern.}
    \label{fig:pol_ang_det}
\end{figure}

Basic Gauss--Hermite patterns are shown in the top left part of Figure \ref{fig:pol_ang_det}. The two orthogonal quadrupole patterns are rotated by 45$^\circ$. The ratio of the two patterns determines the orientation of the combined quadrupole pattern. The completeness and orthogonality of the Gauss--Hermite basis guarantees that all quadrupole information is contained within these two patterns.

Stacked Moon polarization maps for each detector pair were projected to the Gauss--Hermite patterns up to an order of 10, meaning $i + j  \leqslant  10$. Orders greater than two carry power at least one order of magnitude lower than those from the first three. Gauss--Hermite patterns at the order of two contain the quadrupole information. The quadrupole pattern can be fully recovered with the fitted coefficients of two orthogonal Gauss--Hermite quadrupole patterns. The polarization angles were then calculated from the coefficients, which yields the far-field detector polarization angle. In order to estimate the uncertainty of the polarization angle measurement, we used the Moon polarization maps from individual Moon scans as the parent sample and generated 5000 bootstrap samples. Then we stacked each of bootstrap samples to obtain 5000 stacked maps. Finally, we measure the 5000 stacked maps to estimate the uncertainty of the angle measurement. The lower right part of Figure~\ref{fig:pol_ang_det} shows the fitting result of one detector pair as an example.

\begin{figure}
    \centering
    \includegraphics[width=1.0\linewidth]{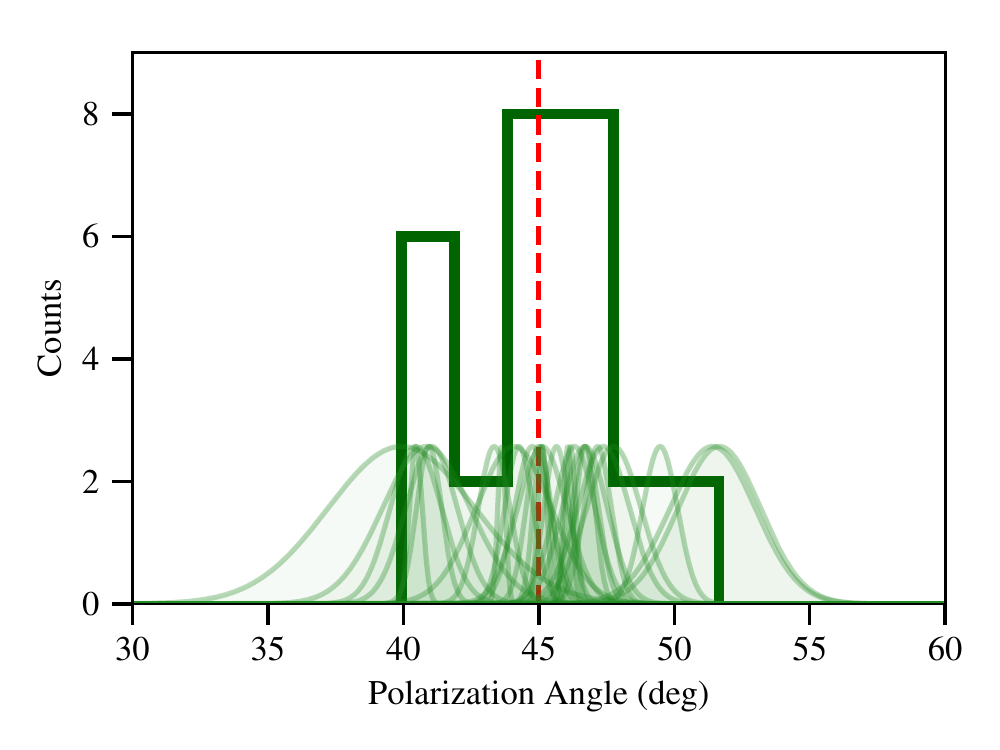}
    \caption{Polarization angle histogram and distribution. The measured polarization angles of different detector pairs are plotted as a histogram, with a 2\dg{} bin width. The nominal 45\dg{} is indicated by a dashed vertical line. The bootstrapped distributions of each measured angle are presented as shaded Gaussian areas. Measurements from different detector pairs are overlaid.}
    \label{fig:pol_ang_dist}
\end{figure}
The same analysis is performed on all of the operational detector pairs; the results are shown in Figure~\ref{fig:pol_ang_dist}. We reached sub-degree-level polarization angle constraints, except for two outliers. The polarization angles center around 45\dg{}. Some deviation from the nominal 45\dg{} is expected in the optical model; the extent depends on the detector pair's location on the focal plane.

\subsection{Temperature-to-polarization Leakage}
\label{subsec:t-to-p_leakage}

Knowledge of the temperature-to-polarization leakage is a critical piece of information for CMB polarization experiments. The Moon is an ideal celestial object for this study since it simultaneously emits bright intensity and faint polarization signals, differing in amplitude by a factor of $>10^3$.

The ``leakage'' from temperature signals in the polarization maps would present a monopole pattern resembling a temperature map. The stacked Moon polarization map does not show a significant monopole component, as shown in Section~\ref{subsec:moon_pol_map}. In order to study the temperature-to-polarization leakage, we need to look at Moon polarization maps for individual dedicated Moon scans. The left two plots in Figure~\ref{fig:moon_sim_meas} show the comparison between a measured Moon polarization map and a simulated map for one detector pair during one dedicated scan. The measured pattern clearly contains a monopole component. This component comes from a combination of the Moon's intrinsic polarization and the temperature-to-polarization leakage. Meanwhile, both maps also contain a quadrupole pattern. Since the leakage from temperature or circular polarization should not contain any quadrupole patterns, only the intrinsic polarization signal from the Moon could account for this signal. 

\begin{figure}
    \centering
    \includegraphics[width=1.0\linewidth]{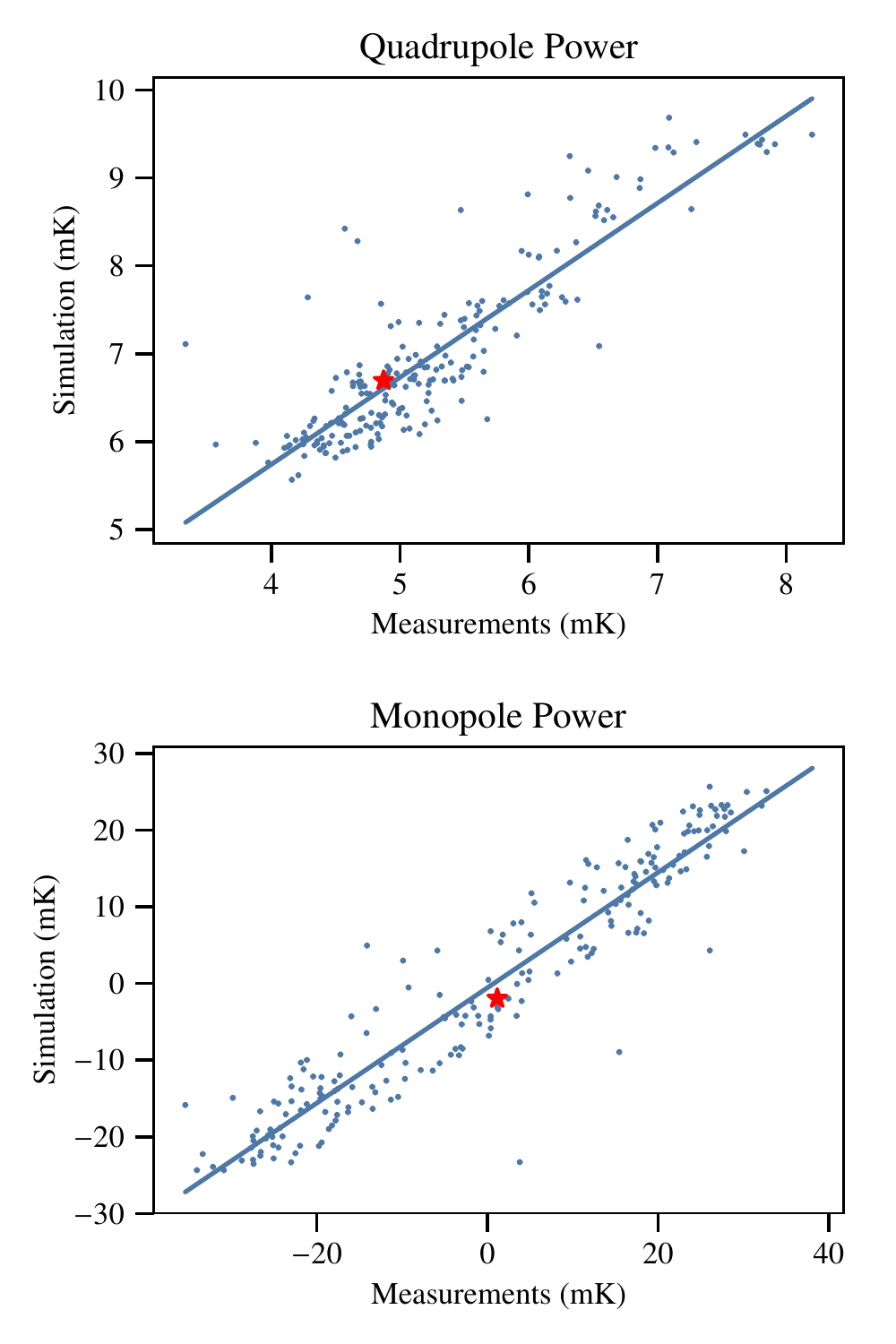} 
    \caption{The two plots show the measured versus the simulated amplitude for the quadrupole pattern and the monopole pattern of the Moon polarization map, respectively. Each data point comes from a dedicated Moon scan, whereas they all come from one detector pair. 
    The scan shown in Figure~\ref{fig:moon_sim_meas} is denoted with red stars. A line is fit to the data in each of the plots.}
    \label{fig:meas_sim_comparison}
\end{figure}

Focusing on the same detector pair used in the previous figures, Figure~\ref{fig:meas_sim_comparison} shows the the measured amplitude versus the simulated amplitude in the quadrupole and monopole patterns for each dedicated Moon scan. Data points for the two plots come from ${\sim}200$ dedicated Moon scans with that detector pair, with the specific scan in Figure~\ref{fig:moon_sim_meas} emphasized. Linear trends are fit to the data. The fitted lines for the two components are
\begin{equation}
\begin{split}
    \textrm{Measurement} &= (1.01\pm0.04) \times  \textrm{Simulation} \\
    &- (1.80\pm0.20)\,\textrm{mK} \quad \textrm{(Quadrupole)}, 
    \label{equ:quad_fit}
\end{split}
\end{equation}

\begin{equation}
\begin{split}
    \textrm{Measurement} &= (1.33\pm0.04) \times \textrm{Simulation}\\
    &+ (0.75\pm0.44)\,\textrm{mK} \quad \textrm{(Monopole)},
    \label{equ:mono_fit}
\end{split}
\end{equation}
where the uncertainties are at a 68\% confidence level. As discussed in Section~\ref{subsec:moon_pol_model}, there is also a $40\%$ systematic uncertainty in the slope due to the uncertainty in the dielectric constant of the lunar regolith. We use the calibration factor from the intensity observation in \citet{appe19} to convert the measured power (in fW) to sky temperature (in mK). The slope for the quadrupole signal is at 1.01, showing the consistency between the intensity and polarization calibration. The nonzero intercept value may come from a combination of imperfections in the Moon polarization model and the fact that the zero value is extrapolated far from the measured data. 

The slope for the monopole is $\sim$30\% greater than one, meaning we are detecting more power than the simulation. However, we found that the measured monopole polarization pattern is not correlated with the variation of the measured brightness temperature amplitude, ruling out the possibility that the nonzero slope is due to temperature-to-polarization leakage. The difference in the slope is most likely from limitations in the Moon model, especially the stratified thermal model of the Moon surface. The monopole component originates from the nonuniform thermal properties, so an error in modeling this nonuniformity directly affects the monopole component. However, the quadrupole component comes from the circular geometry of the Moon, more immune to errors in the stratified model.

The valuable information is in the intercept, which provides strong constraints on the temperature-to-polarization leakage. The monopole amplitude data evenly cover negative and positive values, enabling a reliable fit of the intercept unaffected by the slope value. When the simulated monopole is zero, implying that the intrinsic monopole component is zero, the measured value tells us the level of temperature-to-polarization leakage. If we set the simulation value to zero in Equation~(\ref{equ:mono_fit}), the leakage can be estimated as
\begin{align}
    \textrm{Monopole} = 0.75\pm0.44\, \textrm{mK}\quad (68\%\ \textrm{C.L.}).
\end{align}
The measured Moon brightness temperature is $\sim$17\,K \citep{appe19} with a relative uncertainty much smaller than that of the intercept. Therefore, we only include the uncertainty of the intercept. Thus, the temperature-to-polarization leakage is estimated as:
\begin{align}
\textrm{T-to-P Leakage} &= \frac{0.75\pm0.44 \times 10^{-3}\,\textrm{K}}{17\,\textrm{K}} \\
&= 4.4\pm2.6 \times 10^{-5}\quad (68\%\ \textrm{C.L.}).
\end{align}

\subsection{Forebaffle Blackening}
\label{subsec:forebaffle_blackening}
\begin{figure}
    \centering
    \includegraphics[width=0.9\linewidth]{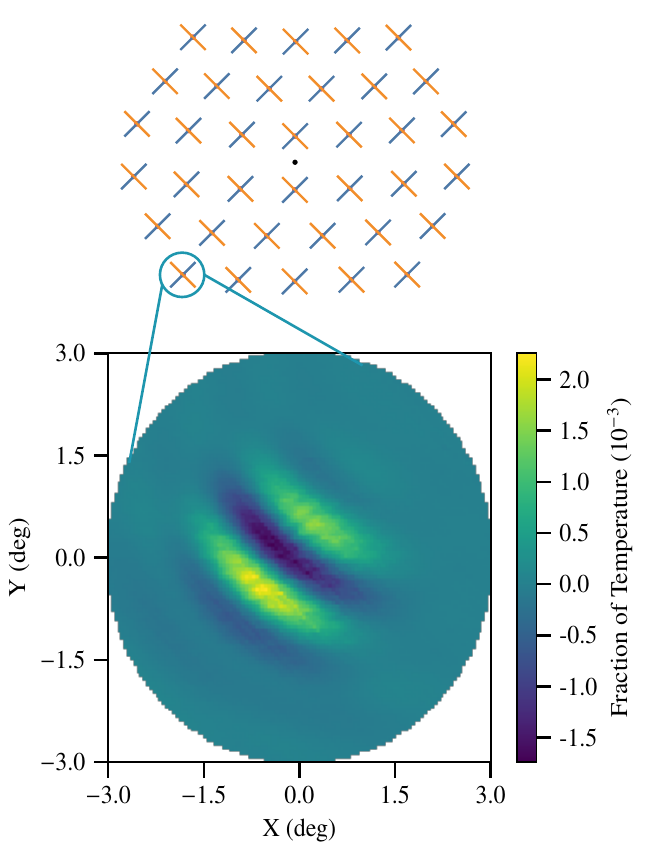}
    \caption{Stripes in the Moon polarization map. The polarization map for one detector pair is shown at the bottom. The position of this detector pair in the focal plane is illustrated in the upper part of this figure. Stripy patterns were observed in this polarization map, with their shape matching the circle of the forebaffle aperture. The amplitude of the pattern was around 0.1\% of the temperature signal. The pattern is believed to be caused by the reflective surface of the forebaffle aperture, since it was eliminated after the forebaffle inner surface was blackened.}
    \label{fig:stripes}
\end{figure}

The high brightness of the Moon enables us to probe low-level systematics, which guides the improvement of the instrument. In the initial stages of the Era~1 observation campaign, the inner surface of the forebaffle was reflective. Meanwhile, in the Moon polarization maps, edge pixels showed stripes resembling the circle of the forebaffle aperture, as shown in Figure~\ref{fig:stripes}.

We soon realized that the striped patterns may have come from the reflective forebaffle converting some of the lunar intensity radiation into polarization signals, as the forebaffle is in front of the VPM. Additionally, GRASP simulations showed that reflection from the forebaffle could create these features (Appendix~\ref{app:grasp_em}). To fix this, we covered the inner surface of the forebaffle with Eccosorb HR-10 sheets from Emerson \& Cuming.\footnote{Emerson \& Cuming \url{http://www.eccosorb.com/}} The striped pattern was then eliminated, revealing the quadrupole patterns expected for the Moon. The polarization analysis results described before this section are mostly from data taken after the forebaffle was blackened, especially for the edge pixels.

\begin{figure*}
    \centering
    \includegraphics[width=0.33\linewidth]{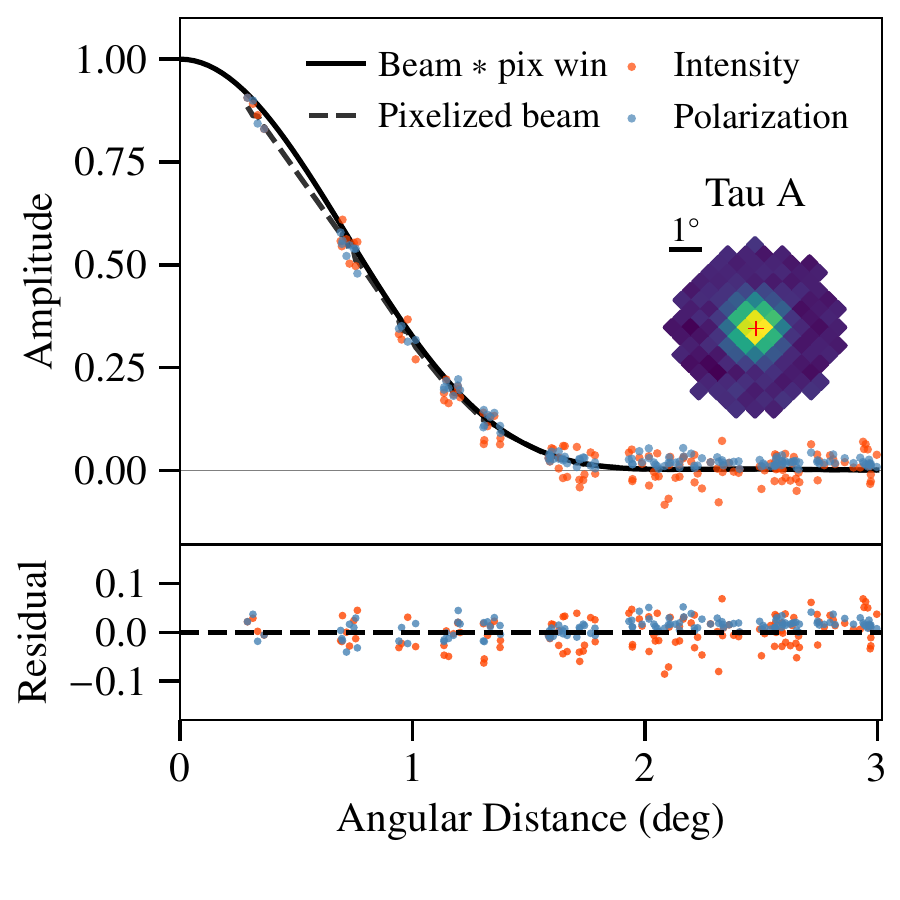}
    \includegraphics[clip=True, trim={.08\linewidth, 0 0 0}, width=0.278\linewidth]{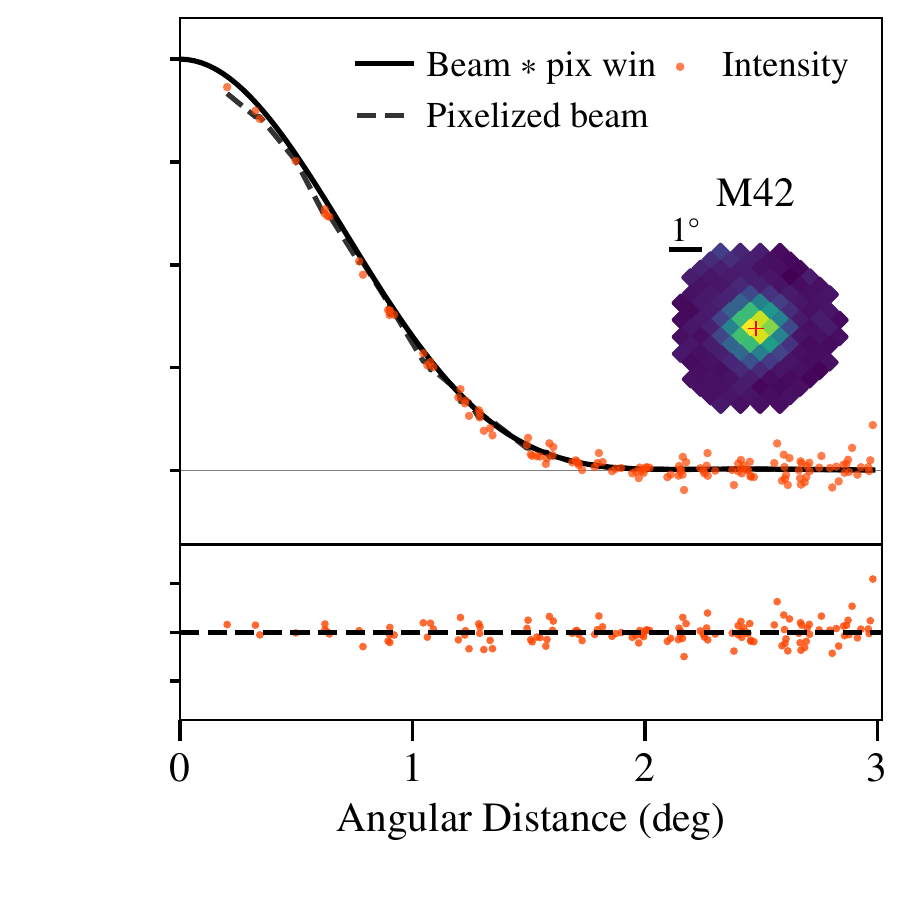}
    \includegraphics[clip=True, trim={.08\linewidth, 0 0 0}, width=0.278\linewidth]{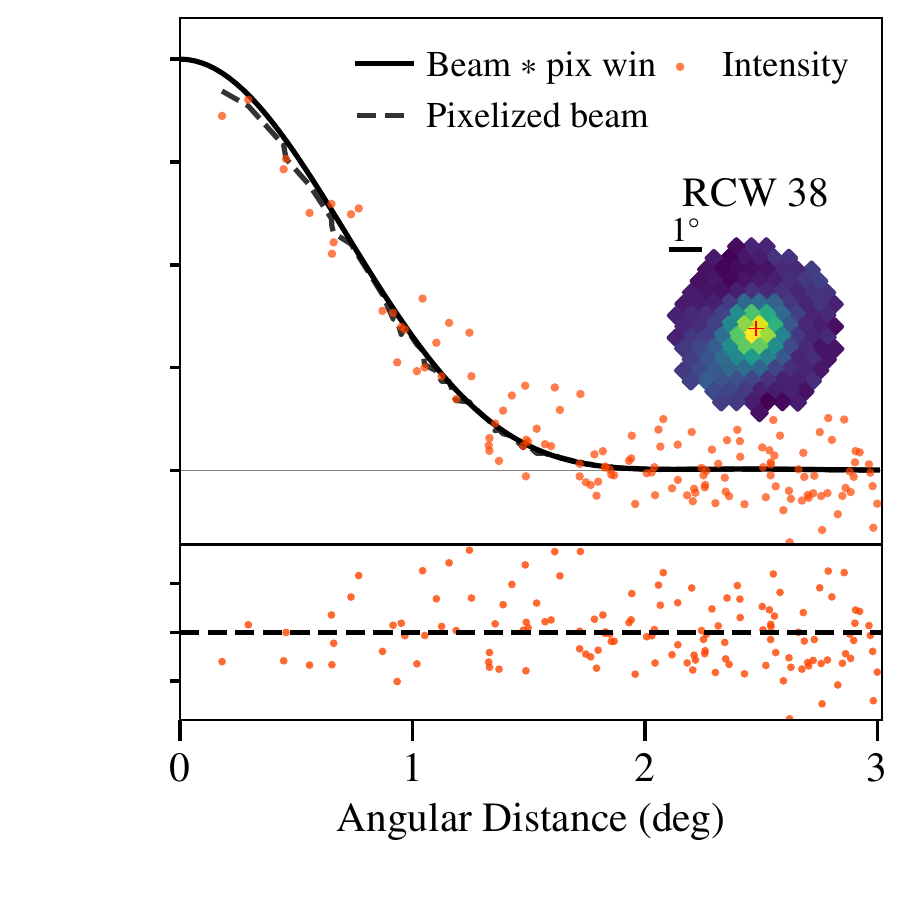}
    \caption{Radial profiles measured from point sources. From left to right are the three brightest off-plane sources, Tau A, M42, and RCW~38. The fit is performed locally on a small patch of the sky centered on the source with a radius of $3^\circ$ (inset plots) to avoid complex structures. For Tau A, we show both the radial profile points in temperature (red) and polarization (blue, $P=\sqrt{Q^2 + U^2}$). No polarized signal is expected or detected in M42 and RCW~38. The solid line shows the beam profile smoothed by the pixel transform. The dashed line shows the beam profiles pixelized as in the survey map. The bottom panels show the residuals with respect to the pixelized beam profile.}
    \label{fig:survey_beam}
\end{figure*}

\section{Comparison to Survey Maps}
\label{sec:comparison_survey_map}
During Era~1, about $60\%$ of calendar time was spent on CMB observations to cover $75 \%$ of the sky, which includes several other bright point sources used for calibration. The data selection and reduction for temperature are similar to those for Moon scans. 
The high-pass filtered TOD for each detector are projected onto sky coordinates and binned into \textsc{HEALPix} \citep{healpix}
pixels with \texttt{NSIDE}=128 to produce the map. 
The polarization signal is then recovered from the demodulated TOD. 
Stokes $Q$ and $U$ parameters are solved for from 28 pair-differenced operational detector pairs and are projected in the same way to make the polarization map. A detailed description of the mapping pipeline can be found in companion papers (Eimer et al. 2020, in preparation, Parker et al. 2020, in preparation).

\begin{figure}
    \centering
    \includegraphics[width=1\linewidth]{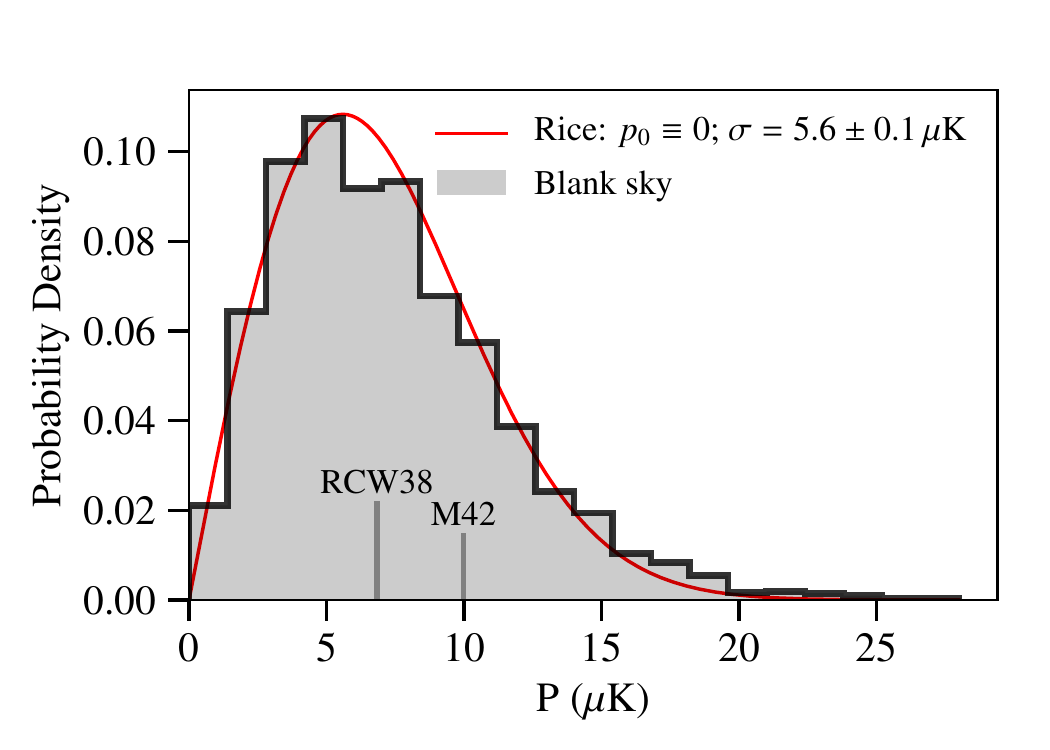}
    \caption{The biased polarization measurements from non-polarized sources. The histogram is the best-fit polarization intensities from randomly chosen blank sky regions with Galactic latitude $b>12^\circ$. The histogram is fit with a Rice distribution (red) with a prior polarization intensity $p_0$ = 0 (fixed) and a standard deviation $\sigma=5.6\pm0.1\,\mathrm{\mu K}$. Polarization of M42 and RCW~38 are measured in the same way and are indicated by the vertical lines. Adopting a flat prior probability distribution of $p_0$, we derive the $90\%$ quantile of the prior polarization. Assuming no intrinsic polarization of M42 and RCW~38, we conclude the temperature-to-polarization leakage of the two sources is smaller than $1.7\times10^{-3}$ and $1.8\times10^{-3}$, respectively, at a 90\% confidence level.}
    \label{fig:rice_hist}
\end{figure}

\subsection{Telescope Beam in Intensity and Polarization}
We check the consistency of the beam profile as seen in the survey map with that from the Moon scans by fitting the radial profile of the brightest point sources in the survey. The profile in map coordinates $(\alpha, \delta)$ is modeled as
\begin{equation} 
    S = A\, b_p(\alpha, \delta) + k_\alpha \alpha + k_\delta \delta + \mathrm{offset},
\end{equation}
where $S$ is modulated over the $I$, $Q$, and $U$ components by the VPM. For each component, we obtain the pixelized beam model $b_p$ by convolving $b(\theta)$ from the Moon analysis with a delta function centered on the source (from SIMBAD\footnote{\url{https://simbad.u-strasbg.fr/simbad/}}), and project it onto the \texttt{NSIDE}=128 \textsc{HEALPix} map. 
The additional variation from extended emission and the filtering are taken into account by the slope factors ($k_\alpha, k_\delta$) and an offset. We show in Figure \ref{fig:survey_beam} the result for the three brightest sources off the Galactic plane: Tau A, M42, and RCW~38, where the dots represent the normalized data after removing the slopes and the offset from the best fit. 

The flux density of Tau A measured in this way is ${308\pm11\,\mathrm{Jy}}$ at $38.4\pm0.2\,\mathrm{GHz}$, in agreement with the \textit{Wilkinson Microwave Anisotropy Probe (WMAP)} time-dependent model \citep{appe19}. The polarization fraction measured from the fit above is $7.78\% \pm 0.11\%$, 
which is consistent with the result from \textit{WMAP}~\citep{2011ApJS..192...19W}. 
M42 and RCW~38 are \ion{H}{2} regions dominated by Bremsstrahlung and are not expected to have polarized signal \citep{Planck2015Catalog}.
We, therefore, assume no polarization from the two sources and use the measured polarization fraction to constrain the temperature-to-polarization leakage of the 40\,GHz telescope. 
The biased estimation of polarization follows the Rice distribution \citep{Rice}, characterized by the intrinsic polarization $p_0$ and the uncertainty in $Q$ and $U$ measurements $\sigma$. 
We fix $p_0=0$ and fit $\sigma$ to the distribution of the measured polarization from randomly chosen blank patches in the map. 
We show in Figure \ref{fig:rice_hist} the best-fit result, along with the measurements of M42 and RCW~38. 
Assuming that the sources are drawn from a Rice distribution with the same $\sigma$ and a flat prior distribution of $p_0$ (from temperature leakage), and integrating over the posterior distribution of $p_0$ with the (biased) polarization measurement from the two sources, we constrain the upper limit of the temperature-to-polarization leakage to be $1.7\times10^{-3}$ and $1.8\times10^{-3}$ at the $90\%$ confidence level for M42 and RCW~38, respectively.

\subsection{Polarization Angles}
\begin{figure}[ht!]
    \centering
    \includegraphics[width=1\linewidth]{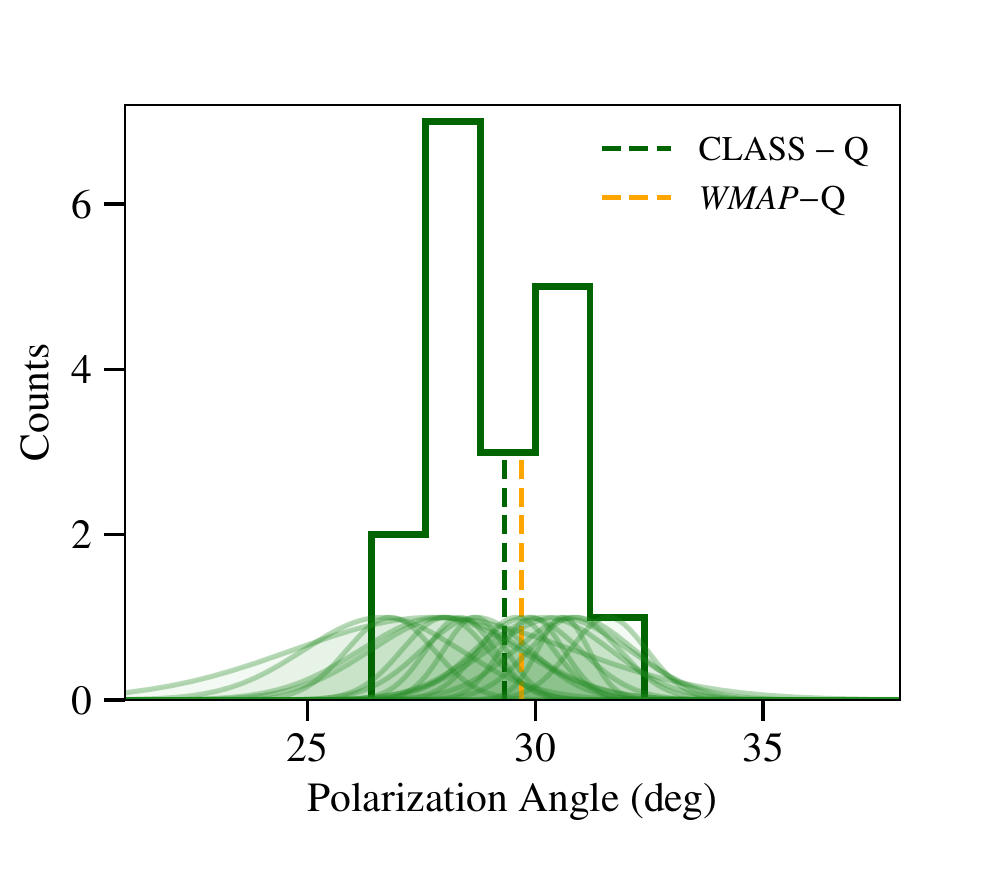}
    \caption{Polarization angle measurements by \textit{WMAP} and CLASS agree well. Tau A polarization angle (equatorial coordinates, CMB convention) measured by 19 detector pairs that cover Tau A in the survey. The green histogram is the distribution of the measurements, while the measurements from individual pairs are shown with errors by the shaded regions. The angle determined from the total map is indicated by the dashed green line, in comparison to the angle measured by \textit{WMAP-Q} band \citep[][orange]{2011ApJS..192...19W}. 
    }
    \label{fig:pol_ang_hist}
\end{figure}

Tau A is the brightest polarized source in the survey and is used for polarization angle calibration. The polarization angle is determined from the best-fit amplitude of the $Q$/$U$ maps,
\begin{equation}
    \psi = \frac{1}{2}\arctan{\frac{U}{Q}}.
\end{equation}
We fit for 19 individual detector pairs that have Tau A coverage, as well as the total map that combines 28 operational pairs. The tightest constraint on the Tau A polarization angle at the 40\,GHz band is given by \textit{WMAP}~\citep{2011ApJS..192...19W}. The comparison between our measurement and the angle measured by \textit{WMAP} is shown in Figure \ref{fig:pol_ang_hist}. The angle measured by each detector pair is consistent with \textit{WMAP} within the fitting uncertainty (${\sim}1^\circ$). The scatter of the angles is consistent with the optical model and the Moon observation result.

\section{Conclusion}
In this paper, we present the optical characterization and calibration of the CLASS 40\,GHz telescope during  Era~1 observations (\EraOneStart{} to \EraOneEnd{}). Our primary calibrator is the Moon, which, at an $\mathrm{S/N}$ ratio of $10^5$, provides precision checks of pointing, beams (including far sidelobes), and temperature-to-polarization (T-to-P) leakage. We present a model adapted from \cite{2012ChangE_data} for the unpolarized and polarized emission of the Moon that accounts for partial illumination. We fit this model to 822 separate Moon datasets, taken from July 2016 to March 2018, both as dedicated observations and during the CMB survey.

The telescope pointing was constantly monitored using both dedicated Moon scans and lunar data collected as the Moon crossed the CMB survey. The telescope boresight pointing deviated around $1.4'$ in reference to corresponding pointing models. Individual detector positions are measured within $2''$ in reference to the telescope boresight pointing. Together, these errors combine to produce 0.3\% smoothing of the beam in the survey maps due to ``beam jitter.'' Although negligible, this broadening is accounted for in the following cosmological analysis. Differential pointing in paired detectors is normally within $0.5'$. The detector array pointing pattern agrees with GRASP physical optics simulations of the original optical design up to an extra $2.5\%$ magnification.

Beam information for each detector is accurately characterized, also from the Moon observations. Per-detector beam maps to angular radius 10\dg{} were stacked from, on average, 600 Moon scans. Measuring the main beam as gives a median FWHM of 1.52$^\circ$ (1.62$^\circ$) for the minor (major) axes. These FWHM values along with rotation angles of the ellipses match GRASP simulations, with a $5\%$ magnification consistent with that seen in the pointing analysis.  Due to the high $\mathrm{S/N}$ ratio of the per-detector beam maps, we are able to detect optical ghosting and electrical cross-talk at the (expected) level of $10^{-3}$ relative to the peak beam response. These features are significantly reduced and symmetrized in the survey maps, which are averaged over all detectors at different boresight angles. 

To compute the beam window function for the survey, individual beam maps were combined with the appropriate boresight angle weightings. A deconvolution procedure was developed to remove the effect of the finite size of the Moon from this composite ``cosmology beam.'' The beam profile and solid angle are calculated from the deconvolved beam map. The beam is symmetric with an FWHM of $1.579^\circ\pm0.001^\circ$. The solid angle is $838\pm6$~$\mu$sr. Additionally, a far-sidelobe map extending to $90^\circ$ in radius is made with a destriping map maker. We combine the far-sidelobe map with the 10\dg{} map, to construct a 90\dg{} beam map. No obvious sidelobe features are observed in the 90\dg{} beam map. Beam window functions are computed as the Legendre transforms of both the $10^\circ$ and $90^\circ$ beam profiles. Consistency between these demonstrates that the far sidelobes have a negligible impact on the beam window function. 

CLASS also observes the polarization of the Moon at a level of 10$^{-3}$ compared to its intensity. We made Moon polarization maps in the native instrument Stokes~$U$ signal. Such a Stokes~$U$ Moon map shows a combination of a monopole pattern, a dipole pattern, and a quadrupole pattern. We observed consistent trends between the simulated and measured amplitudes from the monopole and quadrupole patterns. Detailed analysis on the tends can be used to constrain the physical properties of the Moon regolith in a future work.
Residuals of the observed monopole and quadrupole terms compared to our emission model  constrain T-to-P leakage to be below $10^{-4}$ (95\% C.L.). Furthermore, the observed orientations of the Stokes~$U$ quadrupoles are consistent with the designed polarization angles of the instrument (though additional data from a more strongly polarized source is needed for a more accurate angle measurement).

The beam and polarization properties are also checked with unresolved sources in preliminary temperature and polarization survey maps. 
Intensity profile (from Tau~A, M42, and RCW~38) and polarization profiles (from Tau~A) are checked to match the cosmology beam. 
Polarization angle measurements of Tau~A from a subset of detectors are consistent within a degree of the angle measured by \textit{WMAP}. 
Observations of the unpolarized RCW~38 and M42 constrain T-to-P leakage at the $10^{-3}$ level, again consistent with the stronger constraint from the Moon observations.

This paper is one in a collection covering observations at 40\,GHz during the first two years of CLASS observations (Era 1). Other papers cover overall instrument performance \citep{appe19}, circular polarization \citep{Padilla2019, Petroff2019}, instrument stability (Harrington et al. 2019, in preparation), data pipeline (Parker et al. 2020, in preparation), and overall scientific results (Eimer et al. 2020, in preparation). A major goal of these first publications is to demonstrate the CLASS strategy for recovery of polarization at large angular scales from the ground. Looking ahead, the first 90~GHz telescope has been operational since 2018 June, and the 150/220~GHz telescope commenced observations in 2019 October. Future results with the multifrequency CLASS telescope array will constrain reionization and inflation.

\vskip 5.8mm plus 1mm minus 1mm
\vskip1sp
\section*{Acknowledgments}
\vskip4pt

We acknowledge the National Science Foundation Division of Astronomical Sciences for their support of CLASS under grant Nos. 0959349, 1429236, 1636634, and 1654494. We thank Johns Hopkins University President R. Daniels and Dean B. Wendland for their steadfast support of CLASS. The CLASS project employs detector technology developed in collaboration between JHU and Goddard Space Flight Center under several previous and ongoing NASA grants. Detector development work at JHU was funded by NASA grant No. NNX14AB76A. We thank scientists from NIST for their contributions to the detector and readout systems, including Johannes Hubmayr, Gene Hilton, and Carl Reintsema. B.P. is supported by the Fondecyt Regular Project No. 1171811 (CONICYT) and CONICYT- PFCHA Magister Nacional Scholarship 2016-22161360. P.F. thanks CONICYT for grants Anillo ACT-1417, QUIMAL 160009, and BASAL AFB-170002. P.F. also thanks the TICRA staff for fruitful discussion about the GRASP simulations. R.R. acknowledges partial support from CATA, BASAL grant AFB-170002, and Conicyt-FONDECYT through grant 1181620. R.D. thanks CONICYT for grant BASAL CATA AFB-170002. We acknowledge scientific and engineering contributions from Max Abitbol, Fletcher Boone, David Carcamo, Francisco Raul Espinoza Inostroza, Saianeesh Haridas, Connor Henley, Lindsay Lowry, Isu Ravi, Gary Rhodes, Daniel Swartz, Bingie Wang, Qinan Wang, Tiffany Wei, and Zi'ang Yan. We thank William Deysher, Maria Jose Amaral, and Chantal Boisvert for logistical support. We acknowledge productive collaboration with Dean Carpenter and the JHU Physical Sciences Machine Shop team. Part of this research project was conducted using computational resources at the Maryland Advanced Research Computing Center (MARCC). Part of this research project was conducted using computational resources of the Geryon-2 cluster at Centro de Astroingenier\'ia UC. We further acknowledge the very generous support of Jim and Heather Murren (JHU A\&S '88), Matthew Polk (JHU A\&S Physics BS '71), David Nicholson, and Michael Bloomberg (JHU Engineering '64). CLASS is located in the Parque Astron\'omico Atacama in northern Chile under the auspices of the Comisi\'on Nacional de Investigaci\'on Cient\'ifica y Tecnol\'ogica de Chile (CONICYT). Finally, we thank the anonymous referee of this paper for providing useful comments, which contributed to the final version of this paper.

\software{\texttt{IPython} \citep{ipython}, \texttt{numpy} \citep{numpy}, \texttt{scipy} \citep{scipy}, \texttt{matplotlib} \citep{matplotlib}, \texttt{healpy} \citep{healpix, healpy}, 
\texttt{PyEphem} \citep{pyephem}, \texttt{Astropy} \citep{astropy1, astropy2}, GRASP (\href{https://www.ticra.com/software/grasp/}{www.ticra.com/software/grasp/})}

\bibliography{main}

\appendix 

\section{Beam Parameters}
\label{app:beam_parameters}
Once the stacked instrument beams are available for each detector, basic beam parameters are measured as the fiducial values. Since the beam profile is not exactly Gaussian, we took a cross section of the beam map around the half amplitude and then measured properties of the cross section. The high resolution of the instrument beams provides enough pixels within a small range around the half amplitude. The pixels within the cross section form the shape of an ellipse. By measuring the ellipse, we obtained the well-defined baseline values of $\textrm{FWHM}_{\textrm{major}}$, $\textrm{FWHM}_{\textrm{minor}}$, and $\theta$ for each detector. 

During the analysis of each Moon scan, we used the instrument beams as templates for fitting. The fitting parameters become the scale factors along the major and minor axes and the major axis orientation correction angle (see Section~\ref{subsec:beam_analysis_method}). The deviations for the scale factors are normally $<$ 2\%, so we calculate the $\textrm{FWHM}_{\textrm{major}}$, $\textrm{FWHM}_{\textrm{minor}}$ values by multiplying the corresponding fiducial values by the scale factors. The $\theta$ value is then updated with the fitted correction angle. Those beam parameters are available for all of the detectors across all of the Moon scans. Together with the detector pointing offsets, mean values and uncertainties of the parameters are then estimated from the individual measurements. Table~\ref{tab:detector_info} shows the pointing and beam parameter results for the CLASS 40~GHz telescope in Era~1. The measured beam parameters have been corrected for the convolution of the Moon. 

\startlongtable
\centerwidetable
\begin{deluxetable*}{crrrrrrrr}
\tabletypesize{\footnotesize}
\tablecaption{Detector pointing and beam information.\label{tab:detector_info}}
\tablehead{\nocolhead{} & \nocolhead{} & \nocolhead{} & \twocolhead{$\mathrm{FWHM}_{\mathrm{major}}$ (deg)} & \twocolhead{$\mathrm{FWHM}_{\mathrm{minor}}$ (deg)} & \twocolhead{$\theta$ (deg)} \\
\colhead{Det. No.} & \colhead{$X_{\mathrm{offset}}$ (deg)} & \colhead{$Y_{\mathrm{offset}}$ (deg)} & \colhead{Measurement} & \colhead{Simulation} & \colhead{Measurement} & \colhead{Simulation} & \colhead{Measurement} & \colhead{Simulation}}

\startdata
$0$ & $4.7509 \pm 0.0003$ & $4.5533 \pm 0.0005$ & $1.6089 \pm 0.0003$ & $1.5558$ & $1.5105 \pm 0.0004$ & $1.4470$ & $64.88 \pm 0.08$ & $69.86$ \\
$2$ & $1.5885 \pm 0.0003$ & $4.4131 \pm 0.0004$ & $1.5784 \pm 0.0003$ & $1.5407$ & $1.4876 \pm 0.0003$ & $1.4627$ & $70.33 \pm 0.09$ & $81.76$ \\
$4$ & $1.5846 \pm 0.0003$ & $4.4124 \pm 0.0004$ & $1.5684 \pm 0.0003$ & $1.5412$ & $1.5041 \pm 0.0004$ & $1.4612$ & $82.30 \pm 0.08$ & $79.76$ \\
$5$ & $6.2041 \pm 0.0003$ & $7.4836 \pm 0.0006$ & $1.7238 \pm 0.0004$ & $1.5694$ & $1.5113 \pm 0.0006$ & $1.4150$ & $64.80 \pm 0.07$ & $73.09$ \\
$7$ & $6.1896 \pm 0.0004$ & $7.4756 \pm 0.0006$ & $1.6963 \pm 0.0004$ & $1.5710$ & $1.5002 \pm 0.0006$ & $1.4141$ & $66.71 \pm 0.07$ & $72.45$ \\
$8$ & $3.1282 \pm 0.0004$ & $7.3099 \pm 0.0006$ & $1.5804 \pm 0.0005$ & $1.5572$ & $1.4859 \pm 0.0007$ & $1.4310$ & $77.26 \pm 0.12$ & $80.09$ \\
$9$ & $3.1170 \pm 0.0003$ & $7.3086 \pm 0.0006$ & $1.5800 \pm 0.0004$ & $1.5600$ & $1.4623 \pm 0.0006$ & $1.4300$ & $85.54 \pm 0.07$ & $78.86$ \\
$11$ & $7.8538 \pm 0.0004$ & $4.8480 \pm 0.0006$ & $1.6566 \pm 0.0005$ & $1.5719$ & $1.4655 \pm 0.0007$ & $1.4234$ & $55.86 \pm 0.10$ & $64.79$ \\
$13$ & $0.0018 \pm 0.0003$ & $1.5696 \pm 0.0003$ & $1.5173 \pm 0.0004$ & $1.5216$ & $1.4774 \pm 0.0009$ & $1.4950$ & $70.29 \pm 0.09$ & $95.13$ \\
$14$ & $0.0018 \pm 0.0003$ & $1.5707 \pm 0.0004$ & $1.5353 \pm 0.0003$ & $1.5216$ & $1.5140 \pm 0.0005$ & $1.4950$ & $110.66 \pm 0.09$ & $84.87$ \\
$15$ & $3.2372 \pm 0.0003$ & $1.6546 \pm 0.0003$ & $1.5709 \pm 0.0002$ & $1.5373$ & $1.5275 \pm 0.0003$ & $1.4801$ & $48.82 \pm 0.10$ & $60.05$ \\
$16$ & $3.2384 \pm 0.0003$ & $1.6571 \pm 0.0003$ & $1.5846 \pm 0.0003$ & $1.5396$ & $1.5159 \pm 0.0003$ & $1.4769$ & $46.54 \pm 0.09$ & $59.20$ \\
$17$ & $6.4186 \pm 0.0003$ & $1.9077 \pm 0.0005$ & $1.5925 \pm 0.0004$ & $1.5586$ & $1.4881 \pm 0.0006$ & $1.4487$ & $53.74 \pm 0.09$ & $55.86$ \\
$18$ & $6.4113 \pm 0.0004$ & $1.8996 \pm 0.0004$ & $1.5744 \pm 0.0004$ & $1.5638$ & $1.4922 \pm 0.0006$ & $1.4461$ & $48.32 \pm 0.09$ & $55.58$ \\
$19$ & $9.5218 \pm 0.0004$ & $2.2902 \pm 0.0004$ & $1.6737 \pm 0.0005$ & $1.5856$ & $1.5110 \pm 0.0005$ & $1.4222$ & $41.63 \pm 0.08$ & $55.63$ \\
$21$ & $9.5249 \pm 0.0004$ & $2.3066 \pm 0.0005$ & $1.6699 \pm 0.0006$ & $1.5895$ & $1.4847 \pm 0.0006$ & $1.4193$ & $42.70 \pm 0.08$ & $55.42$ \\
$22$ & $8.2553 \pm 0.0004$ & $-3.8234 \pm 0.0004$ & $1.6437 \pm 0.0003$ & $1.5941$ & $1.5189 \pm 0.0004$ & $1.4404$ & $24.52 \pm 0.08$ & $35.12$ \\
$24$ & $3.2815 \pm 0.0004$ & $-1.5719 \pm 0.0004$ & $1.5873 \pm 0.0004$ & $1.5437$ & $1.5168 \pm 0.0004$ & $1.4848$ & $34.72 \pm 0.10$ & $33.49$ \\
$25$ & $3.2812 \pm 0.0003$ & $-1.5689 \pm 0.0003$ & $1.5654 \pm 0.0005$ & $1.5470$ & $1.5272 \pm 0.0005$ & $1.4828$ & $20.88 \pm 0.10$ & $34.49$ \\
$26$ & $6.5277 \pm 0.0004$ & $-1.2900 \pm 0.0004$ & $1.5768 \pm 0.0003$ & $1.5665$ & $1.4385 \pm 0.0004$ & $1.4546$ & $43.81 \pm 0.10$ & $41.94$ \\
$27$ & $6.5272 \pm 0.0004$ & $-1.2885 \pm 0.0004$ & $1.5492 \pm 0.0003$ & $1.5719$ & $1.4519 \pm 0.0005$ & $1.4520$ & $43.36 \pm 0.10$ & $42.86$ \\
$30$ & $9.7088 \pm 0.0005$ & $-0.8572 \pm 0.0005$ & $1.6062 \pm 0.0005$ & $1.5922$ & $1.4780 \pm 0.0006$ & $1.4303$ & $29.59 \pm 0.10$ & $44.89$ \\
$32$ & $9.7151 \pm 0.0005$ & $-0.8463 \pm 0.0004$ & $1.6253 \pm 0.0005$ & $1.5960$ & $1.4743 \pm 0.0007$ & $1.4269$ & $32.36 \pm 0.10$ & $45.02$ \\
$35$ & $4.9806 \pm 0.0005$ & $-4.2111 \pm 0.0005$ & $1.6403 \pm 0.0007$ & $1.5775$ & $1.5017 \pm 0.0012$ & $1.4603$ & $35.39 \pm 0.13$ & $27.42$ \\
$36$ & $1.6519 \pm 0.0004$ & $-4.4116 \pm 0.0004$ & $1.5955 \pm 0.0003$ & $1.5590$ & $1.5012 \pm 0.0004$ & $1.4783$ & $17.53 \pm 0.07$ & $9.78$ \\
$37$ & $1.6579 \pm 0.0004$ & $-4.4135 \pm 0.0004$ & $1.6138 \pm 0.0003$ & $1.5595$ & $1.5275 \pm 0.0004$ & $1.4784$ & $5.48 \pm 0.07$ & $12.30$ \\
$38$ & $6.7293 \pm 0.0005$ & $-6.7434 \pm 0.0005$ & $1.6659 \pm 0.0005$ & $1.6039$ & $1.5371 \pm 0.0006$ & $1.4449$ & $17.11 \pm 0.10$ & $24.71$ \\
$39$ & $6.7354 \pm 0.0005$ & $-6.7482 \pm 0.0005$ & $1.6729 \pm 0.0005$ & $1.6069$ & $1.5551 \pm 0.0006$ & $1.4422$ & $12.44 \pm 0.09$ & $26.13$ \\
$40$ & $3.3876 \pm 0.0005$ & $-7.0512 \pm 0.0005$ & $1.6118 \pm 0.0005$ & $1.5925$ & $1.5070 \pm 0.0006$ & $1.4556$ & $11.95 \pm 0.08$ & $14.13$ \\
$41$ & $3.3802 \pm 0.0004$ & $-7.0529 \pm 0.0005$ & $1.6682 \pm 0.0006$ & $1.5942$ & $1.5339 \pm 0.0007$ & $1.4548$ & $18.29 \pm 0.07$ & $15.21$ \\
$42$ & $-0.0278 \pm 0.0004$ & $-7.1503 \pm 0.0005$ & $1.6489 \pm 0.0005$ & $1.5879$ & $1.5119 \pm 0.0006$ & $1.4641$ & $176.55 \pm 0.06$ & $179.28$ \\
$43$ & $-0.0356 \pm 0.0005$ & $-7.1545 \pm 0.0005$ & $1.6360 \pm 0.0004$ & $1.5877$ & $1.5161 \pm 0.0006$ & $1.4641$ & $1.59 \pm 0.07$ & $0.76$ \\
$46$ & $-4.9865 \pm 0.0004$ & $-4.2341 \pm 0.0004$ & $1.5685 \pm 0.0003$ & $1.5750$ & $1.4512 \pm 0.0006$ & $1.4622$ & $144.66 \pm 0.09$ & $153.80$ \\
$47$ & $-1.6947 \pm 0.0004$ & $-4.4166 \pm 0.0004$ & $1.5776 \pm 0.0003$ & $1.5595$ & $1.5132 \pm 0.0005$ & $1.4784$ & $170.33 \pm 0.07$ & $167.70$ \\
$48$ & $-1.7009 \pm 0.0004$ & $-4.4177 \pm 0.0004$ & $1.5349 \pm 0.0003$ & $1.5595$ & $1.4688 \pm 0.0004$ & $1.4784$ & $4.15 \pm 0.07$ & $167.70$ \\
$49$ & $-6.7368 \pm 0.0004$ & $-6.7461 \pm 0.0004$ & $1.6056 \pm 0.0003$ & $1.6069$ & $1.5021 \pm 0.0005$ & $1.4423$ & $166.89 \pm 0.07$ & $154.01$ \\
$51$ & $-6.7140 \pm 0.0004$ & $-6.7435 \pm 0.0005$ & $1.6147 \pm 0.0003$ & $1.6039$ & $1.4962 \pm 0.0004$ & $1.4449$ & $159.37 \pm 0.08$ & $155.29$ \\
$52$ & $-3.4153 \pm 0.0004$ & $-7.0564 \pm 0.0005$ & $1.5916 \pm 0.0003$ & $1.5942$ & $1.4827 \pm 0.0009$ & $1.4548$ & $160.42 \pm 0.07$ & $164.79$ \\
$53$ & $-3.4273 \pm 0.0005$ & $-7.0489 \pm 0.0005$ & $1.5238 \pm 0.0002$ & $1.5925$ & $1.4309 \pm 0.0007$ & $1.4556$ & $176.67 \pm 0.08$ & $165.87$ \\
$55$ & $-8.2570 \pm 0.0018$ & $-3.8327 \pm 0.0022$ & $1.6460 \pm 0.0022$ & $1.5971$ & $1.5154 \pm 0.0023$ & $1.4372$ & $153.01 \pm 0.18$ & $144.36$ \\
$57$ & $-0.0120 \pm 0.0005$ & $-1.6563 \pm 0.0006$ & $1.5618 \pm 0.0005$ & $1.5296$ & $1.5377 \pm 0.0007$ & $1.5001$ & $18.11 \pm 0.12$ & $175.05$ \\
$58$ & $-0.0173 \pm 0.0008$ & $-1.6536 \pm 0.0011$ & $1.5706 \pm 0.0008$ & $1.5296$ & $1.5149 \pm 0.0012$ & $1.5001$ & $44.37 \pm 0.13$ & $4.95$ \\
$59$ & $-3.2793 \pm 0.0005$ & $-1.5919 \pm 0.0005$ & $1.6648 \pm 0.0007$ & $1.5470$ & $1.6102 \pm 0.0008$ & $1.4828$ & $136.24 \pm 0.12$ & $145.51$ \\
$60$ & $-3.2867 \pm 0.0004$ & $-1.5968 \pm 0.0004$ & $1.6278 \pm 0.0006$ & $1.5437$ & $1.6041 \pm 0.0006$ & $1.4848$ & $151.33 \pm 0.09$ & $146.51$ \\
$61$ & $-6.5208 \pm 0.0004$ & $-1.3204 \pm 0.0004$ & $1.5865 \pm 0.0004$ & $1.5719$ & $1.4611 \pm 0.0005$ & $1.4522$ & $132.66 \pm 0.09$ & $137.10$ \\
$62$ & $-6.5318 \pm 0.0006$ & $-1.3274 \pm 0.0008$ & $1.5537 \pm 0.0006$ & $1.5665$ & $1.4463 \pm 0.0007$ & $1.4546$ & $134.88 \pm 0.11$ & $138.06$ \\
$63$ & $-9.6781 \pm 0.0011$ & $-0.8473 \pm 0.0014$ & $1.5862 \pm 0.0010$ & $1.5965$ & $1.4736 \pm 0.0008$ & $1.4269$ & $146.90 \pm 0.09$ & $135.05$ \\
$65$ & $-9.6930 \pm 0.0027$ & $-0.8479 \pm 0.0040$ & $1.2789 \pm 0.0018$ & $1.5917$ & $1.1330 \pm 0.0019$ & $1.4304$ & $152.60 \pm 0.12$ & $135.08$ \\
$66$ & $-7.7996 \pm 0.0009$ & $4.8586 \pm 0.0010$ & $1.6241 \pm 0.0009$ & $1.5761$ & $1.4778 \pm 0.0009$ & $1.4195$ & $123.97 \pm 0.07$ & $115.44$ \\
$68$ & $-3.1899 \pm 0.0003$ & $1.6621 \pm 0.0005$ & $1.7439 \pm 0.0015$ & $1.5396$ & $1.6818 \pm 0.0014$ & $1.4769$ & $128.65 \pm 0.10$ & $120.80$ \\
$69$ & $-3.1886 \pm 0.0003$ & $1.6541 \pm 0.0004$ & $1.5726 \pm 0.0005$ & $1.5373$ & $1.5454 \pm 0.0005$ & $1.4801$ & $125.32 \pm 0.06$ & $119.95$ \\
$70$ & $-6.3913 \pm 0.0005$ & $1.9086 \pm 0.0009$ & $1.5395 \pm 0.0007$ & $1.5639$ & $1.4347 \pm 0.0007$ & $1.4462$ & $125.21 \pm 0.10$ & $124.34$ \\
$71$ & $-6.3957 \pm 0.0007$ & $1.8991 \pm 0.0009$ & $1.4978 \pm 0.0007$ & $1.5586$ & $1.4375 \pm 0.0008$ & $1.4487$ & $120.94 \pm 0.08$ & $124.14$ \\
$72$ & $-9.4901 \pm 0.0007$ & $2.3126 \pm 0.0012$ & $1.7453 \pm 0.0013$ & $1.5893$ & $1.5632 \pm 0.0013$ & $1.4189$ & $134.58 \pm 0.10$ & $124.51$ \\
$76$ & $-9.4988 \pm 0.0011$ & $2.3003 \pm 0.0010$ & $1.6973 \pm 0.0017$ & $1.5855$ & $1.5210 \pm 0.0016$ & $1.4225$ & $132.28 \pm 0.06$ & $124.38$ \\
$77$ & $-4.7017 \pm 0.0003$ & $4.5765 \pm 0.0005$ & $1.5434 \pm 0.0004$ & $1.5564$ & $1.4784 \pm 0.0006$ & $1.4448$ & $106.96 \pm 0.09$ & $111.68$ \\
$79$ & $-1.5491 \pm 0.0003$ & $4.4137 \pm 0.0004$ & $1.5538 \pm 0.0003$ & $1.5412$ & $1.4848 \pm 0.0005$ & $1.4612$ & $90.67 \pm 0.09$ & $100.24$ \\
$80$ & $-1.5508 \pm 0.0003$ & $4.4189 \pm 0.0005$ & $1.5299 \pm 0.0003$ & $1.5407$ & $1.4503 \pm 0.0004$ & $1.4627$ & $100.64 \pm 0.09$ & $98.24$ \\
$81$ & $-6.1277 \pm 0.0004$ & $7.5055 \pm 0.0006$ & $1.6578 \pm 0.0005$ & $1.5710$ & $1.4697 \pm 0.0005$ & $1.4141$ & $111.08 \pm 0.09$ & $107.55$ \\
$82$ & $-6.1289 \pm 0.0004$ & $7.5143 \pm 0.0007$ & $1.6664 \pm 0.0006$ & $1.5690$ & $1.4622 \pm 0.0006$ & $1.4152$ & $113.47 \pm 0.08$ & $106.81$ \\
$83$ & $-3.0592 \pm 0.0003$ & $7.3179 \pm 0.0005$ & $1.5820 \pm 0.0004$ & $1.5600$ & $1.4628 \pm 0.0008$ & $1.4302$ & $99.26 \pm 0.08$ & $101.19$ \\
$84$ & $-3.0649 \pm 0.0002$ & $7.3249 \pm 0.0005$ & $1.5826 \pm 0.0004$ & $1.5572$ & $1.4553 \pm 0.0004$ & $1.4310$ & $105.47 \pm 0.08$ & $99.91$ \\
$85$ & $0.0329 \pm 0.0003$ & $7.2403 \pm 0.0005$ & $1.5515 \pm 0.0003$ & $1.5592$ & $1.4349 \pm 0.0005$ & $1.4318$ & $92.96 \pm 0.08$ & $90.76$ \\
$86$ & $0.0405 \pm 0.0003$ & $7.2379 \pm 0.0006$ & $1.5634 \pm 0.0004$ & $1.5592$ & $1.4407 \pm 0.0005$ & $1.4318$ & $83.12 \pm 0.08$ & $89.24$ \\
\enddata
\end{deluxetable*}

\section{Electromagnetic Simulation Using GRASP}
\label{app:grasp_em}
To simulate the 40\,GHz telescope, we used the General Reflector Antenna Software Package (GRASP).\footnote{\url{https://www.ticra.com/software/grasp/}} GRASP is composed of a Computer Aided Design (CAD) interface and an analysis module that solves Maxwell's equations given the CAD model and a source. A rendering of the CAD model for the CLASS 40\,GHz telescope is presented in Figure~\ref{fig:tele_beam}. GRASP uses approximation methods to solve Maxwell equations such as Physical Optics (PO) and Physical Theory of Diffraction (PTD). More accurate solutions can be obtained using Method Of Moments (MoM) methods, at the cost of increased computational requirements. 

To efficiently simulate the telescope, we followed a sequential, time-reversed approach where feedhorns were the primary radiating source, and every optical element was restricted to be illuminated only by the one immediately preceding it. For practical reasons, the optical elements were organized into the re-imaging optics block, warm optics, and the comoving enclosure. Propagation of light through re-imaging and warm optical components was calculated using PO simulation method, while more complex interactions between the VPM mirror and the forebaffle were accomplished using a more advanced approach including a Plane Wave Expansion (PWE) and MoM. 

\subsection{Feedhorn}
As described in \cite{zeng10}, the CLASS 40\,GHz telescope uses smooth-walled feedhorns as beam-forming elements. An accurate model of near- and far-field electromagnetic fields from the feedhorn was obtained from \cite{zeng10}. We validated the predictions from this model at multiple frequencies against a GRASP simulation of the feedhorn (computed MoM) and measurements carried out at the anechoic chamber at the NASA Goddard Space Flight Center. The comparison between the model predictions, MoM simulation, and measurements was performed by comparing the best-fit parameters of the feedhorn far-field beam map to a Gaussian template. This comparison yielded excellent agreement, with the Gaussian beam parameters having relative deviations of less than $2\%$. Given the negligible differences, we used the model described in \cite{zeng10} to obtain beam parameters for the 40\,GHz feedhorn beam at multiple frequencies, covering the the bandpass. This allowed us to efficiently perform a broadband simulation of the 40\,GHz receiver.

The focal plane consists of 36 feedhorns distributed on a flat surface. The positions of individual feedhorns were obtained from the mechanical design of the focal plane. Each feedhorn is electromagnetically coupled to a pair of TES bolometers, which are oriented $+45$\dg{} and $-45$\dg{} with respect to the optical plane of the telescope. This behavior was taken into consideration by rotating the polarization basis of the beam radiated around the feedhorn axis by $\pm 45$\dg{}, depending on the type of detector being simulated. 

\subsection{Re-imaging Optics}
The re-imaging optics in the GRASP model are comprised of two cryogenic lenses and a 4\,K cold stop. The lenses were drawn according to the parameters given in \citet[Table~3]{eime12}. The refractive index of the lenses was set to $1.564$, expected from high-density polyethylene (HDPE) at cryogenic temperatures. The cold stop was modeled as a circular aperture on an infinitely large, perfect electrical conductor (PEC) plane. We used PO to propagate the fields from the feedhorns through the re-imaging optics. Computational constraints prevented us from taking into account internal reflections in the cryogenic camera, so the simulation pipeline models lenses as if they were mounted on a PEC plane with a circular aperture. In practice, the reflected stray light is effectively absorbed by the cryogenic baffling, blackened by conductive powder loaded epoxy.

\subsection{Warm Optical Components}
In a time-reversed way, the cryogenic camera radiates onto the secondary mirror, which reflects the fields onto the primary mirror, which in turn redirects the fields onto the VPM mirror. The primary and secondary mirrors are sections of ellipsoids; see \citet[Table~1 and Table~2]{eime12}, Tables 1 and 2 for parameters. The VPM mirror was drawn as a flat mirror with a circular rim. For simplicity, all mirrors were modeled as PEC. It is important to mention that the interaction between the VPM grid and mirror was not included in the simulations, as the the VPM is a complex electromagnetic system that requires specialized treatment to capture the microinteractions between the wire grid, the mirror, and the rest of the optics. 

\subsection{Comoving Enclosure and Forebaffle}
The forebaffle was designed to limit stray light from bright sources, including the ground, the Sun, etc. The main body is a conic section made of aluminum. At the top of the forebaffle, a flare section is designed to mitigate diffraction at the top. The flare section is not included in the simulation due to computational difficulties. The aperture with a smaller radius (closer to the VPM) interfaces with the telescope comoving enclosure. In GRASP, the forebaffle was modeled as a perfectly conducting conical section. Care was taken to correctly model the interactions between the inner walls of the forebaffle with the rest of the optical elements. This was achieved by performing a PWE at the telescope enclosure-forebaffle interface. This expansion provides the required accurate representation of the near fields. The output of the PWE was used as an input to the MoM solver of GRASP, which calculated the surface currents on the inner walls of the forebaffle. These currents were used to compute the electromagnetic fields in the far field, and, hence, the contribution of the forebaffle ``spill'' to the beam. Finally, the comoving enclosure surrounding the 40\,GHz optics was drawn using the technique described in \cite{pudd19}. Spill of optical elements on the comoving enclosure might cause sidelobes in the beam. Those contributions were calculated by nonsequential PO simulations combined with MoM.

\section{Beam Profile Modeling}
\label{app:beam_modeling}
Together with the two-dimensional map deconvolution procedure in the main text, we also perform beam profile modeling to remove the effect from the finite size of the Moon. We use the cosmology beam in the analysis, assuming rotational symmetry.

To begin with the Moon temperature model, we take a measured Moon map, $\tilde{T}$. This signal can be treated as the convolution between the Moon as a uniform disk of temperature $T$ with angular radius $a$, and a symmetric beam $B$:
\begin{equation}\label{beam-moon_convolution}
\tilde{T} = T \ast B + N\,, 
\end{equation} where a small noise component $N$ has been added.
This convolution can be represented in the $k$-domain by applying the Fourier projector $\mathcal{F}^{-1}(\mathcal{F}(\cdot))$ and using the Fourier representation of a two-dimensional disk, $2\pi a^{2} J_{1}(ka)/ka$ with $J_{1}(x)$ the Bessel function of the first kind; thus, the two-dimensional beam map is expressed as
\begin{equation}
\tilde{T}(\theta,\phi) = \frac{a}{(2\pi)}\int_{\mathbb{R}^2} d\mathbf{k}\, e^{i \mathbf{k}\cdot\mathbf{x}} 
  \frac{J_{1} \left(k a \right)}{k}B(k) + N(\theta,\phi).\\
\end{equation}
To reduce the integral, it is suitable to use the rotational symmetry of the convolved signal by performing the following substitutions: $x = \theta\cos \phi$, $y = \theta \sin \phi$ and $k_x = k \cos \xi$, $k_y = k \sin \xi$; the identity $\cos\xi \cos\phi
+\sin \xi \sin \phi = \cos(\xi-\phi)$; and making a change of variables $\xi-\phi = \psi$. So,
\begin{equation}
 \tilde{T}(\theta,\phi) = \frac{a}{(2\pi)}\int_{0}^{\infty} \int_{0}^{2\pi}   dk\, d\psi e^{ik\theta\cos \psi} J_{1} \left(ka \right)B(k) + N(\theta,\phi).
\end{equation}
Additionally, by using the integral representation of the zeroth-order Bessel function of the first kind, $2\pi J_{0}(z) = \int_{0}^{2\pi} d\psi\, e^{\pm iz \cos \psi}$ and taking the angular average of the noise $\langle N(\mathbf{x})\rangle_{\phi} = N(\theta)$, the observed temperature map is only a function of $\theta$:
\begin{equation}\label{reduced_moon_map}
 \tilde{T}(\theta) = 2\pi a \int_{0}^{\infty} dk\, J_{0}(k\theta) J_{1} \left( ka \right)B_{0}(k) + N(\theta),
\end{equation}
where $B_{0}(k)$ is the zeroth Hankel transform of the beam defined by $B_{0}(k) = \int_{0}^{\infty} d\theta\, \theta B(\theta) J_{0}(k\theta)$. Equation (\ref{reduced_moon_map}) constitutes an analytical expression for the Moon-beam convolution model. Notice that it has been reduced from a two-dimensional convolution (two integrals) to a single one-dimensional integral expression by using the rotational symmetry of both functions. This helps to reduce the computational complexity of the numerical convolution and the signal fitting process.

\subsection{Beam Fitting}
Since the contributions from non-Gaussian components of the symmetrized beam affect the CMB analysis, it is necessary to quantify and parameterize these deviations with some complete basis. The natural way to capture these effects is by projecting the symmetrized beam into the same Hermite basis as the quantum harmonic oscillator since the basis functions parameterize these deviations from Gaussianity and form an orthonormal basis. Thus, the Hermite expansion is given by
\begin{equation}\label{hermite_functions}
 B(\theta) = \sum_{n=0}^{ N_{\textrm{max} }} a_{2n} \frac{H_{2n}\left(\theta/\sigma_{b}\right)}{ \sqrt{2^{2n} (2n)! \sqrt{\pi}}}  \exp\left( -\frac{1}{2}\frac{\theta^2}{\sigma_{b}^2}  \right),
\end{equation}
where $\theta$ is the angular distance from the beam center, $\sigma_{b}$ corresponds to the standard deviation of the Gaussian component, and $N_{\textrm{max}}$ corresponds to the maximum number of Hermite functions implemented. Because of the rotational symmetry, only even Hermite functions are included. Hermite components higher than one parameterize the small features that deviate from Gaussianity. This basis has already been implemented in \cite{Page:2003eu}. Combining Equation~(\ref{hermite_functions}) and Equation~(\ref{reduced_moon_map}), the temperature map can be expanded linearly as:
\begin{equation}\label{t_tilde}
\tilde{T}(\theta) = \sum_{n=0}^{N_{\textrm{max}}} a_{2n} \tilde{T}_{2n}(\theta) + N(\theta).
\end{equation}
The $\tilde{T}_{2n}(\theta)$ corresponds to the temperature contribution of $B_{0,2n}(k)$, which is the 2D-Hankel transformation of $B_{2n}(\theta)$ defined by:
\begin{equation}\label{t_2j}
\tilde{T}_{2n}(\theta) = 2\pi a \int_{0}^{\infty} dk\, J_{0}(k\theta) J_{1} \left( ka \right)B_{0,2n}(k).
\end{equation}
Therefore, if the set of coefficients $a_{2n}$ is found, the beam shape, Equation~(\ref{hermite_functions}), is determined. To compute the Hankel transform of $B_{2n}$ functions and convolve analytically the basis, Equation~(\ref{t_2j}), to get Equation~(\ref{t_tilde}), we can use the fact that the even Hermite polynomials are composed of exclusively even monomials, that is to say, $(\theta/\sigma_{b})^{2j} \subseteq H_{2n}(\theta/\sigma_{b}) $ with $ j\leqslant n$. Then the Hankel transform of the Hermite functions are
\begin{align}
\label{convolved_basis}
\mathcal{H}_{0} &\left[\left(\frac{\theta} {\sigma_{b}}\right)^{2j}\exp\left( -\frac{\theta^2}{2 \sigma_{b}^2} \right)\right](k)\nonumber\\ 
&=
2^{j} \sigma_{b}^2 \Gamma \left(j+1\right) \cdot _{1} F_{1}\left(j+1;1; \frac{-\sigma_{b}^2 k^2}{2} \right), 
\end{align}
where $\mathcal{H}_{0}$ represent the zeroth Hankel transform, and $_{1}F_{1}(a,b,z)$ are the \textsl{confluent hypergeometric functions of the first kind} \citep{bateman1954tables}. This allows us to obtain an analytical expression in Fourier representation for different Hermite modes; as a consequence, each component of the convolved basis, $\tilde{T}_{2n}(\theta)$, is determined exactly as an integral representation of known functions.

\begin{figure}[h]
    \centering
    \includegraphics[width=0.8\linewidth]{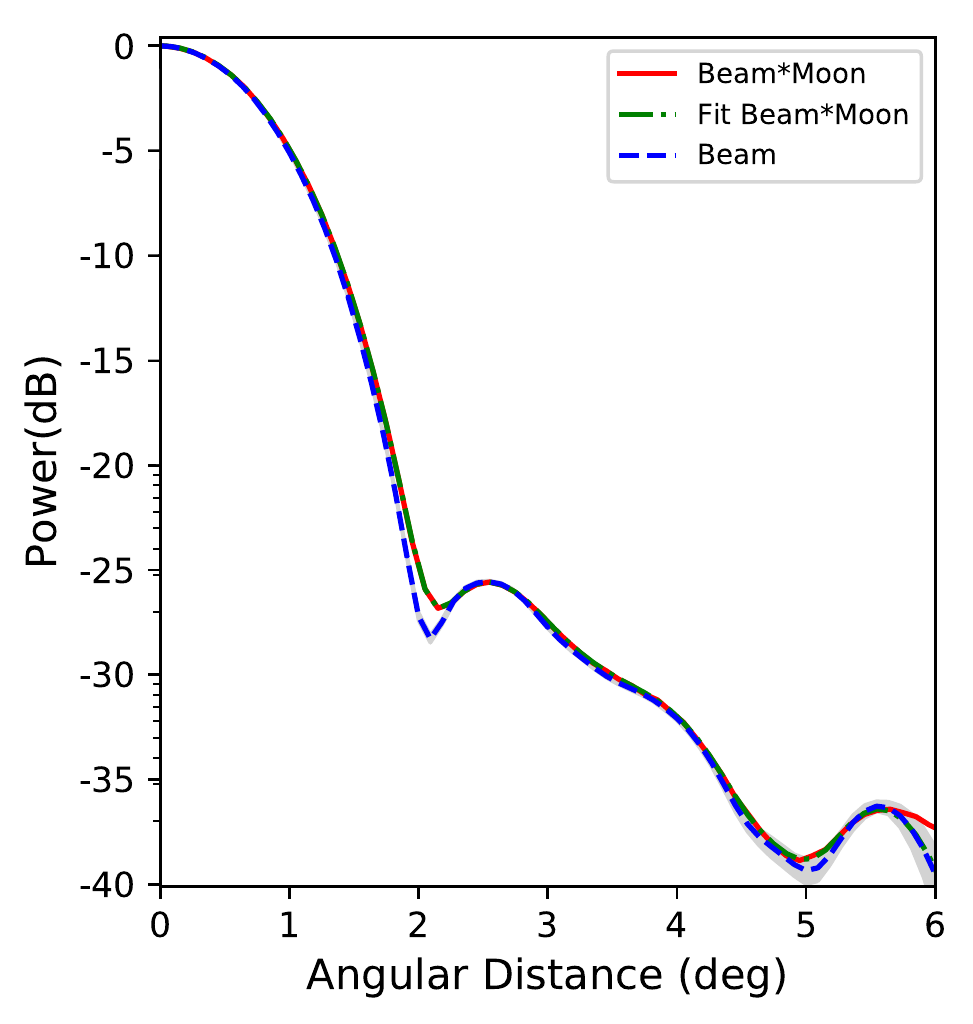}
    \caption{Beam profile: the red line represents the symmetrized convolved signal between the Moon and the beam ($T*B$). The green line is the fit of the convolved signal using Equation~(\ref{normalized_temp_profile}), whereas the blue dashed line represents the deconvolved beam. The gray band shows the uncertainty of the beam profile.}
    \label{beam_deconvolution}
\end{figure}

Given this analytical simplification and in order to obtain the beam shape from Moon scan data while avoiding overfitting, we need to find a finite number of Hermite functions to fit the beam accordingly; this number can be precisely estimated if we note that the $2n$-Hermite functions have a maximum at $\theta_{n,\textrm{max}} =\pm \sigma_{b}\sqrt{2n}$, and that they decay in a Gaussian manner. If we want to fit the profile out to $\sim 7.0$\dg{} with $\sigma_{b}\sim 0.65$\dg{}, it implies $\left \lfloor 2N_{\textrm{max}} \right \rfloor = 44$, imposing an upper limit, $N_{\textrm{max}} \leqslant 22$, for the beam expansion.
Since we are interested only in the beam shape, it is suitable to normalize the stacked map at $\theta=0$ to unity. If $N(0) \ll  \tilde{T}(0)$, then    
\begin{equation}\label{temperature_template}
\tilde{t}(\theta) =  \frac{\int_{0}^{\infty} dk\, J_{0}(k\theta) J_{1} \left(ak \right)B_{0}(k)}{\int_{0}^{\infty} dk J_{1} \left(ak \right)B_{0}(k)} + n(\theta),
\end{equation}
with $n(\theta) = N(\theta)/\tilde{T}(0)$ and $B_{0}(k)$, the 2D-Hankel transform of the beam. 
Using the convolved Hermite basis (Equation~(\ref{t_tilde})), Equation~(\ref{temperature_template}) can be expanded in terms of the components of this basis as
\begin{equation}
\tilde{t}(\theta) = \frac{\sum_{n=0}^{N_{\textrm{max}}} a_{2n}\tilde{T}_{2n}(\theta)}{\sum_{n=0}^{N_{max}} a_{2n}\tilde{T}_{2n}(0)} + n(\theta).
\end{equation}
The above expression is symmetric under a rescaling transformation $\tilde{T} \rightarrow a \tilde{T}$. Therefore, to avoid a scale degeneracy in the set coefficient $\left \{a_{2n}\right\}_{n=0}^{N_{\textrm{max}}}$, it is suitable to choose one of them to be unity, for instance, $a_{0}$, and proceed with the fitting procedure; thus,
\begin{equation}\label{normalized_temp_profile}
\tilde{t}(\theta) = \frac{\tilde{T}_{0}(\theta)+\sum_{n=1}^{N_{\textrm{max}}} a_{2n}\tilde{T}_{2n}(\theta)}{\tilde{T}_{0}(0) + \sum_{n=1}^{N_{\textrm{max}}} a_{2n}\tilde{T}_{2n}(0)} + n(\theta).
\end{equation}
The above expression gives the fitting coefficients for the set $a_{2n}$ and its respective covariance matrix $\Sigma_{a,nn'}$. Figure~\ref{beam_deconvolution} displays the symmetrized convolved beam profile, the fit to this profile, and the deconvolved beam profile.

\subsection{Beam Window Function}
\begin{figure}[h]
    \centering
    \includegraphics[width=1.0\linewidth]{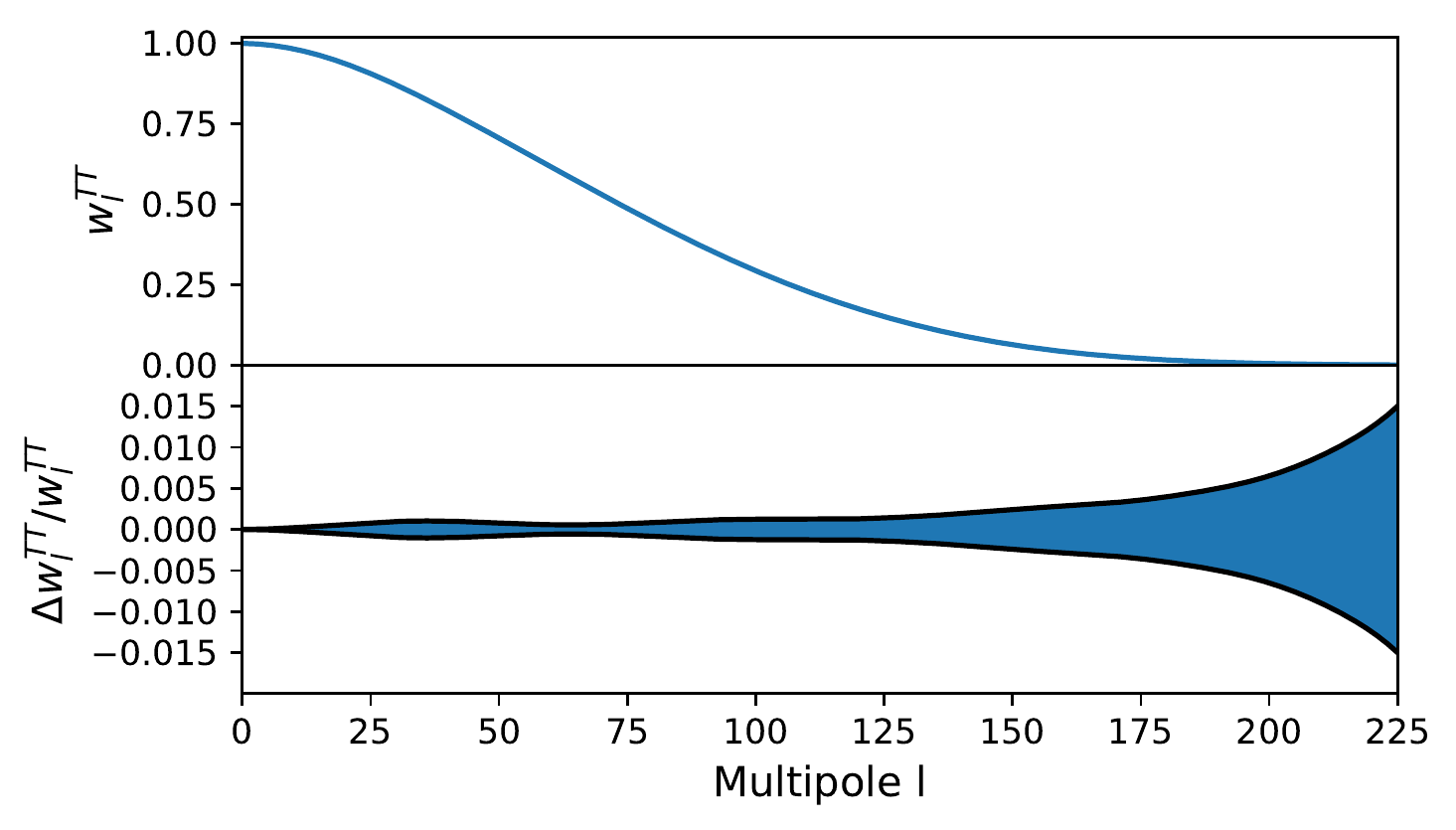}
    \caption{Temperature-temperature window function: The upper panel show the $\ell$ dependence of the window function acting as a low-pass filter. The bottom panel shows its fractional uncertainty.}
    \label{TT_window function}
    \end{figure}

With the beams already characterized, the next step is obtaining their associated beam window functions analytically. For a solid-angle normalized azimuthally symmetric beam $b(\theta)$, its harmonic representation is reduced to

\begin{equation}
    b_{\ell} = \int d\Omega\; b(\theta)P_{\ell}(\cos \theta).
\end{equation}
The above expression defines the beam response function. 
After deconvolving the Moon contributions from the beam, we can construct its beam window function as $b_{\ell}^{2}$. Figure~\ref{TT_window function} shows the temperature window function with its fractional uncertainty included, consistent with the results in the main text.

\end{document}